\documentclass{emulateapj}

\def\puncspace{\ifmmode\,\else{\ifcat.\C{\if.\C\else\if,\C\else\if?\C\else%
\if:\C\else\if;\C\else\if-\C\else\if)\C\else\if/\C\else\if]\C\else\if'\C%
\else\space\fi\fi\fi\fi\fi\fi\fi\fi\fi\fi}%
\else\if\empty\C\else\if\space\C\else\space\fi\fi\fi}\fi}
\def\SP{\let\\=\empty\futurelet\C\puncspace }

\def\deg{$^\circ$\ }

\def\h1{$h^{-1}$}
\def\z{$z\sim~$}
\def\deg{\ifmmode^\circ\kern- .38em\;\else$^\circ\kern- .38em\;$\fi}
\def\solar{\ifmmode_{\mathord\odot}\;\else$_{\mathord\odot}\;$\fi}
\def\amin{\ifmmode^\prime\kern- .38em\;\else$^\prime\kern- .38em\;$\fi}
\def\asec{\ifmmode^{\prime\prime}\kern- .38em\;\else$^{\prime\prime}
    \kern- .38em\;$\fi}
\def\s.{\kern+ .1em\lower 0.0ex\hbox{$\buildrel ^{\prime\prime} \over 
    {\rm .} \kern- .08em$}} %use for arcsec
\def\m.{\kern+ .1em\lower 0.0ex\hbox{$\buildrel ^{\prime} \over 
    {\rm .} \kern- .08em$}} %use for arcmin
\def\grad{\kern+ .1em\lower 0.0ex\hbox{$\buildrel {\rightharpoonup} \over 
    {\nabla} \kern- .08em$}} %use for grad
\def\div{\kern+ .1em\lower 0.0ex\hbox{$\buildrel {\rightharpoonup} \over 
    {\nabla} \kern- .08em$}\cdot} %use for div
\def\ltsim{\lower 0.5ex\hbox{$\; \buildrel < \over \sim \;$}} %use for ltoversim
\def\gtsim{\lower 0.5ex\hbox{$\; \buildrel > \over \sim \;$}} %use for gtoversim

\begin{document}
\title{Galaxy Luminosity Functions to $z \sim 1$: 
DEEP2 vs.~COMBO-17 and Implications for Red Galaxy Formation\altaffilmark{1}} 
\author{S. M. Faber \altaffilmark{2},
  C. N. A. Willmer\altaffilmark{2,3},
  C. Wolf \altaffilmark{4}
  D. C. Koo\altaffilmark{2},  
  B. J. Weiner\altaffilmark{5}, 
  J. A. Newman\altaffilmark{6,7},
  M. Im\altaffilmark{8},
  A. L. Coil,\altaffilmark{9}
  C. Conroy,\altaffilmark{9},
  M. C. Cooper\altaffilmark{9},
  M. Davis\altaffilmark{9}
  D. P. Finkbeiner\altaffilmark{10},
  B. F. Gerke\altaffilmark{9},
  K. Gebhardt\altaffilmark{7,12},
  E. J. Groth\altaffilmark{13},
  P. Guhathakurta\altaffilmark{2}, 
  J. Harker\altaffilmark{2},
  N. Kaiser\altaffilmark{14},
  S. Kassin\altaffilmark{2}, 
  M. Kleinheinrich\altaffilmark{15},
  N. P. Konidaris\altaffilmark{2},
  L. Lin\altaffilmark{2,16},
  G. Luppino\altaffilmark{14},
  D. S. Madgwick\altaffilmark{6, 7, 8},
  K. Meisenheimer\altaffilmark{15},
  K. G. Noeske\altaffilmark{2},
  A. C. Phillips\altaffilmark{2}, 
  V. L. Sarajedini\altaffilmark{17}, 
  L. Simard\altaffilmark{18},
  A. S. Szalay\altaffilmark{19},
  N. P. Vogt\altaffilmark{20},
  R. Yan\altaffilmark{8}}
\altaffiltext{1}{Based on observations taken at the W. M. Keck
  Observatory.}
% which is operated jointly by the University of
%  California and the California Institute of Technology, and on
%  observations made with the NASA/ESO {\it{Hubble Space Telescope}},
%  obtained from the data archives at the Space Telescope Science
%  Institute, which is operated by the Association of Universities for
%  Research in Astronomy, Inc., under NASA contract NAS5-26555, and
%  from the Canadian Astronomy Data Centre.}
\altaffiltext{2}{UCO/Lick Observatory, University of California}
\altaffiltext{3}{On leave from Observat\'orio Nacional, Rio de
  Janeiro, Brazil}
\altaffiltext{4}{Department of Physics, Oxford University}
\altaffiltext{5} {Department of Physics, University of Maryland}
\altaffiltext{6}{Lawrence Berkeley Laboratory, Berkeley}
\altaffiltext{7}{Hubble Fellow}
\altaffiltext{8}{Astronomy Program, School of Earth and Environmental
  Sciences, Seoul National University, Seoul, South Korea}
\altaffiltext{9}{Deparment of Astronomy, University of California, Berkeley}
\altaffiltext{10} {Department of Astrophysics, Princeton University}
%\altaffiltext{11} {Department of Physics,  University of California, Berkeley}
\altaffiltext{12}{Department of Astronomy, University of Texas, Austin}
\altaffiltext{13} {Department of Physics, Princeton University}
\altaffiltext{14} {Institute for Astronomy, Honolulu}
\altaffiltext{15} {Max-Planck-Institut fur Astronomie, Heidelberg, Germany}
\altaffiltext{16} {Department of Physics, National Taiwan University}
\altaffiltext{17}{Astronomy Department, University of Florida}
\altaffiltext{18}{Herzberg Institute of Astrophysics, National Research
  Council of Canada}
\altaffiltext{19}{Department of Physics, The Johns Hopkins University}
\altaffiltext{20}{Department of Astronomy, New Mexico State
  University}

\begin{abstract}
The DEEP2 and COMBO-17 surveys are used to study the evolution of
the luminosity function of red and blue galaxies to $z \sim 1$. Schechter
function fits show that, since $z = 1$, $M^*_B$ dims by $\sim$ 1.3 mag
per unit redshift for both color classes, $\phi^*$ of blue galaxies
shows little change, while $\phi^*$ for red galaxies has formally
nearly quadrupled.  At face value, the number density of blue galaxies
has remained roughly constant since $ z = 1$, whereas that of red
galaxies has been rising.  Luminosity densities support both
conclusions, but we note that most red-galaxy evolution occurs between 
our data and local surveys and in our highest redshift bin, where the
data are weakest.  We discuss the implications of having most red
galaxies emerge after $z = 1$ from precursors among the blue
population, taking into account the properties of local and distant
E/S0s.  We suggest a ``mixed'' scenario in which some blue galaxies
have their star-formation quenched in gas-rich mergers, migrate to the
red sequence with a variety of masses, and merge further on the red
sequence in one or more purely stellar mergers.  E/S0s of a given mass
today will have formed via different routes, in a manner that may help
to explain the fundamental plane and other local scaling laws.

\end{abstract}

\keywords{Galaxies: distances and redshifts -- galaxies: luminosity
  function -- galaxies: evolution}

\section{Introduction}

A major handicap in lookback studies of galaxy evolution is
the inability to follow the evolution of any one galaxy
over time.  Instead, we see only snapshots of the galaxy population
at different epochs, and it is difficult to identify objects
at one epoch with their precursors and descendants
at different epochs.
One of the most important tools to solve this problem
is precision counts of galaxies, which can
quantify the ``flow" of galaxies in parameter
space as masses, morphologies, and stellar populations change.
The luminosity function of galaxies was historically the
first such tool, but the concept is rapidly being
broadened to include counts as a function of mass, internal velocity,
color, and other parameters.  

A further difficulty is caused by the fact that nearly all
functions known to date are {\it unimodal} and lack clear features that
demarcate one class of galaxies from another.  In other words, galaxies
tend to populate one big ``cloud" in most parameter spaces rather than
separate clumps.  This makes interpretation difficult, 
as sub-counts depend on how 
boundaries within these clouds are defined, and it is not clear
whether (or how) these boundaries should be adjusted to
follow galaxy evolution.  As a result,
we often cannot tell whether a change in the number of galaxies 
in any particular
bin is due to a change in the overall number of galaxies
or to the motion of galaxies in and out of
that bin from neighboring bins.
The latter problem is further exacerbated by the fact that samples
are usually size- or brightness-limited and population 
numbers on the other side of these
limits are not known.  Finally, uncertain
errors can smear counts from one bin to another.
All of these problems will be modeled when a full
theory of galaxy formation is available to predict how galaxies evolve in 
in every measured parameter.  In the meantime, it is hard to 
break the population into well motivated sub-populations, and  
we therefore lack the means to obtain more finely divided knowledge.

Amidst this sea of unimodal functions, one function
stands out on account of its
uniquely bimodal character, namely, the color function.  This is visible 
in the color-magnitude
diagram, where early-type E/S0s populate 
a narrow red sequence that is separated from
bluer, star-forming spirals by a shallow valley
(Strateva et al. 2001; Hogg et al. 2003; 
Balogh et al. 2004; Baldry et al. 2004
and references therein).  A similar division 
extends to at least $z\sim 1$ (Im et al. 2002; Bell et
al. 2004b; Weiner et al. 2005, Willmer et al.
2005) and possibly beyond (Giallongo et al. 2005).  
A bimodal distribution is also seen
in other parameters such as spectral class
(Madgwick et al. 2002, 2003)
and morphologies, metallicities, and star formation
rates (Kauffmann et al. 2003a,b), but color is by far
the easiest to measure.
Thus, not only does color sort galaxies cleanly into bins, it is
also highly relevant to the emergence of the Hubble sequence.

However, to exploit this opportunity requires highly accurate
counts, as the expected effects are not large.  
For example, counts of red galaxies in the COMBO-17 survey 
were seen to evolve by only a factor of a few since
$ z = 1$ (Bell et al. 2004b).  Even this small number has vital
implications for galaxy formation (see below), but confirming
and improving the measurement 
clearly requires accuracies of order 10-20\%.
Few previous measurements of distant luminosity
functions have attained this accuracy.  
Red galaxies are especially difficult
because of their high clustering,
which  necessitates 
large samples over a large number of statistically
uncorrelated regions.
We are only just now coming to appreciate how 
formidable the problem of cosmic variance really is 
(e.g., Somerville et al. 2004)

The present paper addresses these challenges by combining two 
large surveys, DEEP2 and COMBO-17, to create the largest database
yet analyzed of galaxies with $z >$ 0.8, 
containing 39,000 galaxies in total, with 15,600  
beyond $z = 0.8$.  Further checks are provided
by pilot measurements from the DEEP1 survey.
The entire sample is large enough and dispersed enough 
over the sky that cosmic variance and Poisson
fluctuations are reduced to 7-15\% per redshift bin.  The
samples were selected and measured in different ways---DEEP2 redshifts
are spectroscopic, while COMBO-17's are photometric---and thus
provide an important check on one another.  Finally,
color bimodality is used to divide red galaxies from blue galaxies
at all epochs.  The red luminosity function is rederived and compared to the
previous results of Bell et al. (2004b, hereafter B04), while the
blue function
is presented in Paper I and here for the first time 
(in a sample of this size) and 
offers a important foil for considering the behavior of the red function.
DEEP2 data and COMBO-17 data are found to agree well
in all major respects, and the
principal conclusions appear to be
robust.  

Our most important result is to confirm the 
recent rise in the number of massive
red galaxies {\it at fixed stellar mass} found by B04.
In contrast, the number density 
of massive blue galaxies has remained essentially
constant since $ z \sim 1$.  A second important conclusion
is that the characteristic $B$-band luminosity
$M_B^*$ of both red and blue populations has dimmed by 
about the same factor: we find $\sim$ 1.3
mag per unit redshift for both red and blue populations
since $ z = 1$.
The rise in the number of massive red galaxies
implies that most early-type galaxies assumed their final form
at relatively late times, below $z = 1$, {\it where the
process can be studied in detail.}  The late emergence of
spheroidal galaxies
disagrees with classic high-redshift, monolithic collapse
models for spheroid formation but seems to be consistent 
with large amounts of other data, as reviewed in
\S6.

The remainder of this Introduction reviews previous
measurements of luminosity functions.
The subject has a venerable history 
(e.g., Binggeli et al. 1988; Tresse 1999; de
Lapparent et al. 2003), with determinations ranging from low to high
redshift using field and cluster galaxies.
Accurate determinations of local field luminosity functions
have finally become available from 
the 2 Degree Field Galaxy Redshift Survey (2dFGRS, Norberg et al. 2002) and
the Sloan  Digital Sky Survey (SDSS, Blanton et al. 2003; Bell et
al. 2003), providing reliable local benchmarks  
against which evolution can be measured.

The dependence of the galaxy luminosity function on the
internal properties of galaxies has been known since Sandage, Binggeli
\& Tammann (1985) showed that the shape and magnitude 
of luminosity functions 
in the Virgo cluster depend on galaxy morphology and
luminosity class. This dependence on internal characteristics 
is also seen in local 
field galaxies when morphologies (Marzke, Huchra
\& Geller 1994; Marzke et
al. 1998; Marinoni et al. 1999), colors (Lilly et al. 1995b; Marzke \&
da Costa 1997; Lin et al. 1999; Blanton et al. 2001) and  spectral
types (e.g., Heyl et  al. 1997; Bromley et al. 1998; Folkes et
al. 1999; Cohen 2002; Magdwick et al. 2002; de Lapparent et al. 2003)
are considered.  

Early studies of the evolution of the galaxy luminosity function to $z \sim$ 1
used samples of a few hundred galaxies (e.g., Cowie et al. 1996; 
Brinchmann et al. 1998; Lin et al. 1999; Cohen 2002; Im et al. 2002;
de Lapparent et al. 2003).
In a landmark paper using the Canada France
Redshift Survey (hereafter CFRS), Lilly et al. (1995b) claimed 
that the evolution
of the luminosity function is coupled to internal
properties, being strongly correlated with color and, to a
lesser extent, with luminosity.  Dividing
red from blue galaxies using the median spectral type of the
sample, the authors claimed a steepening in faint-end 
slope for blue galaxies at redshifts beyond $z > 0.5$, while
red galaxies showed little change in either luminosity or number density over
the redshift range covered, $0.05 \leq z \leq 1$.

A conclusion that evolution depends on internal properties
was also reached by Cowie et al. (1996), based on a sample
reaching to $z\sim 1.6$.  
[OII] fluxes were used to estimate star formation rates and $K$-band
photometry to estimate stellar masses. 
They found that most of the luminosity
evolution since \z1 is due to blue galaxies
with small masses but high star formation rates.   More
massive galaxies were relatively stable in numbers, particularly in the
$K$-band, while the $B$-band showed modest number evolution.
They concluded that the characteristic mass of
galaxies undergoing intense star formation decreases over time,
which they termed ``downsizing.''

Cohen (2002) measured galaxies in a region centered
on the Hubble Deep Field and Flanking Fields and,
in contrast to CFRS, found that the luminosity
functions of several different spectral classes of galaxies showed
no strong evidence of  change in faint-end slope to $z \sim 1$, 
and further that the value of this slope is comparable to the local
value. Galaxies with spectra dominated by absorption lines
at $z=1$ were brighter by $\sim$1.5 magnitudes relative to local ones,
while galaxies with strong [OII] brightened by $\sim$0.75
magnitudes at $z \sim 1$. 

Im et al. (2002) measured evolution in the luminosity function of 
morphologically normal red
early-type galaxies by selecting distant galaxies to match
local E/S0s in both morphology and color.  This survey is related
to the present work, as it utilized DEEP1 redshifts, supplemented 
by photo-z's.
The luminosity function of early-type galaxies showed
a brightening of 1.1-1.9 magnitudes in rest-frame $B$ from
$z=0$ to $z\sim 0.8$, but
number density was relatively static
over the same epoch.  We will have more to say about this work later.

Similar brightening for early-type galaxies was also found by 
Bernardi et al. (2003) in the Sloan Digital
Sky Survey, where a brightening of 
$\sim$1.15 magnitudes per unit redshift back in time
was derived based on a
sample reaching to $z\sim 0.3$.

In a later study going 2 magnitudes fainter than Im et al.,
Cross et al. (2004) studied the faint end of the
luminosity functions of
both red-selected galaxies and morphologically-selected 
early-type galaxies  
using ACS images and photometric redshifts.  
The red-selected luminosity function was found to turn over
steeply at faint magnitudes, whereas the morphologically selected
sample was flat.  The difference was attributed to blue spheroids, which
filled in the counts at faint levels in the morphologically-selected
sample.

The evolution of the luminosity function as a function of color
since $z \sim$ 1
was also investigated by Pozzetti et al. (2003), who
used the near-infrared-selected K20 
survey  of Cimatti et al. (2002b)
and divided the sample using the color of Sa galaxies.
They found a modest rise
of at most 30\% in the number of red galaxies 
after $ z = 1$ and concluded that most bright red galaxies
were already in place by $z \sim$ 1.3.   
%On the other hand, it is
%thought that red galaxies beyond $z = 1$ are 
%increasingly contaminated by
%dusty objects that are not pure E/S0s (e.g., Moustakas et al. 2004), so the
%makeup of a red population chosen purely
%by color at $ z = 1.3$ is uncertain.

The distant surveys just cited clearly
disagree on many points,
including the numbers of galaxies, shapes of luminosity functions,
and degree of fading over time.  However,
these early surveys typically cover only a few tens of square arc minutes
and in retrospect are seen to be subject to large cosmic variance (see below). 
More recent surveys containing several thousands of galaxies are just now
beginning to provide more robust measurements of the luminosity function.
A large survey by Ilbert et al. (2004, VVDS) using the VIMOS spectrograph
on the ESO Very Large Telescope has measured the evolution of the
total galaxy luminosity function to $z\sim2$ using a sample of $\sim$11,000
galaxies to $I_{AB} = 24.0$ with spectroscopic redshifts. The authors
find that  $M_B^*$ for all galaxies has faded by 1.6 to 2.2
mag from $ z = 2$ to $z = 0.05$ but that the number density
of all galaxies has remained nearly constant. They also suggest
a possible steepening in faint-end slope at $z=1$.

The DEEP2 Survey (Davis et al. 2003)
employs spectroscopic redshifts to measure distances and internal kinematics 
for $\sim$40,000 galaxies in four regions of the sky. In 
order to probe galaxies at redshifts $z \sim$ 1, galaxies
are pre-selected in three of the regions to have $z > 0.7$ using $BRI$ colors.
The luminosity function analysis for the
first third of the DEEP2 survey is presented in a companion paper 
to this one by
Willmer et al. (2005, hereafter Paper I), which uses $U-B$ color
bimodality to study how the luminosity functions of blue and red
galaxies change with redshift. The results show significantly
fewer blue galaxies with time at fixed absolute magnitude,
which is well modeled by a fading in  $M^*_B$
together with roughly constant number density.
Counts for red galaxies, in contrast, show little 
change at fixed absolute magnitude and, when fitted with Schechter functions,
show a similar fading in $M_B^*$ with time
but a formally significant rise in number density.

An alternative strategy to create large
samples of galaxies
is the use of photometric redshifts. In spite of
lower precision, photometric redshifts yield a larger
number of redshifts per unit telescope time and enable distances
to be measured for galaxies that are too 
faint for spectroscopy. This approach has been 
pursued using both space-based (e.g., Takeuchi et al. 2000; Poli et
al. 2001; Bolzonella et al. 2002)
and ground-based data (e.g., Fried et al. 2001; Drory et al. 2003;
Wolf et al. 2003, hereafter W03; Chen et al. 2003; Gabasch et al. 2004).
Of the photometric redshift surveys, the ones most comparable to the present
work are COMBO-17 (W03) and the FORS Deep Field (Gabasch et al. 2004, FDF).
The latter used a sample of more than 5,500 galaxies down to
$I_{AB}=26.8$ to measure the total restrame $B$-band 
luminosity function 
from $z~\sim~0.4$ to $z~\sim$~4. Like DEEP2 and VVDS,
they found a constant number of galaxies back
in time but did not find a steepening of faint-end
slope despite the fact that
their sample goes ten times fainter than VVDS.   

COMBO-17 (Wolf et al. 2001, W03) contains $\sim$28,000 galaxies.
Aside from the use of photometric redshifts, it is similar to 
the DEEP2 survey in terms of depth
and coverage.  
%Together with DEEP2, it forms the data sample analyzed
%in this paper.  
The first luminosity-function analysis of COMBO-17
by W03 divided the sample into bins of fixed spectral
type that did not evolve with redshift.  Some of the evolutionary
trends that were discovered may have reflected color evolution between
these fixed spectral
bins rather than changes in overall numbers.
The approach was changed in a follow-up analysis using
the same database by Bell et al. (2004b, B04), who used an evolving
color cut based on bimodality to study 
red galaxies only.  As noted earlier, this work obtained the important new
result that red galaxies were not only brighter in the past
(by $>$1 mag) but were also fewer in number, by
at least a factor of two at $ z = 1$.  This claim based on counts was
buttressed by a separate argument based on the luminosity density
of bright galaxies.  This quantity, $j_B$,
can be measured more accurately than either $L^*$ or number density alone
and was found to hold roughly constant
since $z \sim 1$.  Since stellar
population models predict a
{\it fading} of red stellar populations by 1--2 mag between $z = 1$ and now
(see \S5),
the total stellar mass bound up in (bright) 
red  galaxies must be {\it increasing}
by about the same factor, 
providing additional evidence
for growth in the red galaxy population since that time.

A specific evolutionary scenario proposed
by B04 had the majority of present-day massive E/S0s
moving onto the red sequence after $z \sim 1$.
The stellar populations of such galaxies would age passively once galaxies
were on the red sequence, but individual galaxies would continue
to increase their
stellar masses via mergers along the sequence, 
as predicted by the hierarchical model of
galaxy formation.

Aside from B04, few works 
have used color bimodality to measure the luminosity functions of red
and blue galaxies separately.  One of those that has is by
Giallongo et al. (2005), who used a mixture of deep and shallow
data in four fields containing 1,434 galaxies.  Dividing
galaxies both by $U-V$ color and by star-formation rate, they found that
color bimodality persists to $z \sim 2$ but that the number
density of red galaxies drops steeply beyond $z = 1$.
They do not give numbers for $z = 1$ specifically, which
makes quantitative comparison with our results difficult,
but our conclusions below agree at least qualitatively with theirs.
%The morphological makeup of their distant red populations,
%whether truly spheroidal or dust-reddened, is also not
%established.

The foregoing summary illustrates that information on distant
luminosity functions is still fragmentary and often
contradictory.  In particular, the claim by Bell et al.
(2004b) for the emergence of red galaxies after $ z = 1$ has
not yet been checked.  A major impediment is cosmic variance---many 
of the above samples, especially the early and/or
deepest ones, cover only a few tens of square arc min,
for which the rms cosmic variance is $\sim$50\% per 
$\Delta z = 0.2$ at $z = 1$, being even greater
for red galaxies (Somerville et al. 2004).
Since number-density evolution by factors of two or three is
at issue (B04), definitive
results cannot be obtained
using such small areas, and larger samples are needed.
DEEP2 and COMBO-17 fill that need; each sample is large
enough on its own to give statistically meaningful results,
thus providing significant checks on one another.

The paper is organized as follows: \S2 presents the data, which
include not only COMBO-17 (W03) and DEEP2 (Paper I) 
but also data from the smaller DEEP1 pilot survey, 
which are presented here for the first time.  
\S3 briefly summarizes methods used to measure the
luminosity functions and their evolution, referring
the reader to Paper I and W03 for more details.  Readers wanting
results quickly can skip directly
to \S4, which presents the
luminosity functions, computes values of $M_B^*$, number density,
and $j_B$ from fitting to Schechter functions, and compares
the answers to local and distant values from the literature.
These are our core results on evolution.  A detailed discussion
of possible sources of error is presented in \S5, which can
also be skipped on first reading.
The implications for galaxy formation,
especially of red galaxies, are discussed in \S6, which 
draws heavily on the properties of
local E/S0s as well as distant ones.  
We ultimately favor a stepwise, ``mixed"
scenario  in which (massive) E/S0s are quenched during
a final gas-rich merger event, migrate to the red sequence, and
then undergo a small number of further, purely-stellar mergers 
on the red sequence to
attain their final masses.  This scenario is
similar to the one outlined by B04 and seems to 
have the right mixture of ingredients to explain the boxy-disky
``structure sequence" of local ellipticals plus the
narrow E/S0 scaling relations and the
age-$Z$ anti-correlation that underlies them. 
Speculation on the nature of the quenching mechanism and its possible
downsizing over time ends the discussion.  
A final summary is presented in \S7. 

Throughout this work, a ($H_0, \Omega_M, \Omega_\Lambda)$ = (70, 0.3, 0.7)
cosmology is used. Unless indicated otherwise, all magnitudes and
colors are on the Vega system.  Necessary
conversions to AB magnitudes are given 
for reference in Table 1 of Paper I.  Luminosity functions
are specified per unit co-moving volumes.

%
% Section 2
%
\section{Data}

The main data analyzed in this paper come from the DEEP2 and
COMBO-17 surveys, with supporting data from DEEP1.  
Detailed background information on these surveys appears in
other references, but core information needed to
understand the samples and their selection effects
is provided below.  Basic
properties of the surveys (area, number of galaxies, magnitude and
redshift limits) are summarized in Table 1.
%This section briefly describes the properties of DEEP2 and COMBO-17.

\subsection{DEEP2}		

The DEEP2 survey strategy, data acquisition, and data reduction
pipeline are described in Davis et al. (2003),
Faber et al. (in prep.), and Newman et al. (in prep.).
A detailed summary was provided in Paper I
as background to computing the DEEP2 luminosity functions; the 
functions from Paper I are adopted 
here without change. DEEP2 catalogues are derived from Canada-France-Hawaii 
Telescope images taken with the 12K $\times$ 8K mosaic camera
(Cuillandre et al. 2001)  in $B$, $R$ and $I$ in four different regions
of the sky.  Reduction of the photometric data, object
detection, photometric calibration, and 
construction of the star-galaxy catalogs are described
in Coil et al. (2004).
%Objects were 
%identified using the $imcat$ software written by N. Kaiser,
%using the algorithm described by Kaiser, Squires \& Broadhurst (1995),
%which
%calculates magnitudes and other image parameters.  
$R$-band images  
used to define the galaxy sample have a limiting
magnitude for image detection at $R_{AB}\sim 25.5$. 
Apparent-magnitude cuts of $R_{AB} \geq 18.5$
and $R_{AB}\leq 24.1$ and a surface brightness cut
of $\Sigma_{R} \leq 26.5$ were applied (see Paper I). 
Separation between stars and galaxies is based on
magnitude, size, and color, which were used to
assign each object a probability of being a galaxy;
star-galaxy separation efficiency is 
discussed in Coil et al. (2004) and Paper I.
In Fields 2, 3, and 4, the spectroscopic sample
is pre-selected using $B$, $R$, and $I$ to have
estimated redshifts greater than 0.7, which 
approximately doubles the efficiency of the survey
for galaxies near $ z \sim 1$.  The fourth field,
the Extended Groth Strip (EGS), does not have
this pre-selection applied but instead has roughly equal
numbers of galaxies below and above $ z = 0.7$, which
were selected using a well understood algorithm
versus redshift.  Redshifts were measured spectroscopically
using the DEIMOS spectrograph (Faber et al. 2003)
on the Keck 2 telescope.  Slitlets are placed (nearly)  randomly on
50\% of all galaxies after pre-selection
(70\% without pre-selection in EGS), of which 70\% yield successful
redshifts (80\% in EGS), with a catastrophic failure rate of 1\%.

The DEEP2 sample used here combines data from the
first season of observations in Fields 2, 3, and 4
with 1/3 of the total EGS data, which provides an initial
sample at low redshifts. The total number of
galaxies is 11,284, with 4,946 (44\%) in EGS, 3948 (35\%) in Field 4,
2299 (20\%) in Field 3, and 91 (1\%) in Field 2. Since the photometric
redshift cut at $z \sim$ 0.7 provides a soft boundary for the
selection of galaxies, only EGS is used to probe the
lower-redshift realm $z < 0.8$, 
while data in all four fields are used for $z \ge 0.8$.
Color-magnitude diagrams illustrating the sample binned
by redshift are shown in Paper I.
   
\subsection{COMBO-17}

The COMBO-17 survey consists of multi-color imaging data
in 17 optical filters covering a total of 
1~$\sq\deg$ of sky at high galactic latitudes. The filter set contains five
broad-band filters ($UBVRI$) plus 12 medium-band filters stretching from 400 to
930~nm. 
All observations were obtained with the Wide Field Imager (WFI, Baade et al.,
1998, 1999) at the MPG/ESO 2.2-m telescope on La Silla, Chile. 
The total exposure time is $\sim$160~ksec per field,
which includes a $\sim$20~ksec
exposure in the $R$-band with seeing below 0\farcs8 FWHM. The WFI
provides a field 
of view of $34\arcmin \times 33\arcmin$ on a CCD mosaic consisting of eight
2K $\times$ 4K CCDs with $\sim67$ million pixels providing a scale of $0\farcs
238$/pixel. The observations began during the commissioning phase of the WFI in
January 1999 and are continuing as the area is extended to cover more fields.
The data used here are from three fields covering
an area of 0.78~$\sq\deg$, providing a catalogue of $\sim$200,000 objects
found by SExtractor (Bertin \& Arnouts 1996) on $R$-band images with a
5-$\sigma$ point-source limit of $R\sim26$.

SEDs created from these 17 passbands were used to classify
all objects into stars, galaxies, and QSOs by comparison with 
template SEDs.  Less than 1\% of the sources
have spectra that are peculiar, not yielding an object class (W03),
and star-galaxy separation is highly efficient. 
A first analysis of the COMBO-17 luminosity function was published 
by W03, but the galaxy catalogue has changed 
slightly since then.  Basic details of the
classification algorithm and choice of templates were given in Wolf,
Meisenheimer \& R\"oser (2001). In 2003, an improved set of SED templates
was introduced after it was found that the accuracy of galaxy redshifts was
limited by template mismatch for bright galaxies, 
for which more subtle SED details could be seen. 
The new set of galaxy templates
contains a grid of synthetic spectra based on the PEGASE code 
for population synthesis models (Fioc \& Rocca-Volmerange 1997),
whereas in the past the redshift 
determination relied only on the observed templates by Kinney et
al. (1996).   The new redshifts are accurate to within
$\delta z/(1+z)<0.01$ at $R<21$ and to within 0.05 down to  
$R<24$ (Wolf et al. 2004,
hereafter W04).  The changes from the old redshifts 
are relatively small (see Figure~4 in W04)
and are within the errors of the old estimates, but residual errors
 were reduced by up to a factor of ~3 for galaxies
with selected SED shapes.
Typical catastrophic failure rates for COMBO-17 galaxies are $\sim
1$\% in the magnitude range used for the LFs, as measured using
galaxies in common between W04
and Le F\`evre et al. (2004) in Chandra Deep Field South.
%{\bf SMF Is there any independent check on the
%accuracy of the COMBO-17 redshifts using reliable spectroscopy?
%What exactly is the basis for the \_quantitative\_ claims of accuracy
%being made here?}
The resulting luminosity functions of galaxies are 
also unchanged within
the errors published in W03 if the
same color divisions are used. The luminosity functions of
red-sequence galaxies published by B04
were already based on the new redshift catalogue.

In this paper, we recalculate luminosity
functions from the COMBO-17 galaxy sample using 
different color divisions than before. The red-sequence cut is similar
to the one in B04, with a small difference: B04 measured the mean color of the
red-sequence, which is affected to a small degree by K-correction errors 
that vary in a non-stochastic way with redshift. They then fitted a smooth 
evolution to the measured colors to identify the most likely trend. Here
we use the measured red-sequence colors 
in each individual redshift bin to define the 
color valley, and thus the cut. As stated  
by B04, the difference this makes to luminosity functions is small. However,
we consider it preferable to follow the small but systematic 
variations in the data for the purpose of splitting the population.
The new method is identical to that used for DEEP2 (see Paper I).

In COMBO-17, a galaxy redshift measurement is considered successful when the
error expected from the probability distribution is below a threshold of 
$\sigma_z/(1+z)\la 0.1$. The completeness of successful redshifts depends on
galaxy rest-frame color, and simulated completeness maps are shown
in W03 and W04. The large redshift incompleteness among blue galaxies at the faint
end of the COMBO-17 sample would lead to unreliable completeness
corrections in the last few LF bins just above the survey limit. These
bins have thus been dropped, and all correction factors used in 
the remaining bins (plotted data points) are below 1.5.  
For red-sequence galaxies at $z<1.2$, which
are the special focus of this paper, redshifts
are measured successfully for the {\it entire} sample at $R<24$ used for the
luminosity function.  
This claim can be tested by identifying likely red-sequence galaxies from an
apparent color-magnitude diagram like that in Figure 1$a$.  In
COMBO-17, galaxies with failed redshifts occupy solely the region of   
blue galaxies. This is consistent with (a) the 
known lower COMBO-17 completeness  
for redshifts of blue galaxies close to the faint limit, and (b) the zero   
completeness (by design) of COMBO-17 with respect to $z>1.4$ galaxies. 
Blue galaxies with $z >$ 1.4
are expected to be a much larger part of a flux-limited sample with  
$R<24$ than red-sequence galaxies at $z>1.4$. Owing to their red colors    
and faint near-UV fluxes, only red-sequence galaxies of extremely high    
luminosity could pass the flux limit at high redshift, of which there
are evidently very few.

\subsection{DEEP1}

DEEP1 was a pilot survey for DEEP2 that was conducted
using the LRIS spectrograph on the Keck telescopes in
1995-1999.  
Since the only published luminosity function using DEEP1 data treated
E/S0s (Im et al. 2002) above $I814$ = 22, a more detailed 
description of DEEP1 data will be presented here. This
summary also serves to convey the flavor of our 
treatments of DEEP2 and COMBO-17.
Readers not interested in these details should skip to
\S3.

Background on DEEP1
sample selection and photometry is presented in Vogt et al. (2005),
on photometry and bulge-disk decompositions
by Simard et al. (2002), and 
on spectroscopy and redshifts by Weiner et al. (2005). 
The DEEP1 sample 
is drawn from objects detected in a set of 28 contiguous
Wide Field and Planetary Camera (WFPC2) Hubble Space Telescope
pointings at approximately  14$^h$ 16$^m$ 30$^s$ 
+52$^o$15$'$50$''$ (J2000.0), (PIs E. Groth, GTO 5090; and J. Westphal
GTO 5109).  The solid angle covered by the GSS is 127 square arc minutes.

The DEEP1 photometric catalogue was
created by Groth et al. (1994) and contains several parameters
measured using FOCAS (Tyson \& Jarvis 1981). To this
catalogue were added total magnitudes and colors measured fitting
2-dimensional bulge+disk models using the GIM2D package (Simard et
al. 1999, 2002).
For galaxies without GIM2D measurements,
magnitudes and colors measured by Ratnatunga et al. (1999, MDS) in
the Medium Deep Survey were used, which were shown by Simard et
al. (2002) to have comparable quality to GIM2D measurements. 
When neither GIM2D nor MDS had magnitudes and colors, the FOCAS measurements 
were transformed into the same system as the GIM2D
magnitudes by adding the median offset between GIM2D and FOCAS
magnitudes in each color ($I814$ = $I_{FOCAS}$ -- 0.295 and
$V606-I814$ = $(V-I)_{FOCAS}$ + 0.106) (all
HST magnitudes are on the Vega system.)

DEEP1
spectroscopic data were obtained over several observing runs
using the  Low Resolution Imaging Spectrograph (LRIS, Oke et al. 1995)
on the Keck 1 and 2 telescopes. Most galaxies were observed using two
different gratings, with a blue side ranging from 4500 \AA~to
6500 \AA~and a red side from 6000 \AA~to 9500 \AA. 
%The spectral data were reduced following a pipeline described by
%Weiner et al. (2005). After an initial processing to remove
%instrumental signatures, 
%the 2-D spectra of a given galaxy obtained on different masks
%were combined, and a 1-dimensional spectrum was extracted containing data
%in the region limited by the half-light radius. 
Total exposure times
ranged from about 50 minutes for galaxies observed on one mask up to
$\sim$500 minutes for galaxies placed on several masks.
In the analysis below, the DEEP1 sample is restricted to galaxies
with 16.5 $\leq$ $I814$ $\leq$ 23.5 and with redshift
quality A or B, as explained in Weiner et al. (2005)
(no surface-brightness cut was
applied). Galaxies generally
have more than one identified spectral feature, and the redshift
confidence level is better than 90\%.
The total number of galaxies in the region
that satisfy the apparent magnitude limit
is 2,438, of which 621 have good quality redshifts. 
The typical sampling rate of DEEP1 redshifts is $\sim$40\%, and the typical
redshift success rate is $\sim$70\%, for a final overall sampling
density of $\sim$25\%.

The apparent color-magnitude diagram of DEEP1 galaxies is shown in
Figure 1. The spectroscopic sample was selected from a ``pseudo-$R$''
band magnitude: [($V606+I814$)/2 $\leq 24.0$]; this has been converted to
an approximate $I814$ magnitude shown as the vertical dotted line in all
four panels. Figure 1$a$ shows the full
sample of galaxies, and Figure 1$b$ shows the distribution of galaxies
placed in slits. Because of the $R$-band selection,
shown as the inclined dashed line, there is a dearth of red galaxies
fainter than $I814=23.5$; brighter than this, 
the sample of galaxies placed on masks is a good
representation of the total galaxy sample. 
Figure 1$c$ presents the distribution of galaxies that
were successfully measured; green symbols show galaxies below
the adopted high-$z$ cutoff (see below), while red symbols show galaxies
beyond the cutoff, which were not used.  
Black crosses show galaxies with redshifts from CFRS
(Lilly et al. 1995b) and from Brinchmann et al. (1998), for which DEEP1
measured no spectrum.
Figure 1$d$ shows the
distribution of galaxies with failed redshifts. The few bright cases
of failed redshifts resulted from 
short integrations or spectra of galaxies at mask edges, while
the majority of failures are  of faint and generally blue
galaxies. As in DEEP2 and COMBO-17, most failures 
are likely beyond the adopted high-redshift cutoff of the survey,
here taken to be $z_h =  1.0$.  This is the cutoff for 
the DEEP1 analysis and is the redshift
where O II $\lambda$3727 becomes heavily confused with strong OH sky 
lines in the LRIS data (Weiner et al. 2005).

Figure 2 divides the apparent color-magnitude diagram into magnitude
and color bins. For each bin, the histogram of the distribution of
galaxies as a function of redshift is shown,
where the filled histogram represents successful measures and the open
bar at the right the number of failures inside each bin. The number of
failed redshifts increases at magnitudes fainter than $I814\sim 22.5$.

Figure 3 shows the distribution of restframe
$U-B$ vs.~redshift. The method used to measure $U-B$ is described
in Weiner et al. (2005), using the procedure described in Paper I but
limited to the two observed filters ($V606$ and $I814$).
A bimodal color distribution is clearly
seen, as well as large-scale structure fluctuations due to galaxy 
clustering (vertical stripes).
The number of successful redshifts above  $z = 1$ falls drastically owing to 
OH confusion (see above).

Rest-frame color-magnitude diagrams for different redshift
intervals are shown in Figure 4.  Similar diagrams for
DEEP2 were shown in Paper I and for COMBO-17 in B04.
The solid line in each panel represents
the limiting absolute magnitude that
corresponds to apparent magnitude $I814 = 23.5$ at
the far edge of the bin as a function of
restframe color.  The  color dependence was
calculated using the K-correction code from Paper I.
The changing
slope of the line as a function of redshift is caused by 
the fact that the $I814$-band filter used to select
the sample coincides with 
rest $B$ at $z \sim$ 0.8 but
differs from it increasingly as the redshift is 
either greater or smaller than 0.8.  Intrinsically red
galaxies are included to fainter absolute magnitudes when
observed $I814$ is redder than rest $B$, while intrinsically
blue galaxies are favored when observed $I814$ is bluer than
rest $B$, thus causing the line to swing with redshift.

The upper dashed line in each figure
represents the cut used to separate red from blue galaxies.
This cut is identical to that used by Paper I for DEEP2 since
the restframe colors and magnitudes are on the same system.
The equation for the line is 
\begin{equation}
 U-B =  -0.032 (M_B + 21.52) + 0.454 -0.25,
\end{equation} 
which is taken from the van Dokkum et al. (2000) 
color-magnitude relation for red galaxies in distant clusters, 
converted to the cosmological model used in this paper, and 
corrected downward by 0.25 mag in order to pass through the valley
between red and blue galaxies (Paper I).  
Although the colors of red galaxies
may evolve with redshift, this effect is not strongly seen in either
DEEP1 or DEEP2 colors, 
and a line with constant zero point independent of redshift
is used for all redshift bins. 

We conclude this section by comparing the
strengths and weaknesses of the two major data sets used in
this paper, DEEP2 and COMBO-17.  Both data sets go to nearly
the same apparent magnitude, $R\sim24$, and have comparable numbers
of galaxies beyond $z = 0.8$ (see Table 1).  
The square root of cosmic variances are shown for each sample
in Tables 2-4 by 
galaxy color and by redshift bin. They are comparable for the two
surveys beyond $z = 0.8$ and range between 10-20\% for all redshifts
and color classes. When combined, the two surveys have a total 
(square root) cosmic variance of $\sim$~7-15\% per redshift 
bin at $z \sim 1$.
The strengths of DEEP2
are rock-solid redshifts and high completeness for blue galaxies
all the way to $z = 1.4$ owing to the sensitivity to [O II]
$\lambda$3727, which is strong in distant blue galaxies. 
The strengths of 
COMBO-17 are higher completeness overall at all redshifts, particularly
for distant red galaxies near $z \sim 1$.
This is offset by a tendency to lose redshifts for blue galaxies
towards the faint limits of the survey, which has forced us to cut
off the COMBO-17 All and Blue luminosity functions 
at a shallower point than DEEP2 to keep
completeness corrections small.
The two data sets thus complement each other well at high $z$, 
making a parallel, head-to-head
analysis extremely useful.

%------------------------------------------------------------------------

\section{Methods}

The luminosity function is most frequently
expressed using the Schechter (1976) parameterization, which in magnitudes is:
\begin{equation}
\phi(M) dM = 0.4\; ln\; 10\; \phi^* 10^{0.4(M^*-M)(\alpha+1)}
\lbrace -10^{0.4(M^*-M)} \rbrace dM,
\end{equation}
where $\phi^*$ is a normalizing constant that is proportional
to the total number density of galaxies, 
and $\alpha$ is the slope of the power law that describes
the behavior of the faint end of this relation. Changes in these
parameters with time quantify how galaxy
populations evolve.

The methods used for the DEEP2 (and DEEP1) luminosity functions
are described in Paper I. The methods used for COMBO-17 are described
in W03 and are very similar.  A brief 
overview of all methods is provided here.

Two statistical estimators have traditionally been
used in the calculation of the luminosity
function.  These are the parametric maximum-likelihood method of Sandage,
Tammann \& Yahil (1979, STY; also
Efstathiou, Ellis \& Peterson 1988; and 
Marzke, Huchra \& Geller 1994) and the non-parametric  
$1/V_{max}$ method  of Schmidt (1968; also Felten 1976 
and Eales 1993).  The
STY method fits an analytic Schechter function, yielding
values of the shape parameters $L^*$ and faint-end slope 
$\alpha$, but not
the density normalization $\phi^*$, which is estimated using the
minimum-variance density estimator of Davis \& Huchra (1982). The STY
method also does not produce any visual check of the fit.
In this paper, a visual check both on shape and normalization for each
redshift bin 
is obtained using $1/V_{max}$ since it yields the average number
density of galaxies in bins of redshift and absolute magnitude.
Formulae used for obtaining the STY parameters, 
$1/V_{max}$ points, and the density normalizations are
given in Paper I. 

Since the STY method does not yield $\phi^*$, it is not
suitable for calculating the correlated 
errors between $\phi^*$ and  $M^*$.
For DEEP, these errors were calculated from the 1-$\sigma$ 
error ellipsoid (Press et al. 1992) that results from
fitting a Schechter function to the $1/V_{max}$
data points (Paper I). Although the luminosity functions 
obtained from the  STY
and $1/V_{max}$ methods are not quite identical (see Figure 6 below), 
the differences are small and errors from the $1/V_{max}$ method 
should also apply to the STY method. For COMBO-17, the errors in
parameters were calculated first for $M_B^*$ using STY, and then
the $\phi^*$ errors were calculated using the
field-to-field variations (W03).

Weights are needed for every data set
to correct for missing galaxies.  
The adopted weights need to take into account
the fact that 1) objects may be missing from the photometric catalogue,
2) stars may be identified as galaxies and vice versa,
3) not all objects in the photometric catalogue are targeted
for redshifts, and 4) 
not all targets yield successful
redshifts.  
For the DEEP surveys, factors 1) and 2) are small or zero
(see discussion in Paper I),  
and only factors 3) and 4) need to be
taken into account.  The basic assumption to
deal with 3) is that all unobserved galaxies share the
same average properties as the observed ones in a given color-magnitude
bin.  Factor 4) is dealt with by assigning
a model redshift distribution to the failed galaxies.
We use two such models, as explained in Paper I.  The ``minimal''
model assumes that all failed galaxies lie entirely beyond
the high-redshift cutoff of the survey, which is $z_h = 1.4$ for DEEP2
and $z_h = 1.0$ for DEEP1. As discussed in Paper I, this model 
should provide an
adequate description for blue galaxies.
The second model to treat failed redshifts is the ``average'' model, which
assumes that failed redshifts have the same distribution
as the successful redshifts in the same color-color-magnitude
bin; this is a reasonable assumption for red galaxies.  
The difference in weights between
the minimal and average models is usually 
$\leq$25\% (average weights are higher), 
and most of the large differences occur for galaxies
with extreme colors at faint magnitudes.
Luminosity functions
calculated with the average model are slightly higher than those
using the minimal model, by an amount that averages 10-20\%.
When the combined All-galaxy sample is considered, we use an
``optimal'' model in which red galaxies are modeled using
the average model and blue galaxies are modeled using the minimal model.

Two small alterations to this general scheme
were applied.
In the case of EGS in DEEP2, a final correction (described in Paper I)
was applied to account for the different
sampling strategy used in this field, 
which includes low-redshift galaxies
but de-weights them so that they do not dominate the sample.
In the case of DEEP1, the weights were modified to
use additional size information from HST images, which show
that {\it all} galaxies with angular half-light radius
$r_{hl} \ge 1$ arc sec (from GIM2D) lie
within the legal redshift range $z \le 1.0$.  
Figure 5 illustrates these results for DEEP1 by plotting
sampling rates, redshift-success rates, and weights; analogous
figures are given for DEEP2 in Paper I.  

For the COMBO-17 survey, effect 1) from the list above is small, as
only galaxies very close to very bright stars are lost from the object
catalogue. Effect (3) is zero, as the photo-z code works on the entire  
catalogue. Effects (2) and (4) are linked, since in COMBO-17 both object
classification and redshift estimation are one single process.
In one direction, a few K stars are
misidentified as galaxies, but their number is negligible. 
In the other direction, the misclassification
of galaxies as stars is modeled together with redshift incompleteness  
using simulations as described in Wolf et al. (2001) and W03,
which take into account the photometry S/N, SED, and
redshift and are calibrated using Monte-Carlo simulations.
Weights are calculated as a function of apparent magnitude
and color and are close to unity for all red
galaxies, for which we calculate 
the luminosity to the full sample depth, 
but drop rapidly for blue galaxies 
towards the survey limit. We have not used any 
data points that involve corrections by more than a factor of 1.5, 
and as a result the COMBO-17 luminosity function
points for the Blue and All samples 
do not quite reach luminosities corresponding to apparent
magnitude $R=24$.

%%%%
%--------------------------------------------------------------------------
% Analysis
%--------------------------------------------------------------------------
\section{Analysis}

\subsection{The DEEP and COMBO-17 Luminosity Functions}

This section compares the luminosity functions derived
from DEEP2, DEEP1, and COMBO-17 with one another
and with published data.  The DEEP2 functions are
best estimates from Paper I that use
the optimal missing-redshift model for All galaxies, minimal for Blue
galaxies and the average model for Red galaxies
(see \S3).  
%These are
%appropriate because
%it is believed that most failed blue redshifts lie
%beyond the survey cutoff, while most failed red redshifts lie
%within the survey range, and the combined sample will use both of
%these characteristics, i.e., uses average weights for galaxies that
%are situated in the ``red'' part of the diagram, and minimal weights
%otherwise.  
For the DEEP1 sample we use the minimal
model for all galaxy colors, while COMBO-17 weights are
as described in \S3 and W03.  
Galaxies are analyzed all together (the ``All'' sample) and 
divided into ``Red'' and ``Blue'' sub-samples using color-magnitude 
bimodality.
The method used to divide blue and red
galaxies in DEEP2 and DEEP1 is based on the slanting
line that goes through the color valley in the $U-B$ vs.~$M_B$ 
CM diagram (see Equation 1 and Figure 4).  
The line used for COMBO-17 is
similar to the one used by B04 based on  $U-V$ vs.~$M_V$
except that the smoothly-evolving zero point of the line
through the color valley
is replaced by a zero point adjusted in each redshift
bin to make the line go through
the valley at that redshift.

Figure 6 shows the resulting
luminosity functions for the All data (top row), Blue data
(middle row), and Red data (bottom row).
Redshift increases from left to right across a row.  
Non-parametric 1/$V_{max}$ data points are
shown for DEEP2 by the solid black squares, for DEEP1 by the
grey triangles, and for COMBO-17 by the red circles.  For
all samples, the calculation of the luminosity function 
is truncated at the faint
end using dashed lines analogous to those in Figure 4, 
taking the limiting absolute
magnitude at each color and in each redshift bin into account;
details are given in Paper I.  Blue galaxies were further
trimmed in COMBO-17 as described in \S3, to allow
for greater redshift incompletness.

The error bars on each DEEP2 and DEEP1 point
represent Poisson statistics only.  Cosmic variance 
estimates are shown 
as the separate error bar at the top
left corner of each panel and were estimated
using the procedure of Newman \& Davis (2002) to account for
evolution of the correlation function. The bias factors derived by
Coil et al. (2004) for red galaxies ($b = 1.32$) and blue galaxies ($b
= 0.93$) relative to dark-matter halos are included in these
estimates.  The values plotted are for DEEP2.
To first order, Poisson variance is random from point to point,
whereas cosmic variance should mainly move all points 
in a given bin up and
down together.  Since these effects are different, they
are shown separately. For COMBO-17, the error
bars combine the Poisson errors in $\phi (M)$ with the cosmic variance
estimated from the field-to-field variations.
 
Also shown in the top row of Figure 6 are 1/$V_{max}$
data points by Ilbert et al. (2004, VVDS), represented
by blue diamonds.  This
sample uses $\sim$11,000 spectroscopic redshifts from the
VVDS survey to $I_{AB} = 24$ (7,800 redshifts are termed ``secure'').
Finally, the grey dashed lines show Schechter
fits to local red and blue 
SDSS samples at $z \sim 0.05$ from Bell et al. (2003),
who divided galaxies both by color and by
concentration, getting similar results.  The exact Schechter
parameters used are given in Table 5.

The conclusions from Figure 6 are as follows:

{\it All galaxies (top row):} Measurements
of the All-galaxy luminosity function from all four
surveys agree well out to $z \sim 1$ and down to the apparent
magnitude limit of DEEP2 and COMBO-17 ($R \sim 24$).
Below this, VVDS claim to see a steepening
in faint-end slope from $\alpha \sim -1.2$
at $z=0.05$ to $\alpha \sim -1.5$ at $z = 1$.  
Neither DEEP2 nor COMBO-17 
go deep enough to test this, but, as noted above, 
Gabasch et al. (2004, FDF) go nearly 2.5 magnitudes fainter and do not see it,
getting $\alpha = -1.25$ at all redshifts.
Relative to
the local Schechter total function, the data in
successive redshift bins march to dimmer magnitudes ($M^*_B$)
with time
but stay roughly constant in number density ($\phi^*$).  
This visual assessment is confirmed by Schechter fits below.
In short, for the population as a whole (All sample),
galaxies are getting dimmer with time but
their number density has remained much the same, since $z \sim 1$.

{\it Blue galaxies (middle row):} 
The results found above for the All sample
are replicated for the Blue sample, as expected since
blue galaxies account for the majority of objects at all redshifts.  
Results here are available only from DEEP1, DEEP2, and COMBO-17
since VVDS
do not divide their samples by color.  However, 
these three data sets agree well. Relative to the
local blue Schechter function (dashed grey line), $M^*_B$ dims
with time while 
$\phi^*$ remains constant, again confirmed by Schechter fits below.

{\it Red galaxies (bottom row):} 
Before considering red galaxies, we review the 
conclusions of
Bell et al. (2004b, B04), which offered the first analysis of 
evolution in  $\phi^*$
and $M^*_B$ for red galaxies, 
based on COMBO-17.  The main finding was that  
$M^*_B$ for red galaxies
dims over time by $\sim$1.5 mag from $z = 1$ to 0 and that
$\phi^*$ {\it rises} by at least a factor of two.  
%The change in $\phi^*$ is very
%significant for theories of red-galaxy evolution because it implies that
%half (or more) of all galaxies on the red sequence today 
%arrived there {\it after} $ z = 1$, where the process can
%be studied in detail.
This evidence for
evolution from the luminosity function
was further bolstered by consideration of the
total $B$-band luminosity density of red galaxies, $j_B$, which is 
measured with smaller (formal) errors
than $M^*$ or $\phi^*$ separately.  The quantity $j_B$ was
found to hold nearly constant since $z = 1$.  Since models of stellar
evolution for red galaxies predict a rise in $B$-band stellar
mass-to-light ratio by 1-2 mag since $z = 1$
(see more on this below), constant 
$j_B$ implies that
the total stellar mass contained in red galaxies has 
at least doubled since $z = 1$, providing further evidence for
significant growth and change in 
red galaxies over this epoch.

Bell et al.'s finding of recent strong evolution among red galaxies
disagrees with the classic scenario for red-galaxy
formation in which E/S0 galaxies assembled their mass and formed 
stars very early and have been passively fading ever since (e.g., Eggen,
Lynden-Bell \& Sandage 1962; Larson 1975). 
The monolithic-collapse picture predicts
constant $\phi^*$ accompanied by equal dimming in 
both $M^*_B$ and $j_B$, but neither of these trends was seen by B04.
Checking these
conclusions by remeasuring these quantities with both DEEP2 and COMBO-17
was therefore a major goal of the present study.

The bottom row of Figure 6 presents the new data for red galaxies.
As before, DEEP2 and COMBO-17 agree well.
The most striking impression is the relative {\it lack}
of evolution in the red luminosity function,
especially when compared to the
large shift to brighter magnitudes seen in the blue function.
What evolution there is is quantified below by
Schechter fits, which are shown in Figure 6 as the black lines.
These fits indicate a formal dimming of $M_B^*$
over time, accompanied
by a rise in number density, $\phi^*$.
The sense of these shifts is such that
the data translate {\it nearly parallel} to themselves, leaving the
raw counts at a fixed absolute magnitude relatively
constant.  Since the actual counts are not changing 
a great deal, to first order,
the fitted values of $M_B^*$ and $\phi^*$ must
depend on slight
curvature signals in the data, which could be weak
and unreliable.
We return to this question below, where
the relative constancy
of the red counts is considered
from various points of view.  For now we simply
note that both the raw data and the fitted Schechter function
parameters
from DEEP2 and COMBO-17 agree extremely well, and that the {\it formal} 
values of $\phi^*$ from both
data sets agree with the rise found by B04.

Another important result in Figure 6
is the marked {\it turnover} in the slope of
the Red luminosity function at the faint end.  This turnover is
well established in both DEEP2 and COMBO-17 at
intermediate redshifts and is  seen by Cross et al. (2004) and
by Giallongo et al. (2005) at even higher redshifts.
However, DEEP2 and
COMBO-17 may disagree with one another 
in the lowest redshift bin ($z=0.2$-0.4), where
the number of faint red galaxies continues 
to turn over according 
to DEEP2 but flattens according 
to COMBO-17.  This is noteworthy as only the potential discrepancy
between DEEP2 and COMBO-17, but the error bars on
COBMO-17 are large, reflecting large
field-to-field variations.  Other data sets have also
yielded conflicting values 
for the nearby red faint-end slope.  For example, 
by identifying early-type galaxies in SDSS using both concentration and
color, Bell et al. (2003) found only a modest turnover
at the faint end,
as in COMBO-17, whereas Madgwick
et al. (2002) identified red galaxies spectroscopically
in 2dF and found a strong turnover, more like DEEP2 (see Table 2).
The question of the red faint-end slope and its possible
evolution with redshift is very important for understanding
the processes that created red-sequence galaxies.  We
return to this question in \S5 below when discussing 
errors in the Schechter function
parameters caused by possible 
evolution in the red faint-end slope versus
redshift.

In passing, we note the grey triangles in Figure 6, which show
luminosity functions from DEEP1.
These agree rather well with the functions from DEEP2 and COMBO-17
except in bin $z = 0.8-1.0$,
where total DEEP1/Red is a factor of 1.5 too high. 
Two ``walls'' due to large-scale structure
appear in that redshift bin, one at $z\sim$ 0.81 and
a larger one at $z\sim$ 0.98 (Le F\`evre et al. 1994; Koo et al. 1996;
see also Figure 7).  
%These contain about 11\% of the total DEEP1 sample
%but only $\sim$ 4\% of the DEEP2 EGS sample, so their impact on the 
%counts in the latter survey (which for this redshift bin contains in
%addition fields 2, 3 and 4) is much smaller.  
However, the observed fluctuation is
not much larger than the expected cosmic variance limits ($\sim$30\%).

\subsection {Schechter Fits}

This section presents the results of fitting 
Schechter functions to DEEP2 and COMBO-17
using the STY  method.
Aside from the possible low-redshift flattening of the
Red function in COMBO-17,
we see no variations
in faint-end slopes that are statistically
significant in different redshift bins, motivating 
the use of constant values 
of $\alpha$ obtained from averaging over several bins.  
(In fact, small changes are expected in the shape of 
the All galaxy function with redshift  
because the shapes of the Red and Blue functions differ
and their relative numbers are changing with redshift; however,
this effect is small.)
We decided to average the faint-end slope values found within 
the range $z= 0.2$ to 0.6 for COMBO-17 (because of its larger
number of galaxies in this redshift range), 
which yielded $\alpha=-0.5$ for the 
Red  sample and
$\alpha=-1.3$ for the All and Blue samples
(these values
were also used for DEEP2 in Paper I).  The latter slope
agrees well with the value $\alpha = -1.25$ found for all galaxies
by FDF, while the former is close
to the average value $-0.59$ found for distant red galaxies 
by Giallongo et al. (2005).

Schechter function 
parameters
for both DEEP2 and COMBO-17 
are presented in Table 2 for the All sample
and in Tables 3 and 4 for the Blue and Red samples. Column (1) 
shows the central redshift of the bin; column (2) the number of
galaxies used in the luminosity function calculation in each redshift bin;
column (3) the value of the
adopted faint-end slope $\alpha$; column (4) the value of $M^*_B$, followed by
the upper and lower 68\% errors in columns (5) and (6); the
mean density $\phi^*$ is given 
in column (7), followed by the 68\% errors 
in columns (8) and (9); the square root of the cosmic variance error
is shown in column (10); and 
the luminosity density (see \S4.3), followed by the 68\% errors
in columns (11) and (12).  Column (13)
indicates the the weighting scheme described in \S3.3 adopted  
for the DEEP2 fits.
As explained above, we adopted minimal weighting for the DEEP2 Blue
sample and average weighting for the Red sample because we think
that failed redshifts in the two color classes have different
redshift distributions. The All sample combines each of
these populations with its preferred weighting scheme (called ``optimal" in
Table 2).

For DEEP2, the 68\% errors are Poisson estimates 
for $M^*_B$ and $\phi^*$ and are taken from the
$\Delta\chi^2$ = 1 contour levels in the ($M^*_B$, $\phi^*$) plane,
computed from the $1/V_{max}$ residuals and their errors.
Errors for $j_B$ are conservatively
calculated by adding the fractional Poisson errors for $M^*_B$, 
$\phi^*$, and cosmic variance in quadrature; these latter are overestimates
because they neglect correlated errors in $M^*_B$ and $\phi^*$,
which tend to conserve $j_B$. However, since the biggest error term is
usually cosmic variance, the overestimate is
small.  For COMBO-17, the 68\% errors in $M^*_B$, $\phi^*$, and $j_B$
are rms estimates from field-to-field variations, which 
are particularly large for the redshift bin centered at $z =$ 0.9,
caused by a big downward fluctuation in CDF-South. 
Tbe tabulated cosmic variance errors estimates for both samples
were computed as described
above for DEEP2, taking the volume and field geometries into
account and using separate bias ($b$) values for All, Blue, and Red 
galaxies. 

The resulting Schechter fits for DEEP2
are shown as the solid black lines in Figure 6.
All fits use only the magnitude ranges of the data actually
shown.   
The close match between the fitted curves and all 
data suggests that the Schechter formula,
and in particular the assumed $\alpha$ values, are
a good match to the luminosity function shapes 
over the magnitudes ranges where the data exist. 
The match of the Schechter form
to red galaxies was explored quantitatively in Paper I and is
reviewed again under errors in \S5.

Evolutionary trends in fitted Schechter function parameters
are shown in Figure 7.
Besides DEEP2 and COMBO-17, this figure
adds data from other recent surveys (2dF [Norberg et al. 2002, 
Madgwick et al. 2002]; SDSS [Blanton et al. 2003, Bell et al. 2003];
VVDS [Ilbert et al. 2004]; FDF [Gabasch et al. 2004]; DEEP1 [Im et al. 2002]).
For reference, the parameters from these
surveys are tabulated in Table 5.
Since the various surveys use different values for $\alpha$, changing
them to the same values 
used by DEEP2 and COMBO-17 would cause small shifts in
$M^*_B$ and $\phi^*$.  For example, if local
All and Blues values were corrected to match
DEEP2 and COMBO-17, $M_B^*$ would brighten by $\sim$ 0.2 mag, 
and $\phi^*$
would decline by $\sim$0.1 dex; these would act to {\it reduce} the
gaps visible in Figure 7 between the local and distant values.  
For red galaxies,
the changes are opposite: $M_B^*$ would dim by $\sim$ 0.15 mag while $\phi^*$
would increase by $\sim$0.08 dex, acting to {\it increase} the
gaps.  All thse corrections are small but add somewhat to the 
uncertainties.

Figure 7 contains the principal results of this paper.
The first conclusion (from the top row) is that
$M^*_B$ has dimmed
for all galaxies, and by roughly the same amount for 
All, Blue, and Red samples.  COMBO-17 (red circles) agrees
well with DEEP2 (black squares) in all three color bins,
and VVDS and FDF agree well with them
for All galaxies (the latter 
do not subdivide by color).  The level of
agreement is impressive because the samples were selected
and measured in very different ways:  COMBO-17 and DEEP2 are
$R$-band selected to $R = 24$, VVDS is $I$-band selected to
$I_{AB}=24$, and FDF is $I$-band selected to $I_{AB} = 26.8$.
VVDS and DEEP2 use spectroscopic redshifts, COMBO-17 uses
high-precision photometric redshifts based on 17 filters, 
and FDF uses photo-z's derived from photometry in 9 bands including
$J$ and $K$.  Despite these differences, values of $M^*_B$ for
all four distant surveys typically agree to within $\pm$0.1 mag.
Agreement for the two local surveys as analyzed by Bell et al.
(2003, SDSS) and Norberg et al. (2002, 2dF) is also good
(though Blanton et al. (2003) find SDSS $M^*_B$ dimmer by 0.4 mag).
In short, a consistent picture for the evolution of 
$M_B^*$ for all galaxies since $ z = 1$  is emerging.  

The dashed grey lines in the top row
are an attempt to fit straight lines
to $M^*_B$ versus redshift using all the data.  It is not clear that this
is advisable since the All data in particular seem to show
a leveling out in $M^*_B$ at intermediate redshifts. If this is
ignored, the coefficients (in Table 6) 
show that the total dimming in $M^*_B$ to $z = 1$ is
1.30$\pm$0.20 mag for the Red sample, 1.31$\pm$0.14 mag for the Blue
sample, and 1.37$\pm$0.31 mag for the All sample.
(The last value is not simply a weighted mean of the first two
because the functions for red and blue galaxies have different
shapes.)
Thus, the evolution of $M^*_B$ for both red and blue galaxies
appears to have been very similar since
$z = 1$.    

Based on DEEP2 alone, we wondered in Paper I whether $M_B^*$ for
red galaxies in fact evolved very much, and indeed the slope
derived from DEEP2 (black squares in Figure 7) is
rather shallow.  However, adding the points from COMBO-17
has steepened the slope for the high-redshift
data, and this is bolstered
by the addition of the local values from SDSS
and 2dF.  We return
to this topic in \S5 
when discussing uncertainties in the red fits.

The bottom row of Figure 7 shows evolution in $\phi^*$ for the 
three color classes.  Agreement is again very good among the
data sets, but now red and blue galaxies evolve quite differently.  
The number density of blue galaxies
remains nearly flat to $z = 1$, whereas the number density
of red galaxies appears smaller back in time.  This
rise, already noted in connection with Figure 6,
repeats very closely the
pattern found by B04, whose data  
showed a gradual rise in $\phi^*$ since $z \sim 0.8$ by a factor
of $\sim$2, precededed by a steeper rise before that 
near $z = 1$.  The
new data from DEEP2, which are completely independent,
also show a steep rise near $ z = 1$ 
followed by a shallower rise after that.

Formal values can be calculated for the decline
in $\phi^*$ at $ z = 0.8$
and at $ z = 1$ using
the new DEEP2 and COMBO-17 data together with
the updated values of local $\phi^*$ shown in Figure 7.
The mean value of $\phi^*$ at $z = 1$ is found to be
0.95$\times 10^{-3} \pm$14\%, where the value comes
from interpolating the DEEP2 and COMBO-17 data at $z = 0.9$ and $z = 1.1$
in Table 4
and the error reflects
an assumed uncertainty of 20\% in DEEP2 and COMBO-17 
separately. The local value of $\phi^*$ is taken to be 
3.44$\times 10^{-3} \pm$20\%, where the error is a conservative
estimate for the mean of the two measured values in Figure 7. 
The formal value for the rise in red $\phi^*$ from
$z=1$ to now is therefore 3.6$\pm$24\%
(0.56$\pm$0.09 dex), and the rise since
$z = 0.8$ is 2.3$\pm$24\% (0.36$\pm$0.09 dex).  
These are formal values based on
the fitted values for $\phi^*$;
potential errors and uncertainties are discussed in \S5.

The fall in the number of red galaxies back in time measured here
does not agree 
with the earlier result from DEEP1 by Im et al. (2002)
in which $\phi^*$ for red galaxies was claimed to have
held constant since $ z \sim 1$.
The two redshift bins 
from Im et al. are plotted as crosses in Figure 7,
where they lie both low and remain constant back in time.
%The nearer value is probably not a concern, as the
%number of galaxies in that bin of the survey was low.
%  Furthermore,  
Im et al. applied a very stringent cut to define their sample,
targeting only morphologically normal, spheroid-dominated
E/S0s having red colors that are
consistent with passively fading stellar populations. 
Their numbers therefore have to be corrected
upwards in any event by $\sim30$\% to account for non-E/S0 contamination
on the distant red sequence 
(Bell et al. 2004a, Weiner et al. 2005).  However, 
the real difference between Im et al. and DEEP2 is
nearly a factor of two, based on
counts by DEEP2 over the identical region.
The reason for this bigger discrepancy has not yet been
unravelled and signals that the cut used by Im et al.
to define their sample was even more stringent than thought.  Finally,
Im et al. compared their distant values
to an earlier estimate for the local number density
of spheroidal galaxies that is considerably lower than the 
values used here.
When all of these factors are combined (and coupled with a new and somewhat
larger cosmic variance estimate), it is easy to see why
Im et al. reached they conclusion they did.
However, this case highlights the problems introduced
by selecting distant red galaxy samples in different ways, 
to which we return later below.

Before leaving Schechter fits, we
report a further test that divided the DEEP2
and COMBO-17 blue samples into two equal
halves to see whether ``Moderately Blue''
galaxies evolve differently from ``Very Blue'' galaxies.
This repeats a test reported in Paper I for DEEP2
but now adds COMBO-17.
The method of division used sloping lines
that ran parallel to the red-sequence color-magnitude
relation and bisected the blue sample in each redshift bin into equal 
color halves.
The lines used for DEEP2 are illustrated in Figure 4 of Paper I;
those for COMBO-17 were similar.  
Dynamically adjusted zero points for each redshift bin
were used in preference to a constant color zeropoint because the latter
would yield a spurious evolution if blue 
galaxies were reddening with time, as suggested by the  
motion of the median dividing line by $\sim0.1$ mag for
DEEP2 in Paper I.
With this approach,
$M^*_B$ for Moderately Blue galaxies in DEEP2 was found to
average 0.7 mag brighter than for Very Blue galaxies, 
as expected from
the sloping color-magnitude relation for blue galaxies.
Apart from that, the two blue sub-samples evolve similarly,
with $\phi^*$ 
holding constant for both halves separately
and values of $M^*_B$ retaining a constant offset versus redshift.
COMBO-17 confirms these conclusions.
This sameness of evolution is perhaps surprising---we might
have expected Moderately Blue galaxies to evolve in a
way that is intermediate between Very Blue and Red galaxies.
Work in progress shows, for example, that the
clustering of Moderately Blue galaxies is indeed intermediate 
between the outermost color classes (A. Coil
et al., in prep.).
To the contrary, the data suggest that
blue galaxies are evolving {\it as a bloc} in
the CM diagram (apart from a possible dilation or expansion
in their total color
range, which cannot be tested in the present data). 

We end this section by comparing to other 
published luminosity functions 
divided by color classes.
The discovery of color bimodality
is rather recent, and the study by Giallongo et al. (2005) is one of only two
that divide distant galaxies by restframe color
as we do.  Unfortunately, a quantitative comparison cannot be given
because no results were presented for epoch $ z \sim 1$ specifically,
and our data do not go farther than that. 
Plots in Giallongo et al. 
agree with ours in showing similar dimming for both red and
blue galaxies, a constant number of blue galaxies, and a rise in
the number of red galaxies  since $ z = 1$.  Though the Giallongo et al. 
sample is much smaller than ours, 
it goes roughly two magnitudes fainter and is therefore
valuable for establishing the existence of a {\it turnover} in
the faint red luminosity function at $ z \sim 1$.

A similar turnover is seen in the second study, by 
Cross et al. (2004), who counted red-selected galaxies and galaxies
morphologically selected to be spheroids regardless of color, based on 
ACS images.  Their counts agree well with ours
despite  their small sample size of 72 galaxies.  The main difference 
with us is
an even steeper turnover in faint-end slope near $z \sim 1$ 
in their red-selected
sample, for which they find $\alpha = \sim+0.3$; this is reduced to
$\alpha \sim -0.5$ when blue spheroids are included.

Several other studies have attempted to count galaxies
in various ways
to see whether spheroids are disappearing back in time.
Reviews can be found in Schade et al. (1999) and Im et al. (2001).
Motivated by predictions of semi-analytic models, Kauffmann
et al. (1996) reanalyzed CFRS data and claimed a drop
in spheroid density, but
their conclusions were disputed (Totani \& Yoshii 1998).
Schade et al. (1999) and Menanteau et al. (1999)
counted morphologically normal
$R^{1/4}$-law objects in HST images 
out to $ z \sim 1$ and concluded that there was indeed
no drop.  Sample and field sizes were small in both cases
and no color cuts were applied, with the result that half or more
of all distant objects were blue.  Thus, there is no contradiction
with the present study though the prevalence (again) of
blue spheroids raises interesting questions.  We return to
the topic of blue spheroids in the Discussion section.

Our final reference is to
CFRS, the pioneering study that first attempted
to calculate the luminosity functions
of distant red and blue galaxies separately (Lilly et al. 1995b). 
For blue galaxies, CFRS claimed a steepening in total
faint-end slope back in time to $z = 1$.  As noted, we have
refrained from drawing any strong conclusions about faint-end slope
evolution from our data, despite the fact
that DEEP2 and COMBO-17 have many more galaxies 
and go 1.5 magnitudes deeper than CFRS.  In retrospect,  
the CFRS data do not look strong enough to support that claim.
For red galaxies,
CFRS found no evolution in either $M^*_B$ or $\phi^*$, whereas  we
find a dimming of $M^*_B$ by
$\gtsim$1 mag and a rise in $\phi^*$ by a factor of $\sim$4
(since $z = 1$).
Part of the difference may be that, lacking 
knowledge of color bimodality, CFRS used
a non-evolving color cut that 
did not quite hit the valley at high redshift.
Regardless,
CFRS projected a picture in which red
galaxies have been rather static since $z=1$, while
blue galaxies have significantly changed.  The picture derived
in this paper is that blue galaxies
are rather constant in number (though fading)
over this time interval, while red galaxies are
actively being generated.  The general impression in CFRS of active
blue galaxies versus passive red galaxies is thus essentially 
opposite to what we find.  

\subsection{Luminosity Density}

Luminosity density provides an estimate of the total
amount of light emitted by galaxies per unit volume. The luminosity
density (in Johnson $B$ band) in this work is obtained assuming the 
Schechter form of the luminosity function: 
\begin{equation}
j_B = {\int L \phi(L) dL} = L^*\phi^*\Gamma(\alpha+2),
\end{equation}
where $j_B$ is calculated in solar units using
$M_{B\solar}$=5.48 (Binney \& Merrifield 1998) and
$\Gamma$ is the Gamma function. Use of this expression
entails extrapolation
over faint magnitudes that are not observed, the more so at high
redshifts. 
However, fitting a given bright-end
data set assuming different values of  $\alpha$ 
tends to leave the product
$L^*\phi^*$ unchanged, which means that
most of the uncertainty comes from $\Gamma$.
For example, changing $\alpha$ from $-1.3$ to $-1.7$, 
as suggested by VVDS for their All
sample at $ z = 1.1$, changes $\Gamma$ by 230\%.
This case is extreme, however.  Values of $\alpha$ 
for Blue and All galaxies from nearly all other studies range between
$-1.0$ and $-1.3$, which implies a total change
in $\Gamma$ of only 30\%.
Plausible red $\alpha$'s range in value from $-0.5$ to $-1.0$,
which changes $j_B$ by only 11\%.  We conclude that,
as long as $\alpha$'s remain below $-1.3$, uncertainties
on $j_B$ are small.

The resultant $B$-band luminosity densities 
are plotted versus redshift in Figure 8.  
To the previously shown local points
we have added a second value for
SDSS measured by Blanton et al. (2003).
This latter value has been multiplied by 62\% and
38\% to obtain the fraction of $B$-band light in 
blue and red galaxies separately, based on fractional
light contributions from 
Hogg et al. (2002).  

Local values agree remarkably
well for all three color classes.  Relative to them,
DEEP2, COMBO-17, and FDF 
show at most a mild decline in $j_B$ for all galaxies
with time.
The fall in VVDS is nearly twice as large owing
to their steeper faint-end slope at
high redshift;  
%This demonstrates the importance of assuming
%a constant Schechter-function shape at all redshifts,
%as we have done here. 
%If Ilbert et al. are correct,
%the DEEP2/COMBO-17/FDF estimates here are too low at
%$z~\sim$ 1 by nearly a factor of 2.
however, as noted, FDF 
goes 10 times fainter and does not see such steepening.
If constant $\alpha$ is adopted, as in DEEP2, COMBO-17,
and FDF, the data indicate that $j_B$
for all galaxies has fallen by 
about a factor of $\sim$2 since $z=1$. 
The CFRS survey (Lilly et al. 1996) is also plotted
in Figure 8 and shows
a much steeper decline in $j_B$ after $z = 1$,
like VVDS.   Some of this may
come from their steeper faint-end slope
at high redshift, but part also
comes from their adopted low 
local value (see figure), 
which they took from Loveday et al. (1992). 
The new value measured by 2dF and SDSS is about 30\% higher.

Turning to the individual color
classes, we see that the 
luminosity density of blue galaxies (in  Figure 8$b$) as measured by DEEP2
and COMBO-17 evolves somewhat more than total luminosity density, falling
by a factor of $\sim$3 since $z$ = 1.  This is consistent
with the constant value of $\phi^*$ and the change in $M^*_B$ of 1.3 mag
seen above for blue galaxies.  
Red galaxies are shown in
Figure 8$c$, repeating a similar figure
from B04.  The conclusions are the same---the
$B$-band luminosity density of red galaxies has
remained essentially {\it flat}
since $ z = 0.9$ and was possibly rising before that.
The flat section is caused by the dimming of $M_B$
coupled with the rise in 
$\phi^*$ so that $j_B$ remains constant.  Before $z = 0.9$,
the steep decline in $\phi^*$ wins out, and total
$j_B$ seems to be lower.  

To summarize, DEEP2 
agrees with both old and new analyses of COMBO-17 in showing 
that the $B$-band 
luminosity density for red galaxies has remained 
nearly constant since $z=0.9$, with a possible fall
beyond that.  Despite the fact that only the upper
part of the function is observed at $z \sim 1$,  DEEP2
and COMBO-17 agree within 20\%, and extrapolation errors
must be small because the red function is known to 
turn over, even  at high redshift (Cross et al. 2004,
Giallongo et al. 2005).
Barring actual loss of
galaxies from the samples (see below), 
the constancy of $j_B$ for red galaxies 
after $ z \sim 1$ should therefore be well 
established.
Since stellar mass-to-light ratios are increasing with time (see below),
this constancy implies
that the 
stellar mass bound up in red galaxies has increased 
markedly since $z = 1$, as stressed by B04.
Given the  strong implications of this result
for galaxy formation, it is 
advisable to go back and review the errors and
assumptions, which we do in the next section.
Readers not interested in these details
should skip directly to \S6.

\section{Errors, Assumptions, and Uncertainties}

This section focuses on red-sequence galaxies, although many of the
conclusions apply equally well to the other color classes.
A major issue is whether the apparent fall in the number density of bright 
red galaxies back in time is due to 
galaxies that are being left out in different stages
of the analysis.  A second issue is the extent to
which the conclusions are sensitive to fitting the counts with
Schechter functions
having constant, non-evolving $\alpha$.
These effects and others are discussed below.

1) {\it Completeness of the photometric catalogues:} 
The COMBO-17 photometric catalogue has a 5-$\sigma$
detection limit down to $R_{AB} \sim 26$, nearly two magnitudes 
below what is needed for the luminosity function
surveys ($R_{AB} \sim 24$).  The DEEP2 catalog is shallower
but also adequate
(5-$\sigma$ limit $R = 24.5$, Coil et al. 2004).
DEEP2 makes an additional cut in
surface brightness when designing the DEIMOS masks
that deletes low-surface-brightness 
galaxies in the last half-magnitude bin 
(see Paper I).  However, this is largely taken into account
by calculating weights as a function of color as well as
magnitude.  Furthermore, this cut would not affect early-type galaxies,
which have high surface brightness.   

Errors in star-galaxy separation may result in either too few
or too many galaxies, depending on the errors.
Star-galaxy separation in COMBO-17 is based on 
17-color photometry and is in general highly efficient; red counts
near $ z = 1$ may be $\sim$10\% too high
owing to K-star interlopers (W04), but this would tend
to {\it over}estimate red galaxies.  Star-galaxy
separation for DEEP2 was tested in Paper I 
using high-resolution HST images that cover part of the DEEP2 region  
in the Groth Survey Strip.  Misclassification of red galaxies
as stars amounted to $\sim$10\%, but these were 
nearly cancelled by stars misclassified as galaxies,
so the net effect was nil.  Finally, checks of both
data sets show that almost all galaxies to $R = 24$ have adequate
photometry in all bands, and the few ($\sim$ 1\%) DEEP2 galaxies that
do not have $B$-band 
photometry (``$B$-dropouts'') are corrected for statistically
in the weights (Paper I). 

2) {\it Dividing red galaxies from blue galaxies:}  This is done 
using restframe values of $U-B$ in DEEP2 and $U-V$ in COMBO-17.
Errors in the zero points of these systems do not matter
even if they vary as a function of
redshift, since the dividing line
is adjusted empirically to fit the 
color valley in each redshift bin.  Division of 
local samples into red and blue galaxies
has been done in different ways, but results
are not sensitive to the method used.  
Madgwick et al. (2002) separated 2dF 
galaxies by spectral type,
whereas Bell et al. (2003) separated
SDSS galaxies
based on concentration and optical color.  Since there is very
high correlation among these properties for
local galaxies, it is not surprising that
the results agree well, as shown in Table 5.

3) {\it Errors in $M_B$:} 
Weiner et al. (2005) checked DEEP2 restframe values of $M_B$ 
against values derived
from GIM2D photometry of Groth Strip HST images by
Simard et al. (1999).   GIM2D fitted model  
bulge+disk  profiles to $V$ and $I$ images to find total
magnitudes, whereas
the DEEP2 Hawaii CFHT photometry approximates each object by a
Gaussian profile on the $BRI$ ground-based images.
Despite these different
methods plus
uncertainties in HST WFPC2 photometric zero points and 
charge-transfer-efficiency corrections, the zero points of both 
$M_B$ systems agreed to 0.07 mag.
Furthermore, any mismatch in the magnitude systems 
for distant and local surveys would cause
only an error in the evolution of $M_B$, not $\phi^*$.
The COMBO-17 luminosities have never been independently checked.
However, the detailed SED information allows a precise calculation of
the rest-frame B-band luminosity at all $z<1$ without extrapolation.
The main source of error is the
photo-z error, which translates into a distance error. Most objects
should have luminosity errors between 10\% and 20\%.
Local survey values of $M_B$ are claimed to be accurate to 0.1 to 0.02
mag (for photographic 2dF and CCD SDSS magnitudes respectively).

4) {\it $R$-band selection effect:} 
The use of the $R$-band for selecting DEEP2 and COMBO-17
corresponds to restframe 3300 \AA~at $ z = 1$,
where the SEDs of red galaxies
are rather dim.  It might be thought that red galaxies
are being ``missed" on that account. In practice, this is
completely allowed for by calculating limiting absolute magnitudes 
at each redshift as a function of both redshift {\it and} color using
CM diagrams like those 
illustrated in Figure 4.  The limiting $M_B^*$ magnitude to
which the counts are complete at each redshift and color  is
well understood.

5) {\it Redshift completeness and accuracy:}  Redshift 
completeness has been simulated for COMBO-17 using Monte Carlo
methods (W01, W03, W04).  
From these, it appears that redshifts are highly complete for
red galaxies in COMBO-17 but substantially incomplete for blue galaxies
in the last magnitude bin. This is consistent with the finding that 
nearly all failed galaxies are faint blue galaxies.
Testing the completeness model independently is difficult. However, we
have predicted total galaxy number counts from the best-fit luminosity
functions, including extrapolations to the faint end and to somewhat
higher redshifts, and find remarkable consistency of the prediction with
the observed galaxy number counts in COMBO-17. Of course, the power of
this test to assess the completeness of a sub-sample in any particular
redshift bin is extremely limited.

Redshift incompleteness in DEEP2 
was discussed in Paper I.  Using the minimal versus the average model
for failed redshifts typically results in no change
in $M_B$ and a change in $\phi^*$ of
10-20\%, a small effect compared
to the total measured evolution in $\phi^*$ out to $z = 1$.
Our preferred choice for red galaxies
results in {\it higher} values of $\phi^*$, which minimizes the observed
evolution.
For red galaxies, Paper I also considered a third, extreme model 
in which all failed red galaxies were assumed 
to be located in {\it whatever redshift bin was under consideration}.
It is possible to do this without 
redshifts because red-sequence galaxies near
$ z = 0.7-1.1$ have apparent $R-I > 1.25$ and show up as
a well defined ridge in the apparent CM diagram (see Figure 1 of Paper I).
This test amounts to counting all possible red galaxies
and dumping all of them with unknown redshifts
into a {\it single} redshift bin.  Even this extreme approach hardly affects
results out to $z = 0.9$ (though it does increase
both counts and $j_B$ at $ z = 1.1$).  

Errors for
DEEP2 redshifts used here
are negligible ($< 10^{-4}$ in $z$); catastrophic errors are at
the level of 1\%.  The accuracy of COMBO-17
photo-z's has been studied using simulations,
yielding an estimated rms error of 0.03, which agrees with the
spectroscopic cross-check in W04. The effect of
such errors on the red luminosity function
was simulated by B04 and shown to be small.

6) {\it Formal Schechter fit errors and cosmic variance:} The errors
in $\phi^*$ for red galaxies 
in Table 4 include  Poisson noise and cosmic variance.  In the two most
distant bins, these errors 
are comparable and give an rms error 
in number density of about
20\% per survey, or 14\% for the two together.  The error
in the local zero point of $\phi^*$ for
red galaxies is estimated to be 20\%, yielding a formal
rms error for the total difference between near and far samples
of 24\%, or 0.09 dex.  This is small
compared to the formal rise in $\phi^*$ since
$z = 1.0$ of 3.6, or 0.56 dex.

This completes the list of possible observational errors 
and selection effects.  We turn now to various theoretical
and model assumptions.

7) {\it The assumption of a constant Schechter-function shape 
at all redshifts:}  In practice, this means
1) that the Schechter formula is a good  
match to the bright end of the luminosity function,
and 2) that the shape parameter $\alpha$ can be assumed to hold 
constant with redshift.
A breakdown in either one
of these assumptions will produce a mismatch
between the data and the model, causing errors in
both $M_B$ and $\phi^*$.  If the shape of the real function
is constant with redshift but is not well fitted by the model,
the fitted parameters will drift with $z$
as the data are limited to progressively brighter
magnitudes at higher redshift.  Any real evolution
in shape may cause additional errors.

Inspection of Figure 6
suggests that our Schechter
model (with the adopted value of $\alpha = -0.5$) {\it looks}
like a good fit for red galaxies.  Paper I
tested this quantitatively by truncating DEEP2 data
in nearer bins at brighter magnitudes corresponding to 
the cutoffs in more
distant bins.  For red galaxies, a drift of $M^*_B$ 
of $\sim0.1$ mag toward fainter values was seen with more truncation, 
whereas the measured evolution is 
a brightening of $M^*_B$ by 1.3 back in time. 
The quantity $\phi^*$ drifted upwards by $\sim$0.10 dex,
whereas the measured evolution is a fall of
0.36 dex to $ z = 0.8$ and 0.56 dex to $z = 1.0$ and $z=1.0$.  
The measured evolutions are therefore if anything an
{\it under}estimate owing to errors in assumed Schechter
function shape.  
High-redshift bins cannot be tested in the same way, but visual
inspection indicates that the match between data and model 
remains good.

However, certain recent data hint that
the shape of the red luminosity function  
{\it may} be evolving with time, which would invalidate the
assumption of strictly constant $\alpha$.  For example,
de Lucia et al. (2004) see a deficit in the number of faint  
red galaxies in rich clusters at $ z = 0.8$ compared to 
Coma, and
Kodama et al. (2004) detect a similar deficit of faint red
galaxies in overdense
field regions at $ z \sim1$.  As noted above, Cross
et al. (2004) report a turnover in the counts of distant
field red galaxies at $z \sim1$ that
is stronger than reported for red galaxies locally.  These 
studies at high redshift all go 1-2 mag fainter than
DEEP2 and COMBO-17 and are thus better determinants of
$\alpha$ in distant samples.  Added to this is the potential
flattening of faint-end slope seen by COMBO-17
in its nearest redshift bin (see Figure 6),
which resembles the flattish faint-end slope seen in local SDSS data
by Bell et al. (2003) (though DEEP2 and 2DF [Madgwick
et al. 2002] disagree; see Table 5).  

A flatter 
faint-end slope with time might indicate
that smaller red-sequence galaxies 
formed later than larger ones.  Further support for this
are findings by
McIntosh et al. (2005) that smaller red-sequence
galaxies evolve faster in surface brightness 
back in time than larger ones,
by van der Wel et al. (2005) that the zeropoint of the fundamental
plane evolves faster back in time
for small galaxies than for brighter ones,
by Bundy et al. (2005) that the crossover mass between spheroids
and disk galaxies was larger in the past, by 
Treu et al. (2005a,b) that small spheroidal galaxies
arrived later on the red sequence
than large ones, and by Im et al. (2001)
and Cross et al. (2004) that distant blue spheroidal galaxies
are significantly smaller than red ones and would
preferentially populate the {\it bottom} end of the red sequence
if they were fading towards it.

In short, recent data may be
pointing toward a mild flattening in the faint-end slope of the luminosity
function for red galaxies, with small spheroidal galaxies forming
later than large ones.  However, even if this is occurring, 
the effect on Schechter parameters is not large.  Suppose for example that
$\alpha$ is evolving from $-0.5$ at $z \sim 1$ to $-1.0$ locally,
which is the largest change conceivably allowed by the data.  
Values for DEEP2 and COMBO-17  
use $\alpha = -0.5$, while
local surveys find the average value $\alpha = -0.65$ (Table 5). 
As an experiment, we have 
refitted local data using $\alpha = -1.0$,   
\footnote{For example, using sample data available at
  \url{http://www.mpia-hd.mpg.de/homes/bell/data/glfearlycol.out}},
and find that
$M_B^*$ brightens by 0.3 mag,  $\phi^*$ declines
by 0.18 dex, and $j_B$ declines by 2\%.  
These changes are all small compared to the  
claimed evolution.  We also noted above that $\Gamma$ itself
varies by only 11\% over the entire range $\alpha = -0.5$ to $-1.0$.
The conclusion is that faint-end slope is sufficiently
flat for red galaxies that  
total luminosity density is determined quite well
by data down to $L^*$, as is the case at all redshifts here.

8) {\it Using color as a surrogate for morphological type:}
In focusing on red galaxies, we are implicitly
assuming that restframe color is a good way of 
finding spheroid-dominated
E/S0 types at high redshifts.  The method 
clearly works well at low redshifts,
where only 15-20\% of nearby red-sequence galaxies have Hubble
types later than S0, being mostly edge-on and dust-reddened 
(Strateva et al. 2001,
Weiner et al. 2005).  However, contamination by
non-spheroidal galaxies is larger at higher redshifts,
amounting to 
30\% at $z \sim 0.75$ (Bell et al. 2004a, Weiner et al. 
2005), 
and may increase beyond that (Cimatti et al. 2002a, 2003;
Yan \& Thompson 2003; Gilbank et al. 2003;
Moustakas et al. 2004).  Because contamination
appears to be larger back in time,
our measured rise in the number of red galaxies
is a lower limit to the rise of 
{\it morphologically normal} E/S0s.
Assuming that contamination on the
red sequence at $z = 1$ amounts to 30\% 
would increase the rise in normal E/S0s 
by 0.06 dex (to 0.62 dex) since that time.  The
true correction could be larger since contamination
at $ z = 1$ may be higher than at $ z = 0.75$.

9) {\it Uncertainties in evolving stellar mass-to-light ratios:}
These come into play when converting luminosity density
into the more fundamental quantity 
stellar mass, as shown below.  As noted,
$j_B$ for bright red galaxies above $L_B^*$
is nearly constant out to $z = 0.9$, and may be lower
before that.
Since stellar mass-to-light ratios
are increasing with time, this means that the stellar
mass bound up in red-sequence galaxies 
must also 
increase.  But by how much?  B04 investigated
this question using single-burst, passively evolving models,
but these are only one option.  We have investigated further
possibilities such as $\tau$ models, ``frosting" models with a continuing
low level of star formation (e.g., Gebhardt et al. 2003), 
and ``quenched" models in which star formation is shut down
abruptly at some epoch (J. Harker et al., in prep.).
Models are set up to match the average color of
red-sequence galaxies today and the relatively 
small amount of color evolution that is seen since $z = 1$  
($\Delta(U-B) = 0.15-0.25$ mag;
Bell et al. 2004b, Weiner et al. 2005, Koo et al. 2005).
Recipes that satisfy these constraints all yield fading in $M_B^*$
between 1 and 2 mag.  

Observationally measured brightenings are consistent with these
model estimates.  
The fundamental plane zeropoint  brightens by 1-2 mag
(van Dokkum et al. 2000, Gebhardt et al. 2003, van Dokkum \& Ellis 2003,
Treu et al. 2005a,b, van der Wel et al. 2005),
the magnitude-radius zeropoint brightens by  
1-1.6 mag (Trujillo \& Aguerri 2004; McIntosh et al. 2005),
and $M_B^*$ for red galaxies brightens by 1.3 mag (this paper).
These observational
shifts do not necessarily represent the fading of
stellar populations if galaxies are merging or otherwise evolving in 
radius or $\sigma$ and changing their structure.
Nevertheless, it is
striking that the amount of evolution from the various scaling laws is
very close to the evolution seen in $L_B^*$, and this in turn is
near the middle of the range predicted by 
the stellar population models.
Combining all results together, 
we adopt 1.0 mag (0.4 dex) as the 
{\it minimum} increase in the mass-to-light ratio
of a typical massive red galaxy
since $ z = 1$.  Since luminosity density $j_B$ has remained constant, this
is also the minimum increase in red stellar mass over the same period.

We collect together the following potential
corrections to the above 
estimates of $\Delta\phi^*$ for E/S0 galaxies 
from $ z = 1$ to now.  A positive
sign means that the previously estimated rise
in $\phi^*$ would be even bigger.  The factors are: a possible
mismatch between the adopted Schechter function
shape and the actual bright end of the luminosity function, +0.10 dex;
a possible nearby flattening of $\alpha$ from --0.5 to --1.0,
--0.18 dex; and contamination by distant non-E/S0s, +0.06 dex.
Each effect is small, two of the three are
hypothetical, and collectively they
tend to cancel.  For these reasons, we do not 
apply any corrections to the above-measured rise 
in $\phi^*$, namely, 
0.36 dex since $ z = 0.8$ and 0.56 dex since $ z = 1$.

We have not thus far uncovered
any ``smoking gun" as to why our counts of red galaxies in 
DEEP2 or COMBO-17
should be seriously in error.
Nevertheless, there is a worrisome feature of the data, and that is the fact
that the two surveys, DEEP2 and COMBO-17, 
do not show much {\it internal} evolution
in $\phi^*$ over most of their well-measured
range.  This point was mentioned in Paper I in connection
with DEEP2, and it is visible again in Figure 7, which
plots both DEEP2 and COMBO-17.  In both data sets, there is a 
jump in $\phi^*$ of $\sim0.2$ dex from the distant surveys to the local surveys,
and another jump of $\sim0.3$ dex between $ z = 0.9$ and $z = 1.1$.  In 
between, $\phi^*$ tends to plateau.
This stagnation is illustrated another way in Figure 9, which overplots  
$1/V_{max}$ data points from the lowest bin at $z = 0.3$ 
from both surveys on top of
the data points for distant bins.    As previously noted, the red counts 
at bright magnitudes tend
to translate {\it parallel} to themselves, and one might
even conclude that {\it no evolution} in the luminosity function,
and thus in number density,  
has occurred.  Both DEEP2 and COMBO-17 are
similar in this regard.  This degeneracy 
could be broken by having fainter data, but our two surveys do not
go deep enough to permit this.

At this point, the argument involving stellar mass-to-light-ratios 
assumes great importance.  Imagine replotting Figure 9 versus
{\it stellar mass} instead of $M_B$.  To account for
the evolution in mass-to-light ratio, based on the discussion
immediately above, the counts at $z \sim 1$
would have to be shifted over to the right by 
at least 1 mag, which would 
produce a vertical offset with respect to the low-redshift counts 
by about a factor of four (0.6 dex) near $M_B= -22$, where all
curves superimpose.
This is nearly identical to our previously measured fall-off of 0.56 dex
based on $\phi^*$.
Thus, once mass-to-light ratio evolution is allowed for,
the number of massive red galaxies {\it at fixed stellar mass}
is increasing about as fast as the formal fit for $\phi^*$.
This argument is very similar to the 
one applied by B04 to {\it total} luminosity density, but 
we apply it here to individual
galaxy masses at the top end of the luminosity function.  
The distinction is a small one, but we wish
to emphasize that the present version of the argument relies entirely on
data that are observed.

In summary, the conclusion that the number density of red galaxies
has risen significantly
is supported by three pieces of evidence.
One is the fitted results for $\phi^*$, which yield
formally significant values in number density
that agree very well between 
DEEP2 and COMBO-17.  However, these
may be suspect because of difficulties
in comparing to local surveys and because
both surveys may be prone to unknown
errors at the far edge of their range.  
The next two arguments are therefore 
important and rest on the 
high probability that
stellar mass-to-light ratios of red-sequence galaxies have evolved
by at least one magnitude since $ z = 1$.  If so, the
constancy of total luminosity density $j_B$ implies
a rise in total red stellar mass by 
the same factor.  Finally, the 
constancy of the luminosity function itself (at the bright end) 
implies that a comparable growth in the number of
massive red galaxies (at fixed stellar mass) must also 
have occurred.  In what follows, we adopt the formal
values of $\phi^*$ ($\Delta\phi^* = 0.36\pm0.09$ dex since $ z = 0.8$ and
$0.56\pm0.09$ dex since $ z = 1.0$) as our measures of the 
rise in the number of red-sequence galaxies.

\section{Discussion}

\subsection{A ``Mixed" Scenario
for the Formation of Spheroidal Galaxies}

The most interesting conclusion to
emerge so far from the study of luminosity functions since $ z = 1$
is the growth in red
galaxies {\it over recent times}.  We have argued that this
translates to a similar, or even steeper, rise in
the number of morphologically normal, spheroid-dominated E/S0s.
Barring mergers among a large and 
undiscovered population---which would have
to be tiny and/or highly obscured to avoid detection in our
and other surveys---this discovery means
that the immediate
precursors of most massive E/S0s
{\it must be visible in existing samples} at $ z = 1$ and below.  
The implications of this were discussed
by B04.  We build on their arguments by adding 
data on the Blue luminosity function (measured here
by DEEP2 and COMBO-17) and the properties of local spheroidal
galaxies, which we will argue also have strong implications
for formation scenarios.  Our
discussion focuses on {\it typical} red galaxies
at high redshift
since DEEP2 and COMBO-17 sample all galaxies regardless
of location.
The red sequence in distant clusters has also been extensively
studied, but in general
we do not try to fold these data into the present
picture at this time.

It is well established that residence on the red sequence requires
that the star formation rate be quenched, or at least strongly
reduced.  Stellar
populations become red enough to join the red sequence just 1-2 Gyr
after star formation is 
stopped (J. Harker et al., in prep.), but, in order for them
to stay there, the star formation
rate must remain low.  For example, Gebhardt et al. (2003) explored
a ``frosting model" with an early high rate of
star formation, followed by a slowly
decaying $\tau$ component.  Based on
colors, they found that only 7\% of total stellar mass  
could be formed in the $\tau$ component; more recent
limits based on O II in distant red galaxies
are even lower (N. Konidaris et al., in prep.).
In short, as B04 pointed out, the large 
build-up seen in red
stellar mass after $ z = 1$ could not have arisen from 
star formation within red galaxies themselves.  
Rather, the stellar mass at the bright end of the red sequence
must have {\it migrated} there via one of three processes:
1) the quenching of blue galaxies,
2) the merging of less-luminous already-quenched red galaxies,
or 3) some combination of the two.
In the following discussion,
we focus on the {\it bright} 
end of the red sequence   
because this is where the data are complete.
Galaxies may of course also be migrating to the lower 
end of the red
sequence as well.

It is helpful to visualize this mass migration as the movement
of progenitor galaxies through
the color-magnitude diagram 
or, more fundamentally, the color-mass diagram.
Sample tracks are shown in Figure 10.  
Two parent regions are illustrated, a narrow red locus
corresponding to
the red sequence, and a broader blue clump, which 
we will call the ``blue cloud."  The rather constant
morphology of the color-magnitude diagram since
$z = 1$ suggests that these parent regions are relatively
stable in size and location.  In reality, they are also
moving as galaxies evolve, but this will not be too
important if individual galaxies move
through them more rapidly.  With this
assumption, we show
the clumps as fixed and the galaxies as
moving through them with time.

Each final galaxy today is represented by its most
massive progenitor at any epoch.  Stellar
mass is migrating toward the upper left corner, where
luminous red galaxies reside.  For a galaxy to get
there, two things must happen: the stellar mass composing the final
galaxy must be assembled via gravitational collapse, and star formation 
must be
quenched.  A key question in the formation 
of red-sequence galaxies is therefore {\it when mass 
assembly occurred relative to star-formation quenching:} that is, did 
quenching occur early in the process of mass build-up, midway, or late?
If extremely early, the pieces that would become the final galaxy migrated
to the red sequence while still small, 
producing a large number of small galaxies on the lower red sequence
that must later merge along the sequence
in a series of ``dry," purely  
stellar mergers.  This is Track A.  
If extremely late, the progenitors grew in mass hierarchically 
while still making stars within the blue cloud. 
Upon quenching, the most massive of them
moved to the head of the red sequence and took up 
residence there without any further dry
mergers whatsoever.  This late-stage quenching scenario (for 
various masses) is shown as the tracks labeled B in Figure 10.
Mixed scenarios are also possible,
involving moderate mass assembly during the star-forming stage,
followed by quenching and continued but limited dry merging along
the red sequence.  These are the tracks labeled C.\footnote{Strictly 
speaking, purely stellar mergers 
increase the stellar mass of a galaxy but leave its color
unchanged.  The arrows for an instantaneous merger should therefore
be horizontal in Figure 10, which could add objectionably
to the scatter on the red sequence if much merging
occurred (Bower et al. 1992).  
On the other hand, the total merging process
might last some time, in which case populations
would age and redden as they grow in mass, causing
the track to be tilted upward,  
which is how we have drawn it in Figure 10.  
Evidently, the total scatter induced by stellar
merging depends on the precise timing and amount of
merging.  We return to this question in \S6.3.}

The tracks in Figure 10 assume that quenching is accompanied (and perhaps
triggered) by a major merger.
A related but different scenario would involve the
pure fading of {\it single} blue galaxies without any merging at all.  
We do not consider this, for two reasons given by
B04.  First, distant blue galaxies are disk-dominated (Bell et al.
2004b, Weiner et al.
2005), and fading alone cannot transform disks into 
spheroids---changing the structural morphology requires a merger.
Second, there do not seem to be
enough blue galaxies in the distant color-magnitude diagram
with masses comparable to those of 
massive red galaxies (B04; Weiner et al. 2005; Paper I).  
Hence, in order to boost mass and to create spheroids, a
final episode involving both quenching {\it and} merging seems to be
required, at least for galaxies on the upper red sequence.  On
the other hand, quenching of
pure disks without merging may well feed the {\it lower} 
red sequence.
Indeed, the local red sequence contains many low-luminosity S0s 
(Binggeli et al. 1988) whose disks were probably
quenched by ram-pressure
stripping or other gas-starvation processes not involving 
merging.\footnote{The 
red ``thick-disks" of nearby spiral
galaxies may be yet other examples of quenching without merging.  
A more likely explanation
is that they are old-disk stars that have been 
dynamically heated over time by encounters with
molecular clouds, disk instabilities, and/or 
small satellite galaxies, in
which case no quenching at all is needed
to explain them.}  Because
our focus here is on {\it massive} red galaxies, which lack 
disks, a final episode of quenching plus
merging is assumed.

Yet a third process by which galaxies might migrate to the 
red sequence is {\it unveiling,} whereby a dusty starburst
is cleansed of its interstellar medium and the underlying
galaxy is revealed.  Such a process might cause the galaxy  
to brighten as dust absorption is removed, but also to redden
as the starburst ages.  However, there is no need to discuss this
case separately because it is already subsumed 
under the above cases.  If the starburst 
is an episode in the
life of a single disk galaxy (e.g., Hammer et al. 2005), then the object
today is a late-type spiral and is irrelevant to the red sequence.  
If 
the starburst has been induced by a merger, then the dusty phase is a
temporary stage between the original blue precursor and the final
red remnant, which does not alter our fundamental model of blue
galaxies merging and turning into red galaxies.  
The arrows in Figure 10 
are meant to connect initial and final states, not represent the
detailed track whereby an object moves from blue to red.
The only assumption that we have made concerning that transition 
is that merger remnants move quickly to the red sequence without
lingering very long as bright blue starbursts.  This 
is required by the fact that few if any bright blue starbursts
are visible 
in the CM diagram (Bell et al. 2004b; Weiner et al.
2005; Paper I).  It is also supported by radiative transfer
models of dust in merging galaxies,
which indicate that the burst itself is heavily cloaked by
dust and is optically nearly invisible (Jonsson et al. 2005).
Thus, starbursting galaxies are hard to tell optically from
non-starbursting galaxies, and both types populate the blue
cloud, as we assume.

The above assembly processes could be represented 
equally well by tracks
in the color-magnitude diagram as in the color-mass
diagram, and the former would be closer to existing data.
However, mass 
is the more fundamental parameter, its behavior under merging is
easier to predict than light (because
dust and starbursts are not a problem), and mass estimates for 
samples of distant galaxies are growing and soon will be
a standard tool
(e.g., Drory et al. 2004, Fontana et al. 2004, Drory et al.
2005, Bundy et al. 2005).  With mass as the size variable, the 
motions of galaxies moving onto the red sequence 
are described by vectors moving both upward (redder) and to the left 
(more massive).  The slopes of the vectors in Figure 10 correspond to
equal-mass mergers (i.e., mass doubling);
unequal mergers would have more vertical vectors.

Yet another perturbation to the model
is the possibility that the most-massive
progenitor might take up residence on the red sequence
and then later merge with a smaller gas-rich galaxy.  The resultant starburst
could briefly move the remnant back to the blue cloud,
followed by subsequent decay back 
onto the red sequence (e.g.,. Charlot \& Silk 1994).  However,
such events (while they last) would create massive blue galaxies, which
we have argued are rare.  The events  must therefore
be short-lived and should 
not greatly distort our basic assumption that,
once the most-massive progenitor galaxy enters the red sequence,
the galaxy remains there permanently.  

We make three generic points before considering 
Tracks A, B, and C further.
First, since the number of massive
spheroidal galaxies (and their associated stellar mass)
has been growing over time, the makeup of
the population is not stable, and mean properties such
as average color, stellar age, etc., are constantly
being skewed by recent
arrivals (the so-called ``progenitor bias" phenomenon
of van Dokkum \& Franx (2001)).  The population as a whole
therefore cannot be modeled using classic, single-burst, 
monolithic-collapse models
(e.g., Eggen, Lynden-Bell, \& Sandage 1962, Larson 1975),
even though certain properties, such as $L^*_B$ and color 
evolution, seem to be well fit by such models---this 
similarity is a coincidence and these
models must be abandoned.
A related point pertains to what we mean by the ``age" of 
a galaxy.  In the monolithic picture, the age of spheroidal
galaxies corresponds to
the epoch at which the mass collapsed and 
the stars were formed (both were the same).
In the new picture, each spheroidal galaxy 
has at least {\it three} characteristic ages---the 
epoch of major mass-assembly,
the epoch of major star formation, and the epoch of
quenching---all of which can be different.\footnote{Our 
use of the word {\it age} here is not meant to
obscure the fact that mass assembly and star formation
are both prolonged processes, so that any particular
``age" must be the mean of events that may have
lasted billions of years.}

The final point is the importance of {\it local} E/S0s
in constraining formation models.  
Four local properties are relevant.  The first is 
the basic fact that
morphology correlates closely with average stellar age---
disk stars are younger and blue, whereas spheroidal 
populations are older and red
(e.g., Baade \& Gaposchkin 1963).  This fundamental datum
motivates our basic assumption that the {\it same} process quenching
star-formation also
altered the morphology from disk-like to spheroidal.
This process is believed to involve mergers with other
similar-sized galaxies (Toomre \& Toomre 1972; Toomre 1977;
Mihos \& Hernquist 1994, 1996; Barnes \& Hernquist 1996).  Many local
merger remnants are known whose 
properties are consistent with their evolving
into spheroidal galaxies once the acute merger phase is over 
(e.g., Schweizer 1982, Schweizer 1986, Hibbard 1995).  
The incidence of spheroid-dominated galaxies is also higher 
in groups and clusters of galaxies (e.g., Dressler 1980, Postman
\& Geller 1984, Hogg et al. 2003, Balogh et al. 2004,
Baldry et al. 2004), where mergers
were more frequent.  

Second, 
local E/S0 galaxies populate a rather tight ``fundamental plane" 
linking
radius, luminosity, and velocity dispersion (Faber et
al. 1987, Dressler et al. 1987, Djorgovski \& Davis 1987).
The tightness of this plane implies that stellar mass-to-light
ratio cannot scatter by more than $\pm$15\% at any point
on the plane.  Two other relations, the color-magnitude
relation (Faber 1973, Sandage \& Visvanathan 1978,
Bower, Lucey \& Ellis 1992)
and the Mg-$\sigma$ relation (e.g., Bender, Burstein \& Faber 1992,
Bernardi et al. 1998, Colless et al. 1999, Worthey and Collobert 2003),
are also quite narrow and
further link the properties of stellar populations to those of
their parent galaxies.  

Third,
the mean light-weighted stellar-population
ages of local Es scatter widely,
from over 10 Gyr down to just
a few Gyr (e.g., Gonzalez 1993, Trager et al.
2000a,  J\o rgensen 1999, Terlevich \& Forbes 2002), with
most being younger than classic
single-burst models would predict (11.4 Gyr if
$z_{form} = 3$).  This large number of young ages allows 
room for the late quenching that is 
required by the luminosity function data.  
On the other hand, there is at most a weak
trend in stellar age with mass or $\sigma$ {\it along} the 
red sequence 
(Trager et al. 2001, Terlevich \& Forbes 2002, Bernardi et al. 2005), 
so that stellar ages and metallicities 
scatter substantially at every point on all three relations.
To keep the relations tight, Worthey et al. (1995)
posited that an {\it anti}-correlation must exist between age and metallicity
at constant mass and/or $\sigma$, as later verified
by J\o rgensen (1999), Trager et al. (2000b), and Bernardi et al. (2005).  

The fourth and final point is that
nearby Es populate a {\it structure sequence}, in which small
objects rotate strongly, are flattened by rotation, 
and have disky isophotes and steep central
surface-brightness profiles, whereas massive objects rotate weakly, 
are flattened
by anisotropic velocity dispersions, and have boxy
isophotes and core-type central profiles 
(e.g., Davies et al. 1983, Bender, Burstein \& Faber 1992,
Faber et al. 1997).
At the low-mass end, these properties
connect smoothly with S0s and, through them,
to the remainder of the Hubble sequence (Kormendy \& Bender 1996).
Several authors have suggested that this structure sequence 
can be explained broadly by assuming
that small spheroidals were produced via mergers
of gas-rich, ``wet" progenitors,
while more massive spheroidals were produced
by progressively ``dry," purely stellar mergers of smaller 
spheroidals
(e.g., Bender, Burstein \& Faber 1992, Kormendy \& Bender 1996,
Faber et al. 1997).  This scenario
implies that the most massive of today's spheroidals were created mainly by
purely stellar mergers.

With these points as background, 
we return to the tracks in Figure 10.  The early-quenching
scenario (Track A) has most of its
mass-assembly occurring in dry mergers {\it along} the red sequence.
This can be ruled out on two grounds.  First, to
produce the large amount of stellar mass bound up
in massive red-sequence galaxies would require
a huge reservoir of small, faint galaxies on the lower
red sequence.  This excess 
is not detected at any redshift---the local red-sequence luminosity function
is at most flat ($\alpha \sim -1.0$)
and if anything turns over more steeply at higher
redshifts (Cross et al., 2004, Kodama et al.
2004, Giallongo et al. 2005, Figure 6 here).
The required reservoir of small red galaxies 
therefore does not exist.
Second, building up massive red galaxies 
from purely dry mergers along the red sequence
would yield stellar populations whose metallicities are 
uncorrelated with stellar age and whose ages and
metallicities would converge to
a single value at high masses after many mergers had occurred.
This fails to match the fundamental plane, color-magnitude,
and Mg-$\sigma$ relations, which indicate that age and
$Z$ must be {\it anti}-correlated at each location, nor does it match the strong
spread in age and $Z$ that exists even among massive galaxies
(Trager et al. 2001, Terlevich \& Forbes 2002).
(The upward tilt that we have placed on the arrow 
representing dry mergers on Track A reflects the
probable increase in mean stellar age during dry merging,
not an increase in mean metallicity.)

The late-quenching scenario (Track B) is extreme in the opposite sense 
of having {\it no} dry merging at all along the red sequence.  
In this picture, massive present-day spheroidals were
formed via a single merger of two very massive 
{\it gas-rich} progenitors.
The main reason for ruling out this scenario is the structure
sequence among local Es, which, as noted, implies that  massive Es 
were formed by dry, gas-poor mergers.  The signatures of such
mergers are distinctive because the precursors are dynamically
hot, yielding fuzzy tidal tails without sharp boundaries; 
many examples of such dry mergers can be seen
in local catalogues (e.g., Arp 1966), so it is clear
that they are occurring.

The mixed scenario (Track C) involves early mass assembly and
star formation, followed by quenching and further (but limited)
dry merging.  This scenario seems most naturally
to explain the properties of local E/S0s.  For example,
the final mergers making small spheroidals
would be mostly gas-rich, while later mergers
along the red sequence would be progressively more
gas-poor, as required by the structure sequence.  Furthermore,
the break point between boxy and disky galaxies 
is an upper limit on the masses of blue galaxies that have
migrated onto the red sequence recently.
That break point today is in the range $M_B $ =  --20 to --21, where
boxy and disky galaxies coexist (Faber et al. 1997, Lauer
et al. in progress).  With mean spheroidal ${\cal M}/L_B \sim$ 6 
(from Gebhardt
et al. 2003, adjusted to the $B$-band and $H_0 = 70$), 
this translates to stellar masses
in the range 1-2$ \times 10^{11}$ M$_{\solar}$, or blue
progenitor masses of 0.5-1$ \times 10^{11}$ M$_{\solar}$
for equal-mass mergers.  These are at the upper end of blue masses
today (Bell et al. 2003), as expected.

With one more quite natural assumption, the mixed
scenario might even be able to explain
the narrowness of
the local Fundamental Plane, color-magnitude relation, and Mg-$\sigma$
relations.  
Consider a selection of galaxies at a fixed mass today
on the red sequence.  In the mixed scenario,
these galaxies will have arrived there
via different routes---some will have been produced by
recent gas-rich mergers of two blue galaxies, while
others will have quenched earlier and evolved along
the red sequence via dry mergers for a longer time.
In general, however, we expect there to be a broad
correlation between the mass of a spheroidal galaxy today and
the mass of its latest blue progenitor, massive
spheroidals tending to come from more massive blue galaxies, 
and vice versa.

This ``memory" of progenitor mass then helps to shape
the final metallicity of a galaxy via
the mass-metallicity
relation among star-forming galaxies. 
This relation is strong among nearby galaxies (Tremonti et al. 2004)
and apparently extended well into the past
to beyond $ z = 1$ (Kobulnicky et al. 2003).
If the amount of dry merging on the red sequence is limited,
the original mass-metallicity relation of the 
progenitors will survive to form the {\it backbone} of
the red-sequence scaling relations seen today.  
On the other hand, this backbone will be blurred by
the different star-formation histories of galaxies---galaxies that 
quenched early from low-mass blue progenitors 
and grew later via dry mergers will
have rather low metallicities (reflecting  
their small progenitors), but their average
stellar age will be high (since multiple
dry mergers take time).  In contrast,
galaxies that quenched late and arrived
on the red sequence near their present mass
will have higher metallicities (reflecting 
more massive progenitors), but their
average stellar ages will be younger because they quenched 
recently.  Thus,
the {\it multiplicity of routes} that is inherent in the
mixed scenario might help to 
account naturally for the {\it anti}-correlation between age and metallicity
that is needed to explain the scaling relations.
However, the more dry merging that takes place
along the red sequence, the more the
underlying mass-metallicity correlation of
the blue progenitors will be erased, to be 
replaced by age scatter.  The amount of
scatter in age and $Z$ at fixed mass is
therefore an indicator of the amount of dry merging that could
have occurred,
which might be determined through observations and modeling.

\subsection{Quenching and Downsizing}

This section briefly discusses  
quenching and the related concept of
``downsizing."   To be effective,
quenching requires the removal of
essentially all cold gas within galaxies, and
the prevention of any more falling in. 
The key question is what triggers quenching: why and when
does it happen?  Several processes are
probably involved.  We adopt as a starting point the standard view 
that mergers of gas-rich galaxies can trigger powerful starbursts
(e.g., Mihos \& Hernquist 1994, 1996; Sanders \& Mirabel 1996).  One major
source of gas removal is therefore consumption of gas in the starburst itself.  
The burst also generates internal stellar-driven feedback 
that can remove and/or heat gas, such  as photoionization, 
O-star winds, supernovae, and radiation-driven winds operating
on dust (Murray, Quataert \& Thompson 2005).
An additional source of gas-heating is orbital
energy injected during the merger, which can drive gas
out in a galactic wind (Cox et al. 2005).  Gas can also 
be removed
by external processes such as ram-pressure stripping, tidal stripping, 
and ``harassment" (Moore et al. 1996), which operate
more effectively in the dense environments frequented by  spheroidal
galaxies.  

Despite this abundance of potential gas-removal
mechanisms, it has been suggested that
these alone may not be adequate to keep spheroidal
galaxies gas-freee---the energy 
requirements seem too large, and gas in models continues
to fall in, creating large numbers of
massive blue galaxies that are not seen,
especially at the centers of clusters (e.g., Benson et al. 2003). 
For example, to match the sharp turndown at high masses 
in the galaxy mass function, 
Kauffmann et al. (1999) found it necessary 
in semi-analytic models to truncate gas cooling 
arbitrarily in all halos above a
circular velocity of $V_{circ} = 350$
km s$^{-1}$.  Similar ad hoc
recipes are being tested in 
smaller halos in order to see if they can produce
color bimodality (e.g., R. Somerville et al., in prep.).
Considerable evidence, both theoretical and
empirical, suggests that feedback from AGNs might be the
missing trigger for quenching (e.g., Granato
et al. 2004; Dekel
\& Birnboim 2005; Springel, Di Matteo \& Hernquist 2005),
which is plausible since spheroids are 
precisely those galaxies that possess massive black holes
(Kormendy \& Richstone 1995, Magorrian et al. 1998,
Tremaine et al. 2002).  

Properties of local spheroids may shed further light on 
quenching.  A major clue, as mentioned, is that local galaxies
scatter significantly in age at any location on the red
sequence, which
suggests that galaxies of the same mass have
arrived there via different routes and 
different quenching histories.  
It is also observed that red galaxies
{\it co-exist} with blue galaxies over more than one order
of magnitude in total stellar mass---a
cross-over point exists near $3 \times 10^{10}$ M$_{\solar}$
where the numbers of red and blue galaxies 
are equal (Kauffmann et al. 2003a,b), but
the transition is gradual
(Bell et al. 2003).  Both of these facts 
suggest that the trigger for quenching is {\it not}
simply total stellar mass or any variable closely related to it
(such as luminosity or rotation speed) because any
of those would produce a division
in mass between spheroidals and non-spheroidals
that is too abrupt.
Rather, we seek one (or more) variables that are broadly related
to total mass but with considerable scatter, and perhaps 
also having some relation
to the presence and size of the central black hole,
given the possible need for AGN feedback.  
A natural variable satisfying these requirements is 
{\it stellar spheroid mass}, which increases generally with total
galaxy mass but scatters greatly with respect to it.  Spheroid mass
is also closely linked to the mass of the central black hole (Kormendy
\& Richstone 1995, H\"aring \& Rix 2004), and thus perhaps its
energy output.
Finally, spheroid mass increases discontinuously during a major
merger, and its rise over some threshold might
be the specific trigger for quenching.  This scenario
would meet the requirements of
Balogh et al. (2004), who concluded 
that the blue-to-red transition must be driven primarily by
internal properties rather than by environment.
But environment could play a smaller role
and, through its variation 
from galaxy to galaxy, contribute to  
the scatter seen in stellar ages  
at each point on the scaling relations.

As a final topic,
we consider the matter of ``downsizing."  The basic concept
of downsizing was introduced by Cowie et al. (1996) to explain their
finding that actively star-forming galaxies 
at low redshift are smaller in mass than actively star-forming galaxies at
high redshift, which suggested that star formation  
is stronger at late times in smaller galaxies than large ones.  The essence of
this idea was already latent in the literature.
For example, it was known 
that early-type galaxies are more luminous and more
massive than later-type galaxies (de Vaucouleurs 1977,
Binggeli \& Sandage 1985) 
and that their stellar populations 
are on average older (e.g., Tinsley 1968, Searle, Sargent \& Bagnuolo 1973).
Color and gas fraction were known
to vary systematically along the Hubble sequence
(de Vaucouleurs 1977; Roberts 1969), indicating progressively
slower, less-efficient star formation in later Hubble types.
Finally, the blue end of the Hubble sequence
had been shown explicitly to be a mass sequence 
(van den Bergh 1976, de Vaucouleurs 1977), with low-mass Irr Is
at the bottom having the smallest fraction of stars
and the highest proportion of gas (Roberts 1969).  All
evidence together thus indicated (even then) that
massive galaxies made most of their stars early,
whereas small galaxies
formed theirs relatively later.  
Recent analyses of the star-forming
histories of both local galaxies from SDSS (Heavens et al. 2004)
and distant galaxies from the Gemini Deep Deep survey (Juneau et al. 2005)
have confirmed this basic picture.

Blumenthal et al. (1984) offered a reason 
for this mass-dependent sequence by suggesting 
that early-type galaxies arose from higher-$\sigma$ 
galaxy-sized perturbations
in a cold-dark-matter universe.  Such
perturbations would collapse first and
start making stars early.  Moreover,
because of the non-white nature
of the CDM power spectrum, 
high-$\sigma$ perturbations of galaxy mass 
are embedded preferentially within larger high-$\sigma$ perturbations
(Bardeen et al. 1986), which causes them to merge more 
and eventually wind up in groups and clusters.
The accelerated growth
of high-$\sigma$ perturbations was demonstrated 
in early hydrodynamical simulations
by Cen \& Ostriker (1993), which showed the first
galaxies collapsing 
at the intersections of filaments, forming stars rapidly, 
and assembling later into
groups and clusters.  They identified these 
early-forming objects with E/S0s.\footnote{It 
%Recent observations (notably from
%2DF and SDSS) have gone on to establish the basic character of the
%Hubble sequence as a time sequence in which the rate of star formation 
%correlates strongly with galaxy mass (e.g., Kauffmann et al. ????), 
%as expected if star-formation and mass assembly
%are both driven by the amplitude of early 
%perturbations.
has sometimes been said in the recent literature
that CDM predicts that massive galaxies form late and should therefore
have younger stars, which is opposite to what the Hubble sequence
actually shows.  This remark demonstrates
confusion between the formation of galaxies and their dark-matter
halos.  Massive halos indeed form late, but they are making
clusters of galaxies today, not galaxies.  
Baryonic dissipation has reduced the
collisional cross-sections of galaxies to the point that galaxies 
are merging much
more slowly now than their parent dark-matter halos.
Indeed, it is this reduced merging of galaxies at recent times that 
enables the overall
number density of galaxies, $\phi^*$(All), to remain constant
since $z = 1$, despite the fact that halos are continuing
to merge.  The larger point is that events on the scale
of galaxies are
becoming progressively more decoupled from events on the scale
of halos, including the processes of 
star formation and baryonic mass assembly
(see also Heavens et al. 2004).
But the influence of parent halos (what is 
often termed ``environment") has not yet declined to zero even now.
The halo occupation distribution (HOD)
and related statistics (see review by Cooray \& Sheth 2002) 
are emerging as powerful tools for unravelling the relationship
between galaxies and their dark-matter halos, which 
is crucial to understanding galaxy formation.}

The above concept of downsizing refers to the {\it mean epoch
of star formation}, which was clearly earlier in
massive galaxies than smaller ones.
However, as noted, spheroidal 
galaxies have a second star-formation timescale, 
namely, that of {\it quenching}.
These two timescales might vary 
differently with mass,
thereby generating (in principle) two different kinds of downsizing (or
even {\it up}sizing).  When speaking of downsizing,
it is important to clarify which timescale is meant.

The remainder of this discussion focuses 
on the down(or up)sizing
of {\it quenching}, 
asking whether the typical entry mass 
onto the red sequence has 
increased or decreased with
time.  This amounts to asking whether it is easier (or harder) to keep
galaxies of a given stellar mass free of cold gas
at late epochs.  Three factors suggest that keeping galaxies
gas-free should get
easier with time.  First, gaseous infall generally declines
with time in the Universe,
and gas within galaxies gets converted to
stars; both of these mean that there is less gas overall
that needs to be removed or fended off.  Second, galaxies cluster more and move
more rapidly within clusters, both of which
promote environmentally-driven gas-removal processes
such as stripping and harassment.  
Third, parent dark-matter halos have higher dynamical temperatures
and lower densities
so that any leftover gas outside galaxies is hotter and less likely to cool. 

The sum of these factors suggests
that it is easier to keep galaxies of a given mass gas-free
at late times.  When this is coupled with the observed
fact (not yet fully explained) that quenched
galaxies are larger, we are led to hypothesize
that the typical entry mass onto
the red sequence may be {\it decreasing} with time, such
that progressively smaller galaxies can find their way onto
the red sequence at later epochs.  
We have already mentioned certain observations
that point in this direction,
including the deficit of 
small red-sequence galaxies 
at high redshift (Kodama et al. 2004, Cross et al., 2004,
Giallongo et al. 2005)
followed by possible later infill 
(e.g., Bell et al. 2003; COMBO-17 here), 
the late
arrival of small spheroids on the fundamental
plane
(Treu et al. 2005a,b), the more rapid evolution back in time
of the surface-brightness of smaller spheroids (McIntosh et al. 2005), 
the more rapid evolution back in time of the fundamental plane
zeropoint for smaller galaxies (van der Wel et al. 2005),
and the
possible decrease with time in the crossover mass between
spheroids and disk galaxies (Bundy et al. 2005).  
In aggregate, the evidence may be pointing to a downsizing
of quenching, on top of the downsizing of star formation
in galaxies as a whole.

\subsection{Related Topics}

The finding that the majority of 
spheroidal stellar mass was quenched 
only after $z = 1$ amounts to a paradigm shift
with wide repercussions over a range of issues in
galaxy formation.  This section
lists some important questions that are
raised by the late-quenching picture.

First, does the increase of mass on the red sequence
cause a problem for the observed mass budget of
blue galaxies?  No, because the
increase in red stellar mass is small in absolute terms.  
Red-sequence
galaxies today make up 20\% of the total number of bright
galaxies, and 40\% of the total stellar mass 
(Hogg et al. 2002).  If their stellar
has roughly tripled since $z = 1$,
it would have increased from 0.13 
units to its present level of 0.40 units.
If this increase came at the expense of blue galaxies,
their stellar mass would have declined from 0.87
units to 0.60 units, a fall of only 0.16 dex that if 
translated to $\phi^*$ would be
barely detectable in Figure 7.
Thus, even this extreme scenario in which all new stellar mass
in red galaxies came {\it entirely} from 
the pre-existing stellar mass of blue galaxies
is probably consistent with the data.
But it is more likely that much 
of the new red stellar mass was born via continuing
star-formation in blues after $ z = 1$,
and also in final merger-generated starbursts.
For example, it has been claimed that
50\% of all star formation at $ z = 1$ is occurring
in intermediate-mass LIRGS, many of which are 
mergers and could be the
precursors of spheroidal galaxies (Hammer et al. 2005 and references
therein).  Total stellar mass in all types of
galaxies has also probably increased since $ z = 1$, by between
1.4 and 2.0  (Fontana et al. 2004; Rudnick et al. 2003; Drory et al. 2005;
but see also Bundy et al. 2005).
Either of these increases would be
large enough to maintain blue $\phi^*$ approximately constant
if the new stellar mass were appropriately
distributed over red and blue galaxies.
Thus, the observed constancy of blue number density 
in the face of rising numbers of red galaxies does not seem to be
a problem.

A second point is how galaxy mass functions are predicted to
evolve if the present luminosity function data are correct.
We have already noted that the measured evolution in $M_B^*$ 
for red galaxies is similar to the predicted change in 
their stellar ${\cal M}/L_B$ ratio.  
If these evolutions are identical, the
characteristic mass
${\cal M}^*$ for red galaxies has remained constant.
The same rough equality probably also holds for blue galaxies.
But, since red galaxies are more massive than blue ones, their rise in
number density should cause the
mass function to go up faster at higher masses,
and thus the mass function should change shape.
This does not appear to 
agree with published measurements, 
which show a more even rise over all masses (Drory et al. 2005),
or perhaps even no rise at all (Bundy et al. 2005).  On the other hand,
existing mass functions are measured over small
areas and are subject to large cosmic variance.  Larger samples 
coming soon from DEEP2 (K. Bundy et al., in prep.) and VVDS
may resolve this discrepancy.

A third issue is reconciling the rise in
red-sequence galaxies with the rate of
mergers needed to create them.  Estimated merger rates for
bright galaxies going back to $ z \sim 1$ vary widely in the literature
(see Lin et al. 2004 and references therein).  
Early merger rates from DEEP2 are based on optical pair counts,
not morphologies, and are rather
low: only 9\%  of $L^*$ galaxies are estimated to have suffered a major
merger since $ z = 1.2$  (Lin et al. 2004);
Bundy et al. (2004) obtain similar rates based
on K-band pair counts.  In the present
scenario, merged galaxies are assumed to migrate rapidly 
to the red sequence.
If red galaxies have tripled in number since $z = 1$
and make up 20\% of all galaxies
today (Hogg et al. 2002),
then 2/3 of all red ones---and thus 13\% of all galaxies---must
have merged since $z = 1$.  This fraction
13\% is not far from
the DEEP2 merger fraction of 9\%,
and also does not allow for additional
quenching that may have occured without mergers, such
as in stripped S0s.  In short, the required rate of conversion by
mergers is not excessive and
may even be consistent with the rather low DEEP2 rate.
Higher merger rates could also be accommodated
provided that the extra remnants
wind up as the bulges of spiral galaxies, as is generally
assumed in semi-analytic models (Kauffmann et al. 1996, 
Baugh, Cole \& Frenk 1996, Somerville \& Primack 2001) and
suggested by Hammer et al. (2005) based on the observed 
frequency of LIRGs.

A further puzzle that may now need rethinking
is the existence of non-solar abundance ratios in
early-type galaxies, which display enhanced ratios of 
SNae Type II elements compared to elements generated in Type Ia's 
(e.g., Worthey et al. 1992).  The amount of 
enhancement correlates closely with velocity
dispersion (Kuntschner 1998, Trager et al. 2000b).  It has been
customary to account for these non-solar ratios by appealing
to very rapid 
star formation, which suppresses the iron-peak
elements that are produced more slowly by Type Ia's.
Such early-burst scenarios were natural within the monolithic-collapse
picture.  
However, if most massive ellipticals
were quenched at or after $ z = 1$, they were probably
making stars for at least several Gyr before that, and rapid-burst
models may no longer apply.  Perhaps a correlation existed  
among the {\it precursors} of red galaxies 
between $\sigma$ and average star-formation duration that 
is strong enough to explain the data.
Alternatively, other factors such as galactic wind
strength or IMF variations
(see Trager et al. 2000b) might play a role.

Finally, further work is needed to understand the 
form and scatter of the fundamental
plane, Mg-$\sigma$ relation, and red-sequence
(color-magnitude)
scaling relations.
As noted,
scatter about these relations places strong limits on 
the amount of
stellar merging, which may prove
problematic.  The tilt of the red sequence is also
a mystery: it could come from the underlying
mass-metallicity relation of the blue progenitors or from 
aging during a long series of stellar mergers---both
possibilities have been mentioned here.  In short, the scaling 
relations are sensitive diagnostics.  
We might have attempted
simple estimates of their form and scatter here,
but semi-analytic models seem a much better vehicle for 
such calculations, and we defer these to later papers.

Besides posing difficult questions,
the late-quenching picture also opens up
new opportunities.
If red galaxies indeed emerged 
recently, it becomes feasible to study in detail why 
certain galaxies turn red.  For example, we can
look at the galaxy population just beyond $ z = 1$ and try to predict
which galaxies are about to be quenched.  Given the
close correspondence between red galaxies and dense
environments, it is natural to ask whether galaxies turn
red owing to an increase in the amount of clustering around
them, or whether pre-existing dense environments
suddenly begin to ``spawn" red galaxies near $ z = 1$.
Two studies within DEEP2 are underway
to answer this question (B. Gerke et al., in prep.; M. Cooper et al.,
in prep.), both taking advantage of DEEP2's high 
redshift accuracy, which provides the needed
information on parent groups and clusters.

An important issue going well beyond the confines of this
paper is the impact that the late-quenching picture will have on the
relation between spheroids and black holes.   The masses
of present-day black holes correlate closely with the
properties of their parent spheroids, whether with total
luminosity (Kormendy \& Richstone 1995, H\"aring \& Rix 2004)
or with velocity dispersion (Gebhardt et al. 2000, Ferrarese
\& Merritt 2000).  As long as spheroids were thought
to form early, it was possible to imagine
that the relationship is ancient, with roots going back
to $ z \sim 2$ when spheroids and
black holes were simultaneously
forming (e.g., Richstone et al. 1998).  
However, if most spheroids emerged late, the 
relation could hardly have
existed before $ z \sim 1$, and the massive black holes that were growing 
before that time must somehow have ``known" which spheroids
they would eventually wind up in.  Thus, the late-emergence of
spheroids adds an
important new twist to the black-hole/galaxy co-evolution story.

Although we have clearly come down in favor of 
the late-quenching picture, 
the conclusion is more indirect 
than we would like, resting heavily as it does 
on models for the evolution of 
stellar mass-to-light ratios.
The subject would be on much firmer footing if further
checks could be carried out.  For example, it remains
to be shown whether $K$-band counts and redshift
distributions of near-IR-selected samples
are consistent with the large drop in red
galaxies at $ z \sim 1$ claimed here.  These counts should be reconsidered
using fainter samples with photometric redshifts,
and quenched rather than monolithic-collapse models.
Furthermore, if the red counts 
could be extended just 1.5 magnitudes fainter,
$\phi^*$ could be measured directly at $z = 1$.  
%These data would also set much tighter limits on faint-end slope
%$\alpha$ and constrain the masses of galaxies that are
%migrating to the red sequence as a function of time.
For definitive results, however, 
such data would have to cover $\sim$2 \sq\deg~spaced 
over the sky in several statistically uncorrelated regions.

A related question is whether the emergence of spheroidal
galaxies near $ z = 1$ is consistent with the properties
of red objects earlier than this.
Some studies have searched for red objects 
beyond $ z = 2$ (e.g., van Dokkum, et al. 2003;
Franx et al. 2003), while others have targeted red objects near
$ z = 1.5$ (e.g., Cimatti 2002a).  The results are very different.  
Objects beyond
$ z = 2$ contain old stars but are still star-forming
vigorously; they are red in part 
because they are dusty (F\"orster Schreiber et al. 2004; van Dokkum et al.
2004; Toft 2005).  Although they may be future red-sequence
galaxies, their numbers do not bear on the
question of whether many galaxies quenched
later near $ z = 1$.

In contrast, the number of red galaxies near $ z = 1.5$ is
very relevant.  Despite the fact that a large
fraction of these objects are also dusty
(e.g., Moustakas et al. 2004), a sizeable fraction
also seem to be fully quenched (Longhetti et al. 2005,
Daddi et al. 2005, Saracco et al. 2005).  Number densities
have been variously estimated between 10\% (Daddi et al. 2005)
and 100\% (Saracco et al. 2005) of  
local massive spheroidals, leading different authors 
to conclude that massive Es are, or are not, fully quenched
by $ z = 1$.  However, such surveys
are as yet small and are subject to
large cosmic variance.  Extending this work to
larger areas (and
distingushing dusty galaxies from quenched ones) would
provide the sharpest test of the late-quenching model.

\section{Summary}

The evolution of $B$-band galaxy luminosity functions since 
$z \sim 1$ is determined using a total sample of 39,000
galaxies to $R \sim 24$ mag from the DEEP2 and COMBO-17
surveys.  DEEP2 data come from
Willmer et al. (2005, Paper I), while the COMBO-17 data come originally 
from Wolf et al. (2003) but have been substantially
reworked using using improved photo-z's and new color classes.
Evolution is examined
for blue and red samples separately by dividing galaxies using color
bimodality; this is the first study aside from Willmer (2005)
to compare blue and red galaxies in this way.
Cosmic variance is reduced to 7-15\% 
per redshift bin by combining the results of the surveys.
DEEP2 counts agree remarkably well with COMBO-17 
in all color classes at nearly all
redshifts.  

Luminosity functions of blue and red
galaxies evolve differently with redshift; the blue counts shift to brighter
magnitudes at fixed number density back in time, whereas 
red counts are nearly constant at fixed absolute magnitude.
Both DEEP2 and COMBO-17 agree in this regard.  
Schechter function parameters are fit to the
data assuming non-changing shape (constant $\alpha$), and results
are compared to recent measurements from other 
distant surveys.  Good
agreement is found between DEEP2 and COMBO-17
at all redshifts, and between these 
and other large, recent surveys counting all galaxies.
Results by color are not yet available
from these other surveys.  

Combining the distant Schechter parameters
with local ones, we solve for the fading
over time of characteristic luminosity $M^*_B$ for
All, Red, and Blue galaxies.  All classes fade by nearly the same
amount, showing fadings (per unit redshift) of
1.30$\pm0.20$ mag for red galaxies, 1.31$\pm0.14$ mag
for blue galaxies, and 1.37$\pm0.31$ mag for all galaxies.
In contrast to $M_B^*$, $\phi^*$ evolves differently in different color
classes: formal values for $\phi^*$ 
hold steady for blue galaxies
but rise for red galaxies by 
0.36 dex$\pm$0.09 dex since $ z = 0.8$,
and by 0.56$\pm$0.09 dex since $ z = 1$.
The evolution of luminosity density, $j_B$, also differs with color; 
for blue galaxies it falls by
0.4 dex after $z \sim 1$, while for red galaxies it
remains constant since $z = 0.9$, possibly being
smaller before that. 

The simplest interpretation of these results
is that the number density of blue galaxies has remained 
nearly constant 
since $ z = 1$, whereas the number density of red galaxies
has increased.  The latter conclusion is subjected to close scrutiny,
which is warranted by the fact that {\it most} of the total red evolution 
(in both DEEP2 and
COMBO-17) occurs between the local surveys and our data, and 
in the farthest bin of our data---i.e., in the two places where the
data are weakest.  Although it is possible that 
our formal values of
$\phi^*$ may have unknown errors, we nevertheless 
conclude that substantial evolution
in the number density of red galaxies {\it has} 
occurred, based on strong evidence for
a rise of at least one magnitude in the 
mass-to-light ratios (${\cal M}/L_B$) of red stellar populations
since $ z = 1$.  
When this rise is taken into account, the observed near-constancy of
red luminosity density translates to a rise in overall number
density by at least one magnitude.  A similar argument
applied to the 
red counts at fixed absolute magnitude translates to a rise 
in number {\it at fixed stellar mass} that is also comparable to the 
formal rise 
from $\phi^*$.  Thus, both the new DEEP2 data and the reanalysis
of COMBO-17 together strongly support
the rise in red galaxies since $z \sim 1$ first found in
COMBO-17 by Bell et al. (2004b).  The rise in morphologically
pure E/S0s is even larger if increasing contamination
by non-E/S0s at higher redshifts is allowed for.

The implications of this rise for galaxy formation are examined.
Barring the
existence of  a major, highly obscured and as-yet-unknown
population of galaxies at low redshifts, the immediate
precursors of most modern-day E/S0 galaxies must 
be visible in existing surveys near $ z = 1$.  
The lateness of the rise is inconsistent
with classic, high-redshift single-burst
collapse models for E/S0 formation, which predict
constant numbers of spheroidal galaxies over this epoch.
Instead, it appears that most present-day E/S0s arose from
blue galaxies with ongoing star
formation that were  ``quenched" at or after $z \sim 1$ and then 
migrated
to the red sequence.  The properties of nearby E/S0 galaxies 
support a ``mixed" scenario in which quenched galaxies
enter the red sequence over a wide range of
masses via ``wet," gas-rich mergers, followed 
by a limited number of ``dry," stellar mergers along the
sequence.  The most massive E/S0s are
built up during the last stages of dry merging and are visible
today as boxy, core-dominated ellipticals.  

Some
evidence points to a decline in 
the average entry-mass onto the red sequence
with time, which would amount
to a ``downsizing" of quenching.  This and other processes
might change the shape of the red luminosity
function, violating our assumption throughout of constant
$\alpha$, but the changes are not large enough
to invalidate our major conclusions.
Furthermore, galaxies at a given mass on the red sequence will have
arrived there via different merging and star-formation histories.
Plausible differences among these histories
may account for the 
{\it anti}-correlation that is seen
between age and metallicity residuals on the red sequence
and that is needed to account for the narrowness of
the fundamental plane, Mg-$\sigma$, and color-magnitude relations
of local E/S0s.  

Finally, growing evidence seems to suggest that extra 
feedback from AGN activity might be a key ingredient in
initiating quenching.  Consistent with this is the
fact that quenched stellar populations and massive central
black holes are both uniquely associated with spheroidal
galaxies.  We speculate
that the specific trigger for quenching occurs when
the spheroid stellar mass, and perhaps also the black-hole mass,
exceeds a threshold value during a merger.

\acknowledgments
CNAW thanks G. Galaz, S. Rauzy, M. A. Hendry
and K. D'Mellow for extensive discussions on the measurent of the
luminosity function; E. Bell, J. Brinchmann, A. Gabasch, and G. Galaz for
providing electronic versions of their data;  S. Lilly for 
correspondence on the CFRS luminosity function,
and G. Blumenthal, J. Cohen, and L. Cowie for useful discussions.
SMF thanks R. Somerville, J. Primack, and T. Lauer for
extensive discussions concerning the origin of spheroidal galaxies.
The DEEP team thanks C. Steidel for sharing  unpublished redshift data.
The authors thank the Keck Observatory staff for their constant support
during the several observing runs of DEEP1 and DEEP2; the W. M. 
Keck Foundation and NASA for construction of the Keck telescopes;
and Bev Oke and Judy Cohen for their tireless work on LRIS that 
enabled the spectroscopic observations of DEEP1 galaxies.  
We also wish to recognize and acknowledge the highly significant
cultural role and reverence that the summit of Mauna Kea has always 
had within the indigenous Hawaiian community; it is a privilege to be
given the opportunity to conduct observations from this mountain.

The DEEP1 and DEEP2 surveys were founded under the auspices of the NSF
Center for Particle Astrophysics.  The bulk of the work was supported by   
National Science Foundation grants AST 95-29098 and
00-71198 to UCSC and AST~00-71048 to UCB.  Additional support came from
NASA grants AR-05801.01, AR-06402.01, and
AR-07532.01 from the Space Telescope Science Institute,
which is operated by AURA, Inc., under NASA contract NAS
5-26555.  
The DEIMOS spectrograph was funded by NSF grant ARI92-14621 and by
generous grants from the California Association for Research in
Astronomy, and from UCO/Lick Observatory.
HST imaging of the Groth Strip was planned,
executed, and analyzed by Ed Groth and Jason Rhodes with
support from NASA grants NAS5-1661 and NAG5-6279 from the WFPC1 IDT. 
SMF would like to thank the California
Association for Research in Astronomy for a generous
research grant and the Miller Institute at UC Berkeley for the
support of a Visiting Miller Professorship.  
CW was supported by a PPARC fellowship.
NPV acknowledges support from
NASA grant GO-07883.01-96A and NSF grants
NSF-0349155 from the Career Awards Program and NSF-0123690 via
the ADVANCE Institutional Transformation Program at NMSU.
KG was supported by Hubble Fellowship
grant HF-01090.01-97A awarded by the Association of Universities for
Research in Astronomy, Inc., for NASA under contract NAS5-26555.
JAN acknowledges support from NASA through Hubble Fellowship grant 
HST-HF-01165.01-A awarded by the Space Telescope Science Institute, 
which is operated by the Association of Universities for Research in 
Astronomy, Inc., for NASA, under contract NAS 5-26555.
Computer hardware gifts from Sun Microsystems and Quantum, Inc. are
gratefully acknowledged.  
This research has made use of the NASA/IPAC Extragalactic Database (NED),
which is operated by the Jet Propulsion Laboratory, California
Institute of Technology, under contract with the National Aeronautics
and Space Administration. Finally, we  acknowledge NASA's Astrophysics Data
System Bibliographic Services.

%--------------------------------------------------------------------------
% K corrections
%--------------------------------------------------------------------------

%
% (I, V-I) distributions  uses ~/lf/deep1
%
\begin{figure}
\vspace{170mm}
Figure f1 is available as file 0506044.f1.jpg
\includegraphics{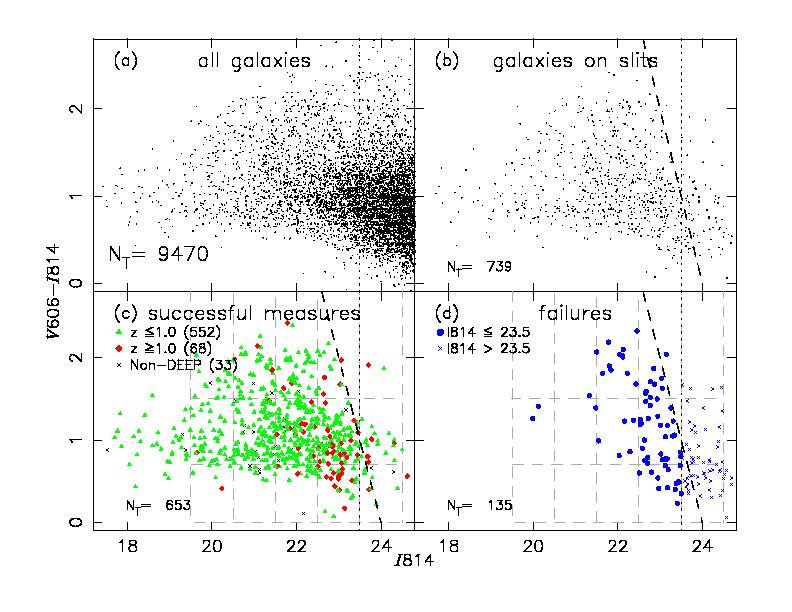}
\caption{Apparent color-magnitude distribution of galaxies in the
  DEEP1 Groth Strip Survey. Panel $a$  shows the full sample, 
  panel $b$ the distribution of galaxies placed on slits,
  and panel $c$ 
  the distribution of successful redshifts, where
  galaxies in the main DEEP1 redshift interval are shown as green
  triangles and galaxies lying beyond
  the upper redshift limit are shown as red
  diamonds.  Galaxies with redshifts coming from Lilly et al. (1995a)
  and Brinchmann et al. (1998)
  are the black crosses. Panel $d$ shows the
  distribution of failed redshifts. 
  The limit $I814 = 23.5$ adopted for the luminosity
  function analysis is shown as the vertical
  dotted line, while the black dashed line represents the limit
  $(V606+I814)/2=24$ actually 
  used to select galaxies for observation. 
  The dashed grey lines show the boundaries
  in color and magnitude used for the redshift histograms in
  Figure 6.
} 
\end{figure}
%
% (I, V-I) distributions  uses ~/lf/deep1
%
\begin{figure}
\vspace{170mm}
\includegraphics{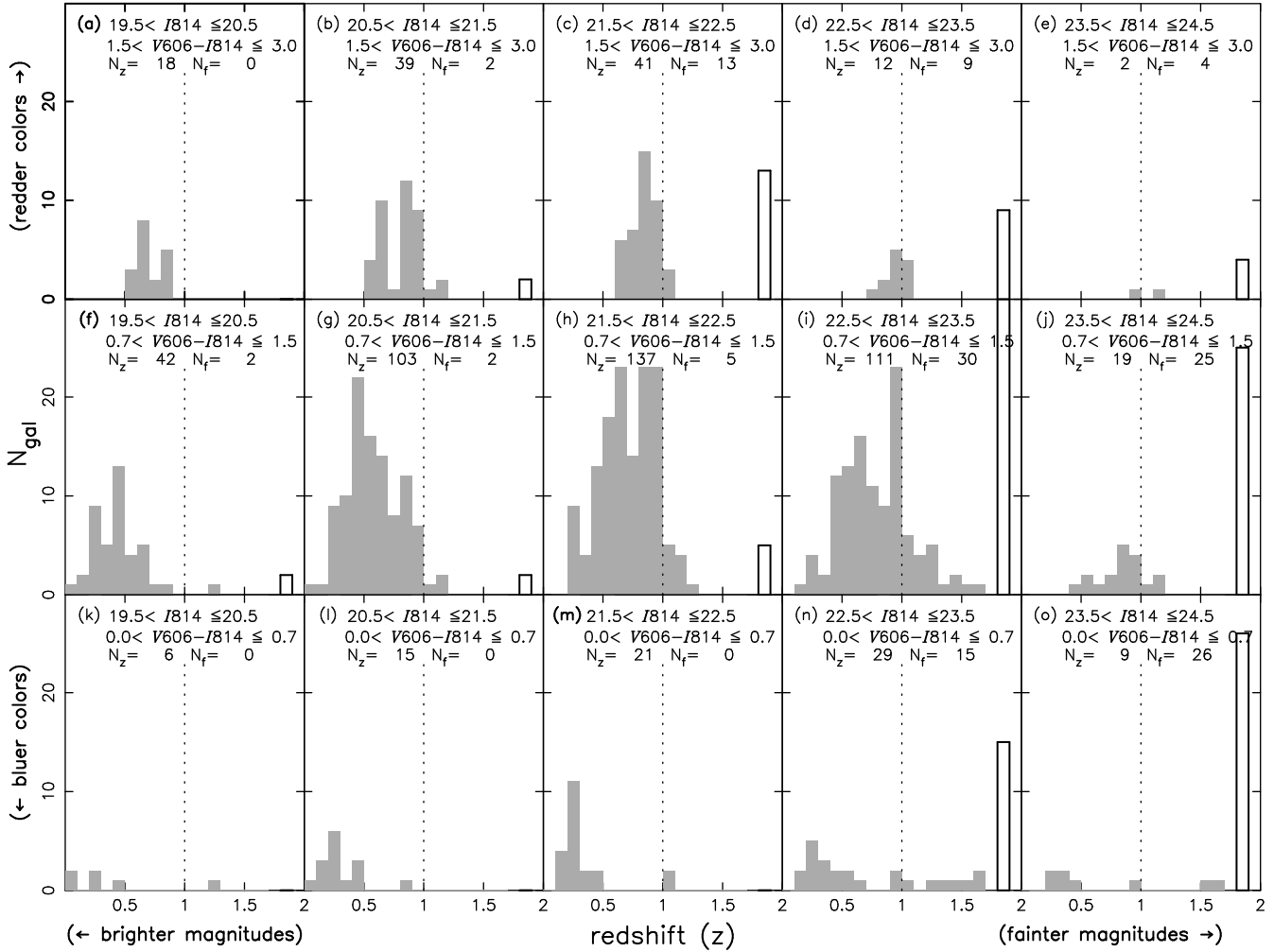}
\caption{Distributions of measured DEEP1 redshifts in the apparent
  color-magnitude bins indicated in Figure 1. The display is such that
  magnitudes become 
  fainter towards the right and colors redder towards the top.
  $N_z$ in each panel is the number of successful redshifts in that bin;
  $N_f$ is the number of attempted galaxies that failed to yield
  successful redshifts.
  The dashed line represents the high-$z$ cut for DEEP1 ($z$=1.0), while the bar
  at the right of each diagram shows the number of failed redshifts. 
  For $I814 \geq 22.5$, the number of
  failures increases significantly, being
  slightly larger for blue galaxies.  
}
\end{figure}
%
% z x (U-B) uses ~/lf/deep1
%
\begin{figure}
\vspace{170mm}
\includegraphics{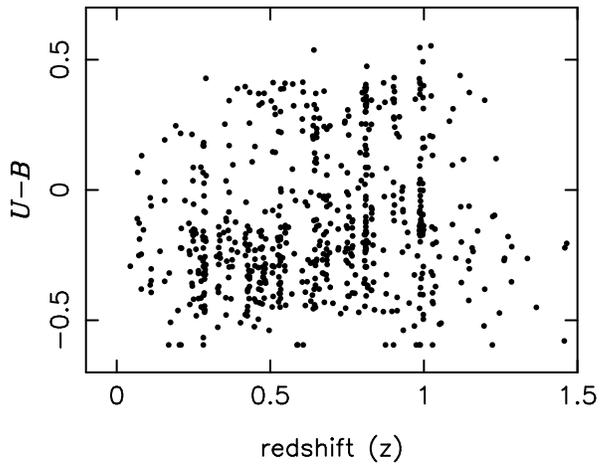}
\caption{Rest-frame $(U-B)$ colors as a function of redshift 
  for DEEP1 galaxies. 
  The bimodal distribution of restframe colors is clearly evident.
  The lack of very low-redshift red galaxies 
  is due to the small volume covered by DEEP1 and the rarity of faint
  red field galaxies.
}
\end{figure}
%
% M_B x (U-B) uses ~/lf/deep1
%
\begin{figure}
\vspace{170mm}
\includegraphics{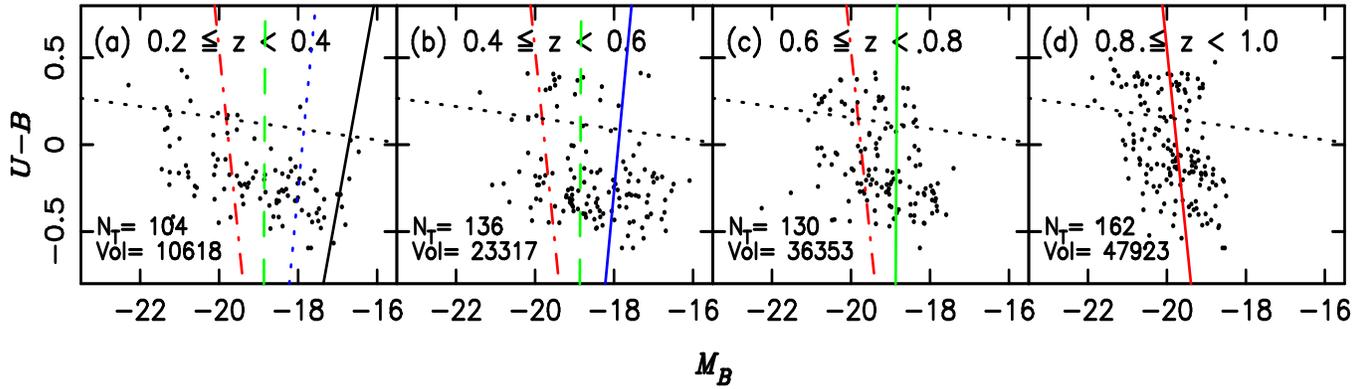}
\caption{Rest-frame color-magnitude diagrams of DEEP1 galaxies
  for four redshift intervals. 
  The solid line in each panel indicates
  the approximate faint absolute magnitude limit as a function of
  intrinsic color and redshift for a sample with a fixed apparent
  magnitude limit at  $I814=23.5$.
  This line represents the faintest galaxy visible as a function of color at
  the upper redshift limit of each panel. The 
  dashed lines repeat these solid lines from other panels. 
  The dotted
  parallel line is the (fixed) cut used to define red-sequence 
  galaxies, calculated in the same manner as for 
  DEEP2 galaxies in Paper I.  The numbers
  at lower left show the number of galaxies plotted and
  the co-moving volume in Mpc$^{3}$ for the ($H_0, \Omega, \Lambda$) =
  (70, 0.3, 0.7) cosmology.
}
\end{figure}
%
%
% DEEP1 sampling rate  uses ~/lf/deep1
%
\begin{figure}
\vspace{160mm}
\includegraphics{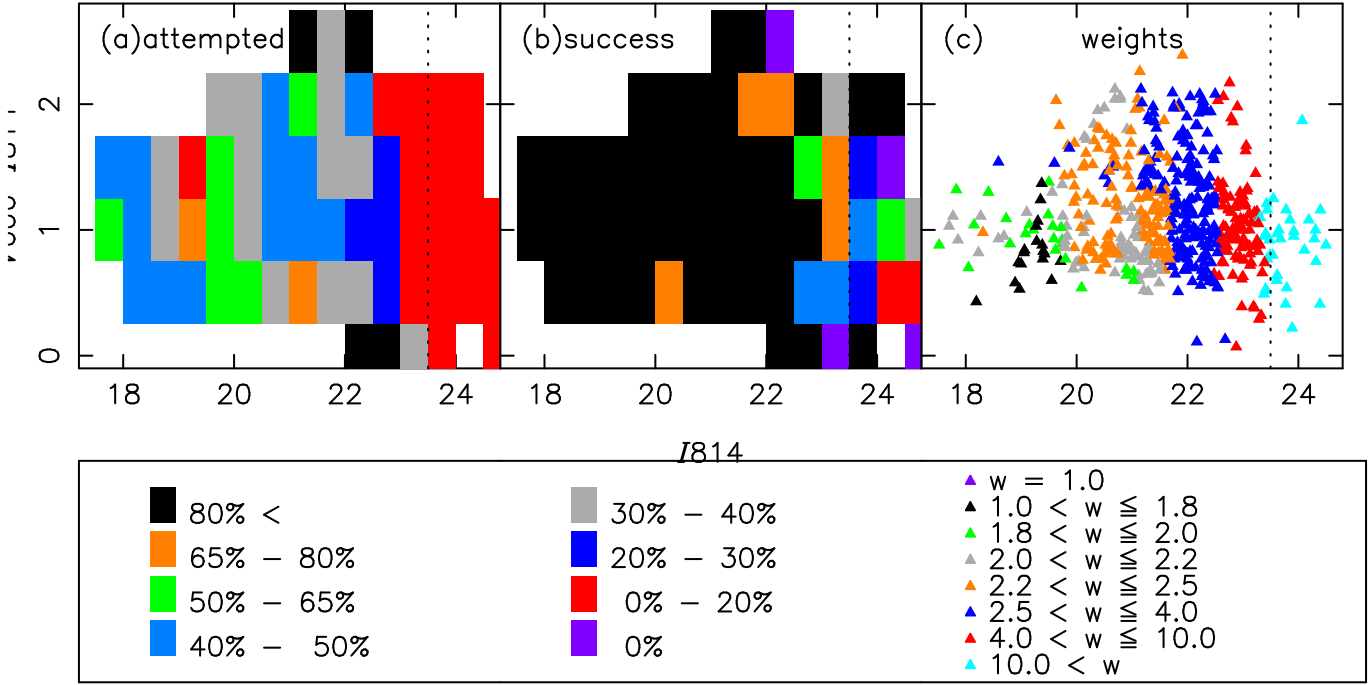}
\caption{ Sampling rate as a function of apparent magnitude and color
  for DEEP1.  The vertical 
  dotted line represents the $I814 =23.5$ magnitude limit.
  Panel $a$ shows the percentage of
  galaxies placed on slits relative to the total sample, panel $b$
  shows the success rate  for obtaining good redshifts among those
  attempted,  and
  panel $c$ shows the weight of each galaxy used to correct for
  incomplete sampling as a function of color and apparent magnitude.
  The weighting scheme for all of DEEP1 uses the minimal model 
  in which all failures
  correspond to galaxies beyond the upper redshift limit 
  of the survey ($z>1.0$).
}
\end{figure}
%
% ldensplot
%
\begin{figure}
\vspace{140mm}
\includegraphics{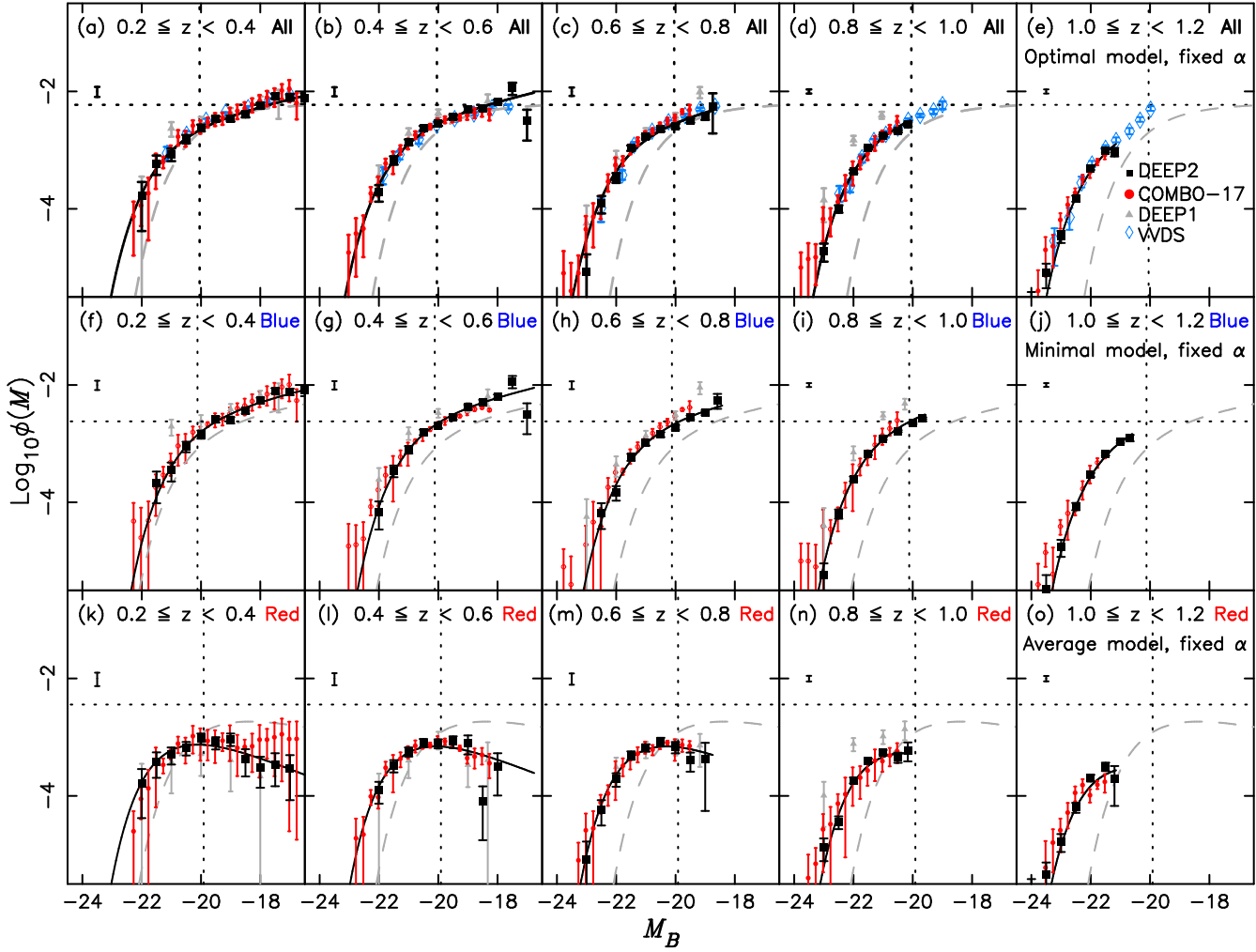}
\caption{Luminosity functions measured in different redshift bins for
  ``All'' galaxies (top row), ``Blue'' galaxies (middle row), 
  and ``Red'' galaxies (bottom row).  Points determined using the
  $1/V_{max}$ 
  method are shown as black squares for DEEP2, grey triangles for DEEP1, and
  red dots for COMBO-17.  COMBO-17 data come from new calculations by
  C. Wolf using the revised bimodality method to separate blue and red galaxies.
  Similar points for All galaxies from VVDS
  (Ilbert et al. 2004) are shown as blue diamonds.  Error bars for DEEP2 
  are 68\% Poisson values only; those for COMBO-17 combine
  Poisson and cosmic variance.  Cosmic variance for DEEP2 is shown at
  the top left of each panel based on theoretical values using 
  the method described in the
  text; cosmic variance for COMBO-17 based on actual field-to-field
  measurements is similar (see Tables 2-4).
  The solid black lines represent Schechter functions
  fitted to  DEEP2 data using the STY method.  For these, 
  $\alpha$'s were kept fixed at the values measured
  from the COMBO-17 ``quasi-local'' sample in bins extending
  from $z = 0.2$ to 0.6.
  The values used are $\alpha = -1.3$ (All), $\alpha = -1.3$ (Blue),
  and $\alpha = -0.5$ (Red).
  Schechter functions for local samples are shown
  as the dashed grey lines, using results from SDSS measurements by
  Bell et al. (2003) as tabulated in Table 5.
  The dotted lines serve as a visual reference
  and are plotted at the values of  $M^*_B$ and $\phi^*$ for the local
  data.
  Overall the agreement between DEEP2, COMBO-17, and VVDS is very
  good where data overlap.
}
\end{figure}
%
% Evolution of M* and phi* as a function of z ; ldensplot
%
\begin{figure}
\vspace{140mm}
\includegraphics{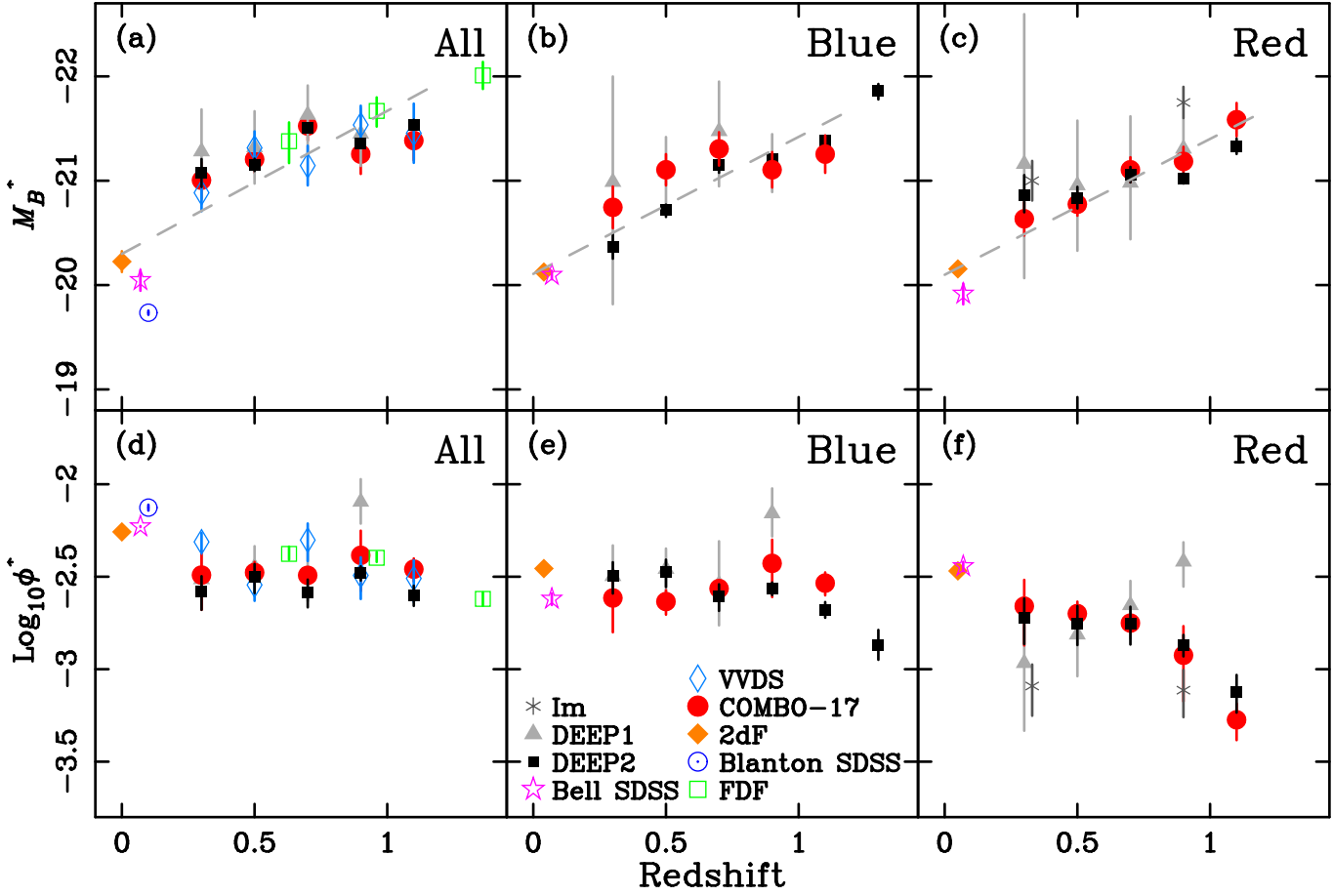}
\caption{Evolution with redshift of the Schechter function parameters
  $M_B^*$ (top panels) and $\phi^*$ (bottom panels) for the All, Blue
  and Red galaxy samples. Data values come from Tables 2-4.
  For DEEP2 and COMBO-17, these parameters were
  calculated keeping the faint-end slope parameter $\alpha$ fixed to
  the ``quasi-local'' COMBO-17 values (--1.3 [All], 
  --1.3 [Blue] and --0.5 [Red]).  Error bars on DEEP2 and COMBO-17
  are Poisson 68\% values for $M_B^*$ and Poisson errors convolved
  in quadrature with cosmic variance for $\phi^*$ (see text).
  Also shown are the Schechter parameters from the other works
  summarized in Table 5.
  Estimates for distant galaxies come from
  Ilbert et al. (2004, VVDS, panels $a$ and $d$) and Gabasch et
  al. (2004, FDF, panels $a$ and $d$).   Previous DEEP1 values
  from Im et al. (2002),
% transformed to the same cosmology as this paper,
  are discussed in the text.
  Local values come from Bell et al. (2003) and Blanton et al. (2003)
  using SDSS and  
  Norberg et al. (2002) and Madgwick et al. (2003) using 2dF. 
  The trend of $M_B^*$ becoming fainter towards lower redshifts is seen
  in all samples. The
  dashed grey lines represent linear  fits to the evolution of $M^*_B$
  as a function of $z$, with coefficients shown in Table 6; 
  $M^*_B$ at $z\sim$ 1 is $\sim$1.3 magnitudes brighter than at
  $z = 0$ for all colors. Changes in galaxy number density 
  ($\phi^*$) vary with color; Blue number densities remain
  roughly constant to $z$ = 1, Red number densities rise strongly with
  time, and the All sample is a blend of the two. 
}
\end{figure}
%
% ldensplot
%
\begin{figure}
\vspace{140mm}
\includegraphics{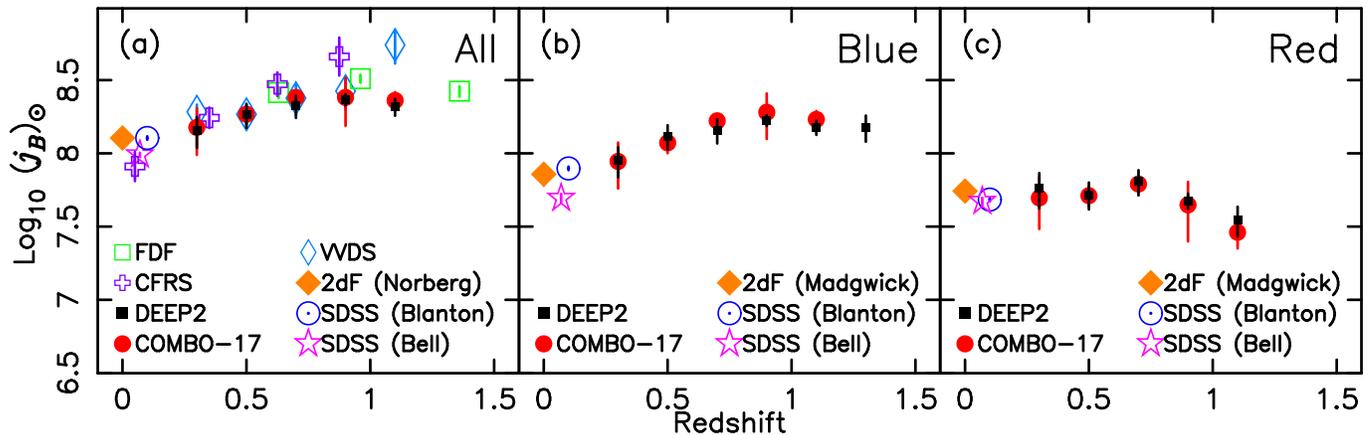}
\caption{
Evolution of the 
co-moving $B$-band luminosity density 
in units of solar luminosities per Mpc$^{-3}$, versus redshift.
Data values come from Tables 2-4.
%Though calculated using Equation 3 over all magnitudes, this quantity is
%conservatively interpreted as referring only to galaxies above
%$L^*_B$, where the data are complete at all redshifts (see text).
References and symbols are the same as in Figure 7 except for CFRS
(blue crosses), which comes from
Lilly et al. (1996).  The blue and red estimates from Blanton et al. (2003) 
use the total luminosity density from that paper, corrected by
a contribution due to red galaxies of 38\% from
Hogg et al. (2003).  The luminosity density of blue galaxies decreases
with time by about a factor of 3 since $ z = 1$, 
while that for red galaxies has remained roughly
constant since $z$ = 0.9, with a possible fall-off before that, in accordance
with Bell et al. (2004a).}
\end{figure}
%
% ldensplot
%
\begin{figure}
\vspace{140mm}
\includegraphics{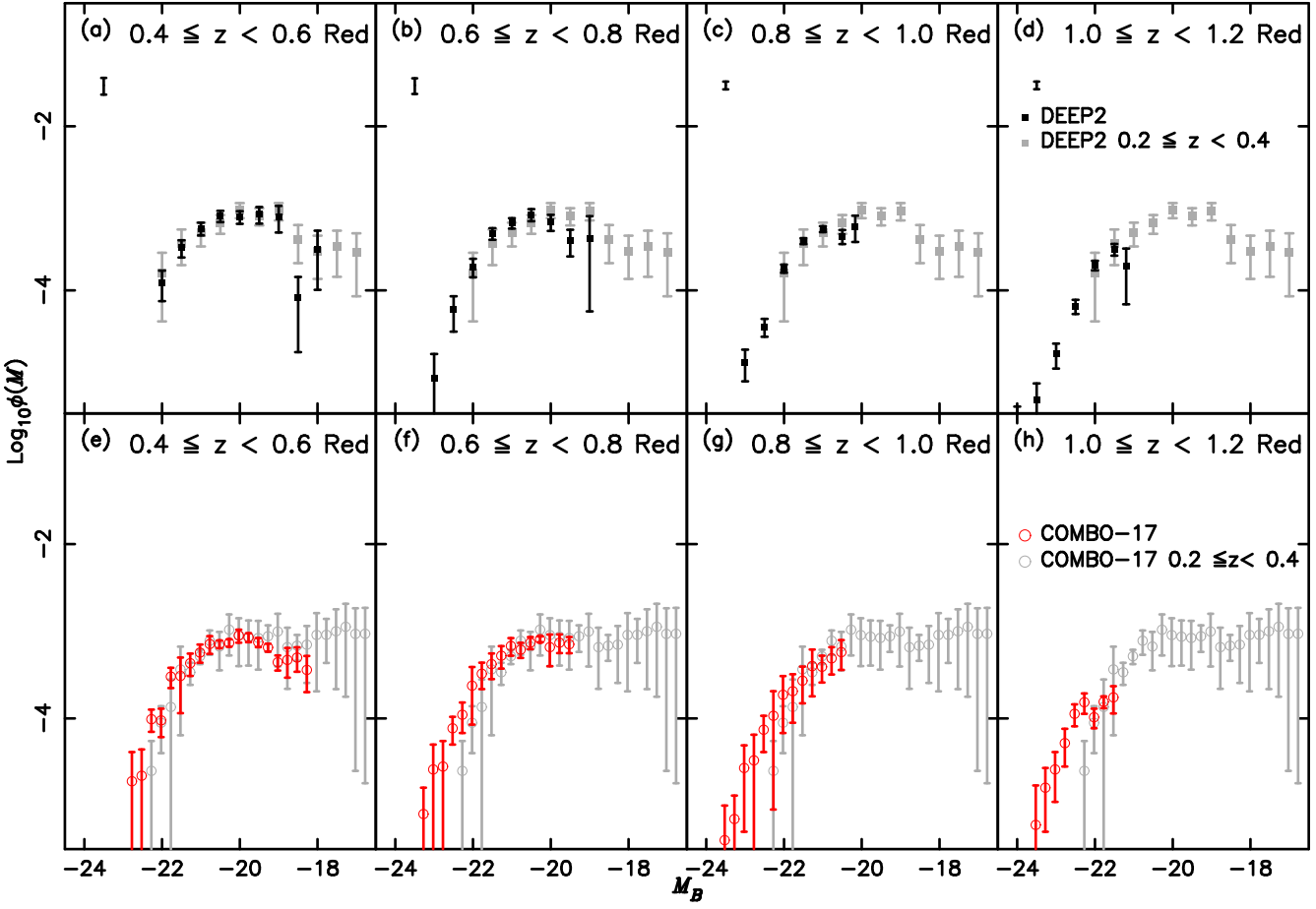}
\caption{
This figure overplots Red counts from the
lowest-redshift bin (grey symbols) on top of counts
at high redshift, for both DEEP2 and COMBO-17.
The purpose is to illustrate how the Red counts
do not evolve very much
internally to each survey.   The
luminosity functions either
stay fixed or translate parallel to themselves
such that there is little change in numbers at constant
absolute magnitude (over the magnitude range probed
by the data).   Both
DEEP2 and COMBO-17 are the same in this regard.
Fitted values of $\phi^*$ are decreasing and
$M_B^*$ are increasing throughout this range; though formally
significant, these values clearly
depend on subtle curvature in the data.
However, stellar mass-to-light ratios are
also increasing and, if taken into
account, lead to the conclusion that
the number density of galaxies
{\it at fixed stellar mass} 
is falling approximately as $\phi^*$.}
\end{figure}
\begin{figure}
\vspace{140mm}

The panels of this figure are available as files 0506044.f10.jpg, 0506044.f11.jpg and 0506044.f12.jpg 
\includegraphics{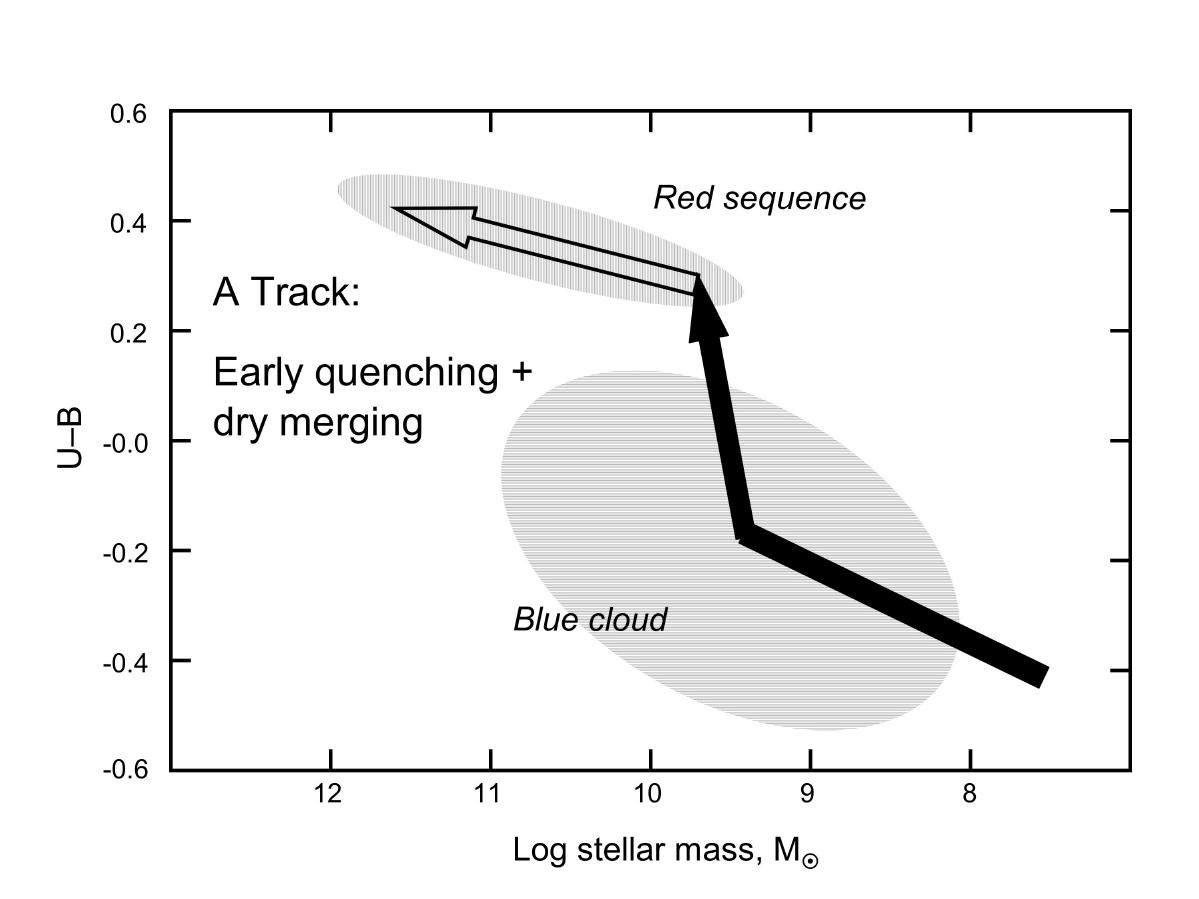}
\includegraphics{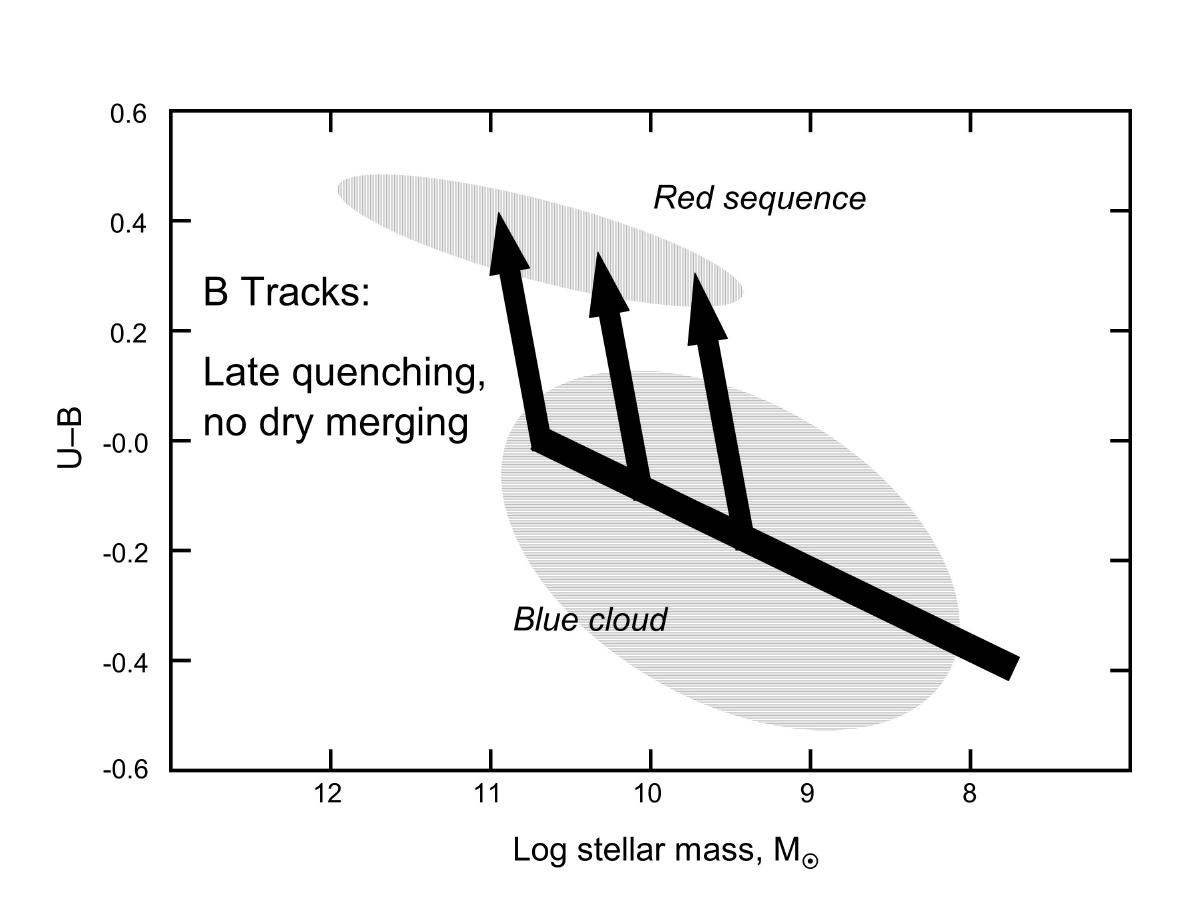}
\includegraphics{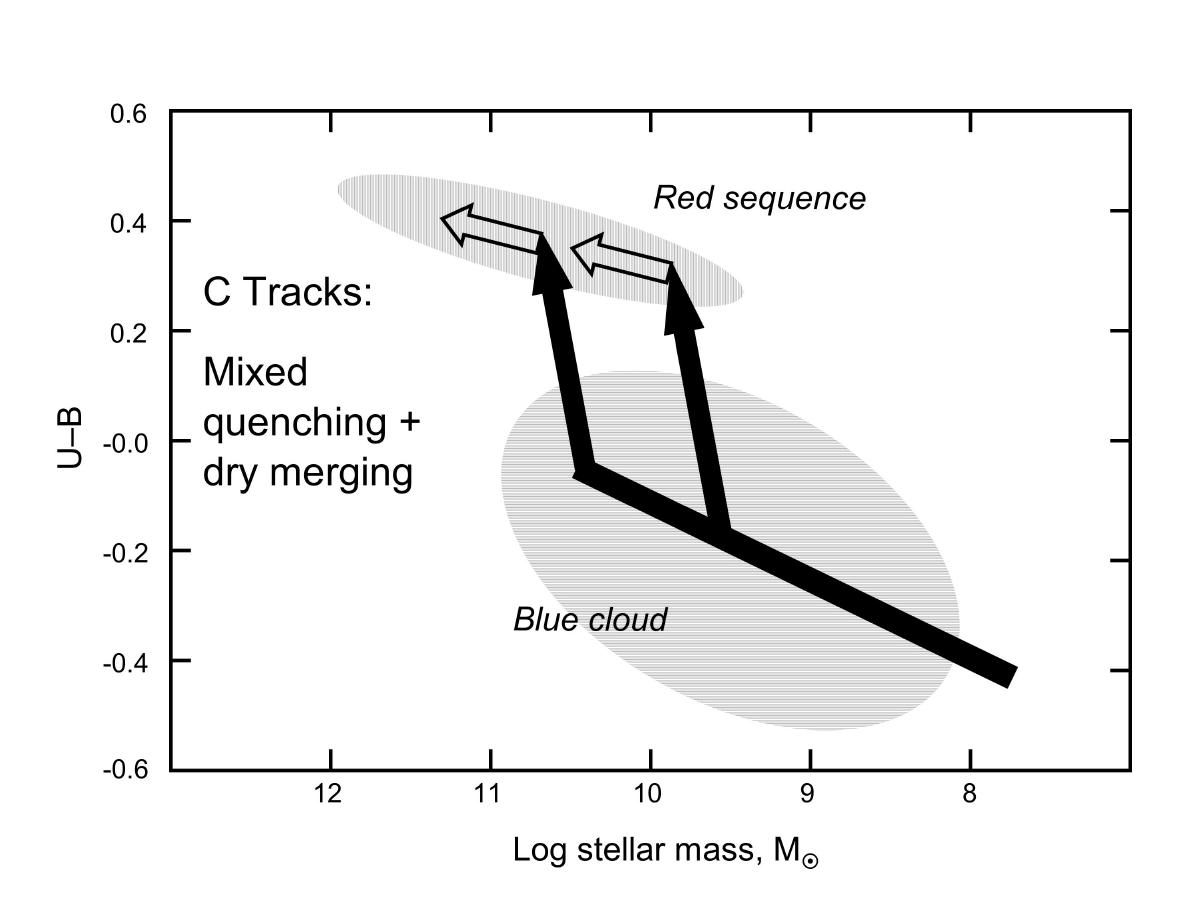}
\caption{
Schematic arrows showing galaxies migrating 
to the red sequence under different formation scenarios.
Evolutionary tracks are plotted in the color-mass diagram.  It
is assumed that red galaxies arise from blue galaxies
when star formation is quenched during a merger.
This merger causes the galaxy to increase in mass   
(mass-doubling in equal-mass mergers is illustrated) while the stellar
population reddens, as shown by the nearly vertical
black arrows.  These mergers would be gas-rich (or ``wet'')
because the merging galaxies are making stars and
hence contain gas.  Once a galaxy arrives on the red sequence,
it may evolve further along it through a series of
gas-poor, or ``dry'' mergers.  These are shown as
the white arrows.  They are tilted upward
to reflect the gradual aging of the stellar
populations during dry merging.  A major question in
red-galaxy formation is the time of mass assembly
versus the time of quenching.  Three schematic
possibilities are shown in the three panels.  Track
A represents very early quenching while the fragments
of the galaxy are still small.  In that case, most
mass assembly occurs in dry mergers along the red sequence.
Track B is extreme in the other sense, having maximally
late quenching.  In that case, galaxies assemble most
of their mass while still blue and then merge
once to become red, with no further dry merging.
Track C is intermediate with a mixture of both
histories.  The properties of local E/S0 galaxies  
favor this ``mixed'' scenario.
}
\end{figure}

\begin{table*}
\caption{Survey Characteristics}
\begin{center}
\begin{tabular}{lrrrrrrrrrl}
\hline
\hline
Survey            &
Area               &
$N_{field}$       &
$N_{gal}$         &
$N_z$             &
$N_z~>~0.8$       &
$m_l$             &
$m_u$             &
$z_{min}$         &
$z_{max}$         &
System            \\
{}                &
$\sq\deg$         &
{}                &
{}                &
{}                &
{}                &
{}                &
{}                &
{}                &
{}                \\
(1)               &
(2)               &
(3)               &
(4)               &
(5)               &
(6)               &
(7)               &
(8)               &
(9)               &
(10)              &
(11)              \\
\noalign{\smallskip} \hline \noalign{\smallskip}
EGS           & 0.28 &1 & 9115  & 4946 & 2026 &18.5 & 24.1 & 0.2 & 1.4 & $R_{AB}$   \\
Fields 2+3+4  & 0.85 &3 & 18756 & 6338 & 4820 &18.5 & 24.1 & 0.8 & 1.4 & $R_{AB}$   \\
DEEP1         & 0.04 &1 & 2438  &  621 &  241 &16.5 & 23.5 & 0.2 & 1.0 & $I814_{Vega}$   \\
COMBO 17      & 0.78 &3 & 40210 & 27947& 8792 &17.0 & 24.0 & 0.2 & 1.2 & $R_{Vega}$ \\
\noalign{\smallskip} \hline
\end{tabular}
\end{center}
\ \ Note. --
The meanings of columns are:  (1) surveyed region;
(2) area in square degrees; 
(3) number of non-contiguous fields in  surveyed region;
(4) number of galaxies in source catalogue;
(5) number of good quality redshifts;
(6) number of good quality redshifts above $z$ = 0.8;
(7) bright apparent magnitude limit;
(8) faint apparent magnitude limit;
(9) lower redshift limit;
(10) upper redshift limit;
(11) apparent magnitude system of catalogue.
\end{table*}

\begin{table*}
\scriptsize
\caption{Schechter function parameters for All galaxy samples                                                             }
\begin{center}
\begin{tabular}{crcrrrrrrrrcl}
\hline
\hline
$\langle z \rangle$ & N$_{gal}$ & $\alpha$ &\multicolumn{3}{c}{$M_B^*$} &\multicolumn{3}{c}{$\phi^*$} &$\sqrt{Var}$&\multicolumn{2}{c}{$j_B$} & Weights \cr
 {} & {} & {} & \multicolumn{3}{c}{ } &\multicolumn{3}{c}{$\times$ 10$^{-4}$ Gal Mpc$^{-3}$} & {} &\multicolumn{2}{c}{$\times$ 10$^{8}$ $L_{\odot}$}&{} \cr
 (1) & (2) & (3) & (4) & (5) & (6) & (7) & (8) & (9) & (10) & (11) & (12) & (13)\cr
\noalign{\smallskip} \hline \noalign{\smallskip}
      0.30 &   734 &  -1.30 &  -21.07 & (+   0.13 & -   0.13) &   26.39 &(+    1.81 & -   1.62) &    0.20 &    1.43 & $\pm$   0.33 & DEEP2 optimal         \cr
      0.50 &   983 &  -1.30 &  -21.15 & (+   0.06 & -   0.06) &   31.39 &(+    0.97 & -   1.04) &    0.18 &    1.83 & $\pm$   0.32 & DEEP2 optimal         \cr
      0.70 &   914 &  -1.30 &  -21.51 & (+   0.03 & -   0.03) &   26.07 &(+    1.39 & -   1.14) &    0.16 &    2.11 & $\pm$   0.34 & DEEP2 optimal         \cr
      0.90 &  2561 &  -1.30 &  -21.36 & (+   0.01 & -   0.02) &   33.04 &(+    0.90 & -   1.11) &    0.08 &    2.33 & $\pm$   0.20 & DEEP2 optimal         \cr
      1.10 &   844 &  -1.30 &  -21.54 & (+   0.04 & -   0.04) &   24.94 &(+    2.20 & -   2.63) &    0.08 &    2.08 & $\pm$   0.27 & DEEP2 optimal         \cr
{} &{} &{} &{} &{} &{} &{} &{} &{} &{} &{} &{} \cr
      0.30 &  6205 &   -1.30 &  -21.00 & (+   0.17 & -   0.17) &   32.26 &(+   11.32 & -  11.32) &   0.11&   1.50 & $\pm$   0.53 & COMBO-17  \cr
      0.50 &  5828 &   -1.30 &  -21.20 & (+   0.13 & -   0.13) &   33.32 &(+    4.73 & -   4.73) &   0.10&   1.85 & $\pm$   0.26 & COMBO-17  \cr
      0.70 &  7122 &   -1.30 &  -21.52 & (+   0.14 & -   0.14) &   32.16 &(+    2.91 & -   2.91) &   0.10&   2.41 & $\pm$   0.22 & COMBO-17  \cr
      0.90 &  5795 &   -1.30 &  -21.25 & (+   0.19 & -   0.19) &   41.26 &(+   14.84 & -  14.84) &   0.09&   2.41 & $\pm$   0.87 & COMBO-17  \cr
      1.10 &  2997 &   -1.30 &  -21.38 & (+   0.19 & -   0.19) &   34.72 &(+    4.97 & -   4.97) &   0.09&   2.30 & $\pm$   0.33 & COMBO-17  \cr
\noalign{\smallskip} \hline
\end{tabular}
\end{center}
\ \ Note. --
The meanings of columns are:  (1) central 
redshift of bin; (2) number of galaxies in bin;
(3) the value of the adopted faint-end
 slope; (4) the value of $M^*_B$, and upper (5) and 
lower (6) 68\% Poisson errors; (7) mean density
 $\phi^*$ followed in columns (8) and (9)
by the 68\% Poisson errors for DEEP2 and
combined Poisson with cosmic variance
estimates for COMBO-17;
(10) square root of cosmic variance, 
based on field geometry, bin volume and galaxy bias 
($b$) as a function of color (see text)
(11) luminosity density,  followed in (12) by a 
conservative error that combines 
Poisson errors in $M^*_B$ and $\phi^*$ 
with cosmic variance in quadrature;
 see text for further explanation
(13) indicates the weighting scheme used.  Optimal weights mean that
 minimal weights were used for blue galaxies and average weights were 
 used for red galaxies.  Both these and the COMBO-17 weights are
 explained in \S3.
\end{table*}

\begin{table*}
\scriptsize
\caption{Schechter function parameters for Blue galaxy samples                                                            }
\begin{center}
\begin{tabular}{crcrrrrrrrrcl}
\hline
\hline
$\langle z \rangle$ & N$_{gal}$ & $\alpha$ &\multicolumn{3}{c}{$M_B^*$} &\multicolumn{3}{c}{$\phi^*$} &$\sqrt{Var}$&\multicolumn{2}{c}{$j_B$} & Weights \cr
 {} & {} & {} & \multicolumn{3}{c}{ } &\multicolumn{3}{c}{$\times$ 10$^{-4}$ Gal Mpc$^{-3}$} & {} &\multicolumn{2}{c}{$\times$ 10$^{8}$ $L_{\odot}$}&{} \cr
 (1) & (2) & (3) & (4) & (5) & (6) & (7) & (8) & (9) & (10) & (11) & (12) & (13)\cr
\noalign{\smallskip} \hline \noalign{\smallskip}
      0.30 &   627 &  -1.30 &  -20.36 & (+   0.13 & -   0.11) &   31.78 &(+    2.15 & -   1.87) &    0.11 &    0.89 & $\pm$   0.20 & DEEP2 minimal         \cr
      0.50 &   812 &  -1.30 &  -20.72 & (+   0.05 & -   0.07) &   33.40 &(+    1.39 & -   1.77) &    0.10 &    1.31 & $\pm$   0.23 & DEEP2 minimal         \cr
      0.70 &   764 &  -1.30 &  -21.15 & (+   0.07 & -   0.07) &   24.67 &(+    1.35 & -   1.58) &    0.09 &    1.44 & $\pm$   0.26 & DEEP2 minimal         \cr
      0.90 &  2644 &  -1.30 &  -21.21 & (+   0.00 & -   0.03) &   27.27 &(+    0.35 & -   0.42) &    0.09 &    1.68 & $\pm$   0.13 & DEEP2 minimal         \cr
      1.10 &  1224 &  -1.30 &  -21.38 & (+   0.04 & -   0.05) &   20.84 &(+    1.08 & -   1.58) &    0.09 &    1.50 & $\pm$   0.16 & DEEP2 minimal         \cr
      1.30 &   448 &  -1.30 &  -21.86 & (+   0.07 & -   0.08) &   13.44 &(+    2.00 & -   2.71) &    0.07 &    1.51 & $\pm$   0.31 & DEEP2 minimal         \cr
{} &{} &{} &{} &{} &{} &{} &{} &{} &{} &{} &{} \cr
      0.30 &  5109 &   -1.30 &  -20.74 & (+   0.20 & -   0.20) &   24.26 &(+    8.42 & -   8.42) &   0.11&   0.88 & $\pm$   0.31 & COMBO-17  \cr
      0.50 &  4649 &   -1.30 &  -21.10 & (+   0.15 & -   0.15) &   23.20 &(+    3.48 & -   3.48) &   0.10&   1.18 & $\pm$   0.18 & COMBO-17  \cr
      0.70 &  5691 &   -1.30 &  -21.30 & (+   0.16 & -   0.16) &   27.27 &(+    3.12 & -   3.12) &   0.09&   1.67 & $\pm$   0.19 & COMBO-17  \cr
      0.90 &  4903 &   -1.30 &  -21.10 & (+   0.17 & -   0.17) &   37.32 &(+   12.80 & -  12.80) &   0.09&   1.91 & $\pm$   0.65 & COMBO-17  \cr
      1.10 &  2741 &   -1.30 &  -21.25 & (+   0.18 & -   0.18) &   29.19 &(+    4.14 & -   4.14) &   0.09&   1.71 & $\pm$   0.24 & COMBO-17  \cr
\noalign{\smallskip} \hline
\end{tabular}
\end{center}
\ \ Note. --
The meanings of columns are:  (1) central 
redshift of bin; (2) number of galaxies in bin;
(3) the value of the adopted faint-end
 slope; (4) the value of $M^*_B$, and upper (5) and 
lower (6) 68\% Poisson errors; (7) mean density
 $\phi^*$ followed in columns (8) and (9)
by the 68\% Poisson errors for DEEP2 and
combined Poisson with cosmic variance
estimates for COMBO-17;
(10) square root of cosmic variance, 
based on field geometry, bin volume and galaxy bias 
($b$) as a function of color (see text)
(11) luminosity density,  followed in (12) by a 
conservative error that combines 
Poisson errors in $M^*_B$ and $\phi^*$ 
with cosmic variance in quadrature;
 see text for further explanation
(13) indicates the weighting scheme used.  Minimal weights were used for
 blue galaxies and mean that all failed redshifts were assumed to lie
 beyond the upper limit of the survey, $zh=1.4$ for DEEP2.  Both these and
 the COMBO-17 weights are explained further in \S3.
\end{table*}

\begin{table*}
\scriptsize
\caption{Schechter function parameters for Red galaxy samples                                                             }
\begin{center}
\begin{tabular}{crcrrrrrrrrcl}
\hline
\hline
$\langle z \rangle$ & N$_{gal}$ & $\alpha$ &\multicolumn{3}{c}{$M_B^*$} &\multicolumn{3}{c}{$\phi^*$} &$\sqrt{Var}$&\multicolumn{2}{c}{$j_B$} & Weights \cr
 {} & {} & {} & \multicolumn{3}{c}{ } &\multicolumn{3}{c}{$\times$ 10$^{-4}$ Gal Mpc$^{-3}$} & {} &\multicolumn{2}{c}{$\times$ 10$^{8}$ $L_{\odot}$}&{} \cr
 (1) & (2) & (3) & (4) & (5) & (6) & (7) & (8) & (9) & (10) & (11) & (12) & (13)\cr
\noalign{\smallskip} \hline \noalign{\smallskip}
      0.30 &   109 &  -0.50 &  -20.86 & (+   0.16 & -   0.17) &   18.89 &(+    1.89 & -   1.85) &    0.15 &    0.58 & $\pm$   0.18 & DEEP2 average         \cr
      0.50 &   173 &  -0.50 &  -20.83 & (+   0.12 & -   0.09) &   17.71 &(+    1.03 & -   1.13) &    0.14 &    0.52 & $\pm$   0.13 & DEEP2 average         \cr
      0.70 &   196 &  -0.50 &  -21.05 & (+   0.06 & -   0.06) &   17.63 &(+    1.29 & -   1.50) &    0.13 &    0.64 & $\pm$   0.15 & DEEP2 average         \cr
      0.90 &   535 &  -0.50 &  -21.02 & (+   0.04 & -   0.02) &   13.47 &(+    0.60 & -   0.82) &    0.12 &    0.47 & $\pm$   0.06 & DEEP2 average         \cr
      1.10 &   178 &  -0.50 &  -21.33 & (+   0.08 & -   0.07) &    7.51 &(+    1.31 & -   1.52) &    0.12 &    0.35 & $\pm$   0.08 & DEEP2 average         \cr
{} &{} &{} &{} &{} &{} &{} &{} &{} &{} &{} &{} \cr
      0.30 &  1096 &   -0.50 &  -20.63 & (+   0.16 & -   0.16) &   21.91 &(+    8.48 & -   8.48) &   0.15&   0.50 & $\pm$   0.19 & COMBO-17  \cr
      0.50 &  1179 &   -0.50 &  -20.77 & (+   0.11 & -   0.11) &   19.97 &(+    3.21 & -   3.21) &   0.14&   0.51 & $\pm$   0.08 & COMBO-17  \cr
      0.70 &  1431 &   -0.50 &  -21.10 & (+   0.12 & -   0.12) &   17.75 &(+    0.70 & -   0.70) &   0.13&   0.62 & $\pm$   0.02 & COMBO-17  \cr
      0.90 &   892 &   -0.50 &  -21.18 & (+   0.14 & -   0.14) &   11.89 &(+    5.20 & -   5.20) &   0.12&   0.45 & $\pm$   0.19 & COMBO-17  \cr
      1.10 &   256 &   -0.50 &  -21.58 & (+   0.16 & -   0.16) &    5.32 &(+    1.19 & -   1.19) &   0.12&   0.29 & $\pm$   0.07 & COMBO-17  \cr
\noalign{\smallskip} \hline
\end{tabular}
\end{center}
\ \ Note. --
The meanings of columns are:  (1) central 
redshift of bin; (2) number of galaxies in bin;
(3) the value of the adopted faint-end
 slope; (4) the value of $M^*_B$, and upper (5) and 
lower (6) 68\% Poisson errors; (7) mean density
 $\phi^*$ followed in columns (8) and (9)
by the 68\% Poisson errors for DEEP2 and
combined Poisson with cosmic variance
estimates for COMBO-17;
(10) square root of cosmic variance, 
based on field geometry, bin volume and galaxy bias 
($b$) as a function of color (see text)
(11) luminosity density,  followed in (12) by a 
conservative error that combines 
Poisson errors in $M^*_B$ and $\phi^*$ 
with cosmic variance in quadrature;
 see text for further explanation
(13) indicates the weighting scheme used.  Average  weights were used for
 red galaxies and mean that all failed redshifts were assumed to follow
the same redshift distribution as successfully observed galaxies.
Both these and the COMBO-17 weights are explained further in \S3.
\end{table*}

\begin{table*}
\scriptsize
\caption{Schechter function parameters from the literature}
\begin{center}
\begin{tabular}{lcccccccccll}
\hline
\hline
Sample & $z_{med}$ & $\alpha$ & \multicolumn{3}{c}{$M_B^*$}
&\multicolumn{3}{c} {$\phi^*$} & $j_B$ & Reference & Notes\cr
{} & { }    &\multicolumn{3}{c}{} & & \multicolumn{3}{c}{$\times$ 10$^{-4}$ Gal Mpc$^{-3}$} &$\times$
10$^{8}$ $L_{\odot}$ & {} & {}\cr
 (1) & (2) & (3) & (4) & (5) & (6) & (7) & (8) & (9) & (10) &(11) &(12) \cr
\noalign{\smallskip} \hline \noalign{\smallskip}
A &0.07 &-1.21 $\pm$ 0.03 & -20.22 &(+0.10 &-0.10) &55.22&(+ 4.46& -4.46)& 1.27$\pm$0.12 &1 &\tablenotemark{a}\cr
A &0.07 &-1.03 $\pm$ 0.10 & -20.04 &(+0.10 &-0.10) &59.00&(+ 0.34&- 0.34)& 0.98$\pm$0.03 &2&\tablenotemark{b}\cr
A &0.10 &-0.89 $\pm$ 0.01 & -19.73 &(+0.02 &-0.02) &74.77&(+ 2.74&- 2.74)& 1.27$\pm$0.04 &3& {}\cr
A &0.30 &-1.16 $\pm$ 0.03 & -20.88 &(+0.18 &-0.18) &48.74&(+ 5.87&- 5.87)& 1.92$\pm$0.23 &4&\tablenotemark{c}\cr
A &0.50 &-1.26 $\pm$ 0.05 & -21.31 &(+0.16 &-0.16) &28.54&(+ 5.14&- 5.14)& 1.85$\pm$0.33 &4&\tablenotemark{c}\cr
A &0.70 &-1.10 $\pm$ 0.11 & -21.14 &(+0.19 &-0.19) &49.87&(+11.46&-11.46)& 2.38$\pm$0.55 &4&\tablenotemark{c}\cr
A &0.90 &-1.30 $\pm$ 0.11 & -21.53 &(+0.18 &-0.18) &32.10&(+ 8.16&- 8.16)& 2.66$\pm$0.68 &4&\tablenotemark{c}\cr
A &1.10 &-1.70 $\pm$ 0.17 & -21.45 &(+0.28 &-0.28) &30.94&(+ 7.85&- 7.85)& 5.50$\pm$1.40 &4&\tablenotemark{c}\cr
A &0.63 &-1.25 $\pm$ 0.17 & -21.38 &(+0.21 &-0.18) &42.00&(+ 4.00&- 3.00)& 2.60$\pm$0.22 &5& {} \cr
A &0.96 &-1.25 $\pm$ 0.17 & -21.67 &(+0.15 &-0.13) &40.00&(+ 3.00&- 2.00)& 3.24$\pm$0.20 &5& {} \cr
A &1.36 &-1.25 $\pm$ 0.17 & -22.01 &(+0.13 &-0.13) &24.00&(+ 2.00&- 2.00)& 2.66$\pm$0.22 &5& {} \cr
 {} & {} & {} & {} & {} & {} & {} & {} & {} & {} &{} &{} \cr
B &0.04 &-1.24 $\pm$ 0.01 & -20.12 &(+0.05 &-0.05) &34.99&(+ 0.69&- 0.69)& 0.72$\pm$0.12 &6&\tablenotemark{a}\cr
B &0.07 &-1.24 $\pm$ 0.10 & -20.09 &(+0.04 &-0.04) &23.96&(+ 1.73&- 1.73)& 0.49$\pm$0.04 &2&\tablenotemark{d}\cr
 {} & {} & {} & {} & {} & {} & {} & {} & {} & {} &{} &{} \cr
R &0.04 &-0.54 $\pm$ 0.02 & -20.15 &(+0.05 &-0.05) &33.96&(+ 1.72&- 1.72)& 0.55$\pm$0.04 &6&\tablenotemark{a}\cr
R &0.07 &-0.76 $\pm$ 0.10 & -19.91 &(+0.10 &-0.10) &35.96&(+ 2.06&- 2.06)& 0.47$\pm$0.03 &2&\tablenotemark{e}\cr
R &0.33 &-1.00            & -21.00 &(+0.19 &-0.19) & 7.10&(+ 2.20&- 2.20)&   \nodata &7&\tablenotemark{f}\cr
R &0.90 &-1.00            & -21.75 &(+0.15 &-0.15) & 6.80&(+ 1.90&- 1.90)&   \nodata &7&\tablenotemark{f}\cr
\noalign{\smallskip} \hline
\end{tabular}
\end{center}
The meanings of the columns are: (1) Sample : All galaxies (A),
Blue (B) or Red (R);
(2) median redshift of sample; (3) Schechter function
faint-end slope and error; (4), (5) and (6) characteristic luminosity and
errors ; (7), (8) and (9) Schechter function normalization and errors;
(10) luminosity density and error; (11) reference; (12) additional notes.
All quantities are converted to the ($H_0, \Omega_M, \Omega_\Lambda)$
cosmology of this paper and are calculated for $B_{Johnson}$ using the
transformations listed in the individual footnotes. Those involving
SDSS magnitudes and colors come from B04, using average colors
(g-r) =(0.82, 0.69, 1.01) for (All, Blue, Red) galaxies respectively.

References: (1) Norberg et al. 2002 ; (2) Bell et al. 2003; (3) Blanton
  et al. 2003.; (4) Ilbert et al. 2004; (5) Gabasch et al. 2004 ; (6)
  Madgwick et al. 2002; (7) Im et al. 2002.

$^a$ $B_{Johnson} = b_j$ + 0.21.

$^b$ $B_{Johnson} = g$ + 0.115 + 0.370 $\times$ 0.82.

$^c$ $\phi^*$ inferred from fits to VVDS data points fixing $\alpha$ and $M_B^*$.

$^d$ $B_{Johnson} = g$ + 0.115 + 0.370 $\times$ 0.69.

$^e$ $B_{Johnson} = g$ + 0.115 + 0.370 $\times$ 1.01.

$^f$ Fixed $\alpha$.
\end{table*}

\begin{deluxetable}{lrrrr}
\small
\tablewidth{0pc}
\tablecaption{Evolution of $M^*$ from SDSS, 2dF, COMBO-17 and DEEP2}
\tablehead{
\colhead{sample}         &
\colhead{$M^*$($z$=0.6)}   &
\colhead{$\pm$}   &
\colhead{$Q$}     & 
\colhead{$\pm$}   
}
\startdata
All  & -21.12 & 0.08 & 1.37 & 0.31 \\ 
Blue & -20.89 & 0.05 & 1.31 & 0.14 \\
Red  & -20.87 & 0.05 & 1.30 & 0.20 \\
%moderately blue & -20.66 & 0.08 & 0.80 & 0.07 \\
%very blue       & -19.99 & 0.09 & 0.88 & 0.10 \\
\enddata
\tablecomments{
The equation fitted is $M^*(z-0.6)~=~M^*(z-0.6)~+~Q(z-0.6)$.
The fits are obtained weighting each value of $M^*$ 
by its error, $\delta M^*$. The midpoint of the $z$-range was chosen
as the horizontal zero-point to minimize correlated errors between
$M^*$ and $Q$.}
\end{deluxetable}

\end{document}